\pdfoutput=1
\documentclass[11pt]{article}

\usepackage[margin=1in]{geometry}
\usepackage[T1]{fontenc}
\usepackage{lmodern}
\usepackage{microtype}
\usepackage{amsmath,amssymb,amsthm,mathtools}
\usepackage{booktabs}
\usepackage{array}
\usepackage{enumitem}
\usepackage{xcolor}
\usepackage{tikz}
\usetikzlibrary{arrows.meta,positioning,fit,calc,backgrounds,decorations.pathreplacing,shapes.geometric}
\usepackage{pgfplots}
\pgfplotsset{compat=1.18}
\usepackage{subcaption}
\usepackage[hidelinks]{hyperref}
\hypersetup{pdfauthor={Tomer Solberg},pdftitle={Sable: A Small-Field Sumcheck Argument for Bit-Exact Inference of Large Language Models}}
\usepackage[capitalise,noabbrev]{cleveref}

\setlist[itemize]{itemsep=0.2em,topsep=0.3em}
\setlist[enumerate]{itemsep=0.2em,topsep=0.3em}

\newcommand{\F}{\mathbb{F}}
\newcommand{\Z}{\mathbb{Z}}
\newcommand{\Fp}{\F_p}
\newcommand{\Fq}{\F_{p^4}}
\newcommand{\Fo}{\F_{p^8}}
\newcommand{\bits}{\{0,1\}}
\newcommand{\eq}{\mathrm{eq}}
\newcommand{\mle}[1]{\widetilde{#1}}

\newcommand{\Tab}{\mathcal{T}}
\newcommand{\FO}{F_{\mathsf{O}}}
\newcommand{\Hash}{\mathsf{H}}
\newcommand{\Sha}{\mathsf{SHA256}}

\newcommand{\gad}[1]{\mathsf{#1}}

\DeclareMathOperator{\clip}{clip}

\theoremstyle{plain}
\newtheorem{theorem}{Theorem}[section]
\newtheorem{lemma}[theorem]{Lemma}

\theoremstyle{definition}
\newtheorem{definition}[theorem]{Definition}
\newtheorem{remark}[theorem]{Remark}

\title{Sable: A Small-Field Sumcheck Argument\\for Bit-Exact Inference of Large Language Models}
\author{Tomer Solberg\\[0.3em]
\normalsize MoonMath.ai\\
\normalsize\texttt{research@moonmath.ai}}
\date{}

\begin{document}
\maketitle

\begin{abstract}
\noindent
We present Sable, an argument system and single-GPU prover for the statement that a committed, integer-quantized transformer produced a given token sequence from a given prompt, bit for bit.
Sable fixes an integer function \(\FO\) without any floating-point step, in which every rounding, normalization, softmax and sampling decision is determined by its inputs, and proves it with a constraint system whose relations provably never wrap modulo the 31-bit KoalaBear prime.
Within one \emph{domain}, the proof batches zerochecks and the leaf layer of LogUp-GKR lookups into a single sumcheck.
Lookups are split into chunks of fewer than \(p\) lookups, which LogUp needs for soundness in a prime field.
Every witness tensor and weight lives in one multilinear commitment, built from a rate-\(1/2\) Reed--Solomon code, BLAKE3 Merkle trees of mixed height, and a mixed-length BaseFold opening over the degree-8 extension.
The prover streams the model layer by layer in two passes, recomputes the witness where it would otherwise store it, and runs on one GPU with a device-resident Fiat--Shamir transcript and CUDA-graph replay of identical layers.
On one NVIDIA RTX PRO 6000 Blackwell GPU, Sable proves all 64 layers of Qwen2.5-32B in 92.5\,s for one sequence of 4{,}096 tokens, 253\,s for four such sequences, and 398\,s for one sequence of 16{,}384 tokens.
The same prover covers Qwen3-32B, which adds QK-norm, and the mixture-of-experts Qwen3-30B-A3B, proving 4{,}096 tokens in 121.8 and 84.7\,s.
A proof takes 41 to 50 times as long as computing the same tokens in our serving engine for \(\FO\), or 6.4 cents for a 4{,}096-token request at public GPU prices, and 40 to 64 times less time per parameter and token than zkLLM.
The price of exactness is accuracy: on WikiText-2, \(\FO\)'s greedy token matches the BF16 model's at 86.3\% of positions and its perplexity is 8.3\% higher; we trace about a third of this gap to the 8-bit queries and keys. On Qwen3, choosing per key-value group how the score exponent is split between queries and keys, which costs nothing in the proof, removes 53\% of the perplexity gap of Qwen3-30B-A3B and 23\% of that of Qwen3-32B, and on both models \(\FO\) is within 1.6 points of BF16 on MMLU and 1.7 points on GSM8K.
We reduce the work of the verifier, which a recursive compression must arithmetize, from \(17.8\) to \(4.2\) million BLAKE3 compressions for a 4{,}096-token proof of the full model, and a verifier with a compiled core checks the proofs of all three models on one CPU core in \(19\) to \(56\)\,s.
We also give a BLAKE3 circuit in the same proof system, whose constraint proof costs \(2.8\,\mu\)s per compression.
In the random-oracle model and with proven proximity bounds, the interactive proof has round-by-round soundness error \(2^{-103}\), counting a 30-bit proof of work on the queries.
The proof is not zero-knowledge, so the system as built is meant for open-weight models and for verifiers, such as auditors, that can afford a proof of 106 to 477\,MiB.
The prover, the verifier and the scripts behind every measurement are available under the MIT license at \url{https://github.com/moonmath-ai/Sable}.
\end{abstract}

\tableofcontents
\newpage

\section{Introduction}
\label{sec:intro}

A client that sends a prompt to a hosted language model receives text and nothing else.
It cannot tell which model produced the text, whether the weights were quantized more aggressively than advertised, whether the answer was cut short, or whether the sampling decisions were made with the randomness it was promised.
\emph{Verifiable inference} asks the server to attach evidence that the output \(y\) equals \(F(W, x, r)\) for the prompt \(x\), a sampling seed \(r\) the server did not choose, and weights \(W\) fixed by a public commitment \(C_W\).
The evidence can come from trusted hardware, from re-execution by other parties, from statistical fingerprints of the computation, or from a cryptographic proof.
Only the last gives a guarantee that rests on cryptographic assumptions alone and can be checked by anyone, at any later time, without the weights.

Cryptographic proofs of neural network inference have so far been costly.
Systems based on sumcheck and GKR proved convolutional networks~\cite{zkcnn} and later language models with up to \(13\) billion parameters in under fifteen minutes per inference~\cite{zkllm}.
Circuit-based systems compile models into general-purpose SNARKs~\cite{zkml,ezkl}.
Three obstacles separate these results from proving the inference that a server actually performs on a production-size model.

\paragraph{The function must be exact.}
A proof establishes a relation between integers or field elements.
Floating-point inference is not a function of its inputs in that sense: results change with the kernel, the batch shape and the order of reductions~\cite{he2025nondeterminism}, and a single changed argmax or sampling decision changes every later token.
A proof system therefore needs a precisely defined integer function \(F\), and the server must serve that function.
Approximate checks of the floating-point computation, as in~\cite{toploc}, answer a different question.

\paragraph{The constraints must not wrap.}
Small prime fields make provers fast, since a base-field element fits a machine word and arithmetic needs no multi-precision code~\cite{plonky3,circlestark,bdt24}.
They also make every intermediate quantity of a transformer large compared to \(p\): a dot product of \(2^{15}\) products of 8-bit integers reaches \(2^{29}\), and a softmax denominator over \(2^{20}\) keys reaches \(2^{30}\), against \(p \approx 2^{31}\).
A relation that holds modulo \(p\) but not over the integers lets a prover change the computation and still pass.
Every relation must therefore be accompanied by bounds that keep both sides below \(p\).

\paragraph{The proof must scale.}
For one 16{,}384-token sequence through Qwen2.5-32B, the trace has about \(6 \times 10^{11}\) committed field elements and \(7 \times 10^{11}\) lookups.
This trace does not fit in GPU memory or in host memory, and the number of lookups exceeds \(p\), so the multiset argument of LogUp~\cite{logup} is unsound unless the lookups are partitioned.

\subsection{Contributions}

\begin{itemize}
\item \textbf{An exact integer transformer} (\cref{sec:function}).
We define \(\FO\), an integer function for Qwen2-style decoder-only transformers, and extend it to the QK-norm and mixture-of-experts layers of Qwen3.
Activations are 24-bit, matrix products take 8-bit inputs with power-of-two scales placed off the reduction axis, and attention and the MLP use fused rounding steps.
Sampling, greedy or by inverse CDF, reads a seed derived from a client nonce.
A deterministic program maps a BF16 checkpoint to \(\FO\).

\item \textbf{A wrap-free constraint system over a 31-bit field} (\cref{sec:constraints}).
Every gadget comes with integer bounds, under which the relation over \(\Fp\) is equivalent to the relation over \(\Z\) (\cref{lem:nowrap}).
Every gadget also has exactly one satisfying output (\cref{lem:det}).
Range and table checks go through one tagged table of \(2^{26}\) rows.

\item \textbf{A streamed proof built from three primitives} (\cref{sec:protocol}).
A \emph{domain} proves the zerocheck and the leaf layer of the LogUp-GKR trees of a set of columns in one sumcheck (\cref{lem:domain}).
Lookups are split into chunks of at most \(2^{30}\) lookups, each with its own multiplicity column, and one GKR over the table proves all chunks at once (\cref{lem:chunks}).
Claims flow backward from the output to the embedding, so the prover makes two passes over the model and never holds more than one layer's witness.
All tensors and weights live in one multilinear commitment whose openings are batched across sizes and whose Merkle paths are shared across the tensors of a layer.

\item \textbf{A single-GPU prover} (\cref{sec:gpu}).
Sumcheck round kernels are generated per constraint program and run their first rounds on base-field tables.
The Fiat--Shamir transcript runs on the GPU, one warp per BLAKE3 compression, so the prove pass has no host round trips.
One middle layer's prove pass is captured as a CUDA graph and replayed for the other layers.
\(\Fq\) products use one modular reduction per coordinate.

\item \textbf{A verifier that a recursion can afford} (\cref{sec:recursion}).
Public columns are evaluated in closed form, fixed tables are committed, final openings are recorded without being hashed into the transcript, and tensors share columns and Merkle paths.
Together these steps cut the verifier's hashing by \(4.2\times\), and the proof for 4{,}096 tokens from \(360\) to \(134.8\)\,MiB.
We also build the first component of the recursive circuit: BLAKE3 compressions as rows of a domain, with a LogUp memory that wires the rows together.

\item \textbf{Evaluation and analysis} (\cref{sec:security,sec:eval}).
We measure the prover on all 64 layers of Qwen2.5-32B and on Qwen3-32B and Qwen3-30B-A3B, relate its time to the inference it attests and to prior systems, measure what exactness costs in accuracy and which integer steps cause the loss, and give a soundness analysis with every term of the error budget.
The analysis of tuple compression in LogUp (\cref{lem:tuple}) shows that compressing a tuple of arity \(k\) with the powers of one challenge costs \(\log_2 k\) bits, so Sable draws one challenge per tuple coordinate.
\end{itemize}
The implementation, its tests and the scripts that produce every measurement in this paper are released under the MIT license at \url{https://github.com/moonmath-ai/Sable}.

\subsection{Scope}

The main proof is not zero-knowledge: it reveals evaluations of the trace to its verifier.
It is large, \(134.8\)\,MiB for one 4{,}096-token sequence through the full model, and a verifier with a compiled core takes \(19\)\,s on one CPU core to check it.
As built, Sable therefore produces an auditor's proof for open-weight models: a non-zero-knowledge proof that reopens the committed weights would leak closed weights after about \(2 \times 10^4\) proofs.
\Cref{sec:threat} states the threat model, the trust anchors of the verifier and these limits.
A recursive layer that proves the verifier, with zero-knowledge outer layers, is the intended form for every client; \cref{sec:recursion} reports the parts of it that exist.
The statement concerns \(\FO\) itself, and the BF16 checkpoint it was derived from plays no role in it, so the accuracy of \(\FO\) relative to the original model is part of what a deployment must report (\cref{sec:eval-acc}).

\section{Related work}
\label{sec:related}

\paragraph{Proofs of machine-learning inference.}
Early systems proved small networks with general-purpose SNARKs or with interactive proofs specialized to matrix products.
zkCNN~\cite{zkcnn} introduced a sumcheck-based argument for convolutions, building on the GKR line of work.
Mystique~\cite{mystique} used VOLE-based interactive zero-knowledge proofs with conversions between arithmetic and Boolean values.
ZKML~\cite{zkml} and ezkl~\cite{ezkl} compile models to Halo2 circuits and optimize the circuit layout.
For language models, zkLLM~\cite{zkllm} combines sumcheck for tensor operations with a lookup argument (tlookup) for non-arithmetic operations and a dedicated protocol for attention; it proves models up to 13B parameters and outputs short proofs through a pairing-based commitment.
zkGPT~\cite{zkgpt} targets GPT-2-sized models with optimized constraint layouts for non-linear functions and proves GPT-2 in about 22\,s on a CPU.
\Cref{sec:eval-prior} compares proving times per parameter and token.
zkLLM and zkGPT work over large prime fields (BLS12-381 or BN254 scalar fields), where a single field element holds any intermediate value, and both approximate non-linear functions without fixing the integer semantics of the served model.
Sable differs in three respects: it works over a 31-bit field and pays for that with explicit no-wrap bounds; it defines the served function bit for bit, including normalization, softmax, sampling and stopping; and it targets models and contexts large enough that the trace must be streamed.

\paragraph{Concurrent work.}
FlashProve~\cite{flashprove,flashproveblog}, a concurrent system from Anthropic, proves inference of Llama~4 Maverick, a mixture of experts with \(400.7\)B parameters of which a token uses \(16.15\)B, from a third-party 4-bit release of its weights.
It replaces the model by an integer function with table-based non-linear steps, as we do, and compiles interactive oracle proofs over the 64-bit Goldilocks field with Reed--Solomon codes, BLAKE2b Merkle trees and zero-knowledge proximity tests; the proof hides the weights, it is about \(18.8\)\,MB per request, and the soundness, completeness and zero knowledge of the protocol are proved in Lean, while its C++ and CUDA programs are tested.
On one H100 it proves 1{,}024-token windows at about 111 tokens per second in batches of 24; under its default policy a proof checks one token, the greedy successor of the window, and an optional policy checks every token of a stored reply against a looser criterion than greedy choice.
Sable differs in what it proves and where it is measured.
A Sable proof covers a whole reply: every generated token, the stopping rule, and, at a positive temperature, sampling from a seed fixed by the statement, FlashProve does not check sampling, and a reply cut short verifies there as a shorter reply.
Sable proves contexts of up to 16{,}384 tokens, where attention, quadratic in the length, is about half of the work, against 1{,}024 there.
Our commitment to the weights is a plain Merkle root that anyone can recompute from an open checkpoint and that serves any number of proofs, while FlashProve's hiding commitment bounds the number of batches it can serve, at most 688{,}127.
We report the accuracy of the integer model on full task benchmarks (MMLU, GSM8K) against BF16 and two deployed 8-bit schemes; FlashProve reports how often its greedy token differs from the 16-bit model's on WikiText-2 (\(6.8\%\) and \(9.4\%\) of positions on two text sets) and runs no task benchmark.
Its prover needs about 260\,GB of host memory, against at most 188\,GB (175\,GiB) for ours.
Per active parameter and token, FlashProve's mixture of experts takes about as long as Sable's dense proofs and 6 to 11 times less than Sable's proofs of Qwen3-30B-A3B (\cref{sec:eval-prior}), with zero knowledge included and at a shorter context.
FlashProve's zero knowledge, proofs 6 to 27 times smaller than ours, faster proofs of a mixture of experts with about twelve times as many parameters as our models, and machine-checked protocol mark directions of future work for Sable: zero-knowledge and recursive layers that shrink the proof (\cref{sec:recursion,sec:discussion}), proofs of larger mixtures of experts at a lower cost per active parameter, and a machine-checked proof of the protocol.

\paragraph{Other approaches to trustworthy inference.}
Trusted execution environments on GPUs attest to the code that ran, under a hardware trust assumption.
Refereed delegation~\cite{crr11} and optimistic schemes resolve disputes by re-execution, which requires deterministic computation and an honest party with the weights.
TOPLOC~\cite{toploc} hashes features of the floating-point activations so that a verifier with the model can check, approximately and statistically, that a claimed computation was performed.
Our statement is exact and our verifier does not need the weights, at a much higher proving cost.

\paragraph{Sumcheck, GKR and lookups.}
The sumcheck protocol~\cite{lfkn} and the GKR protocol~\cite{gkr} underlie provers whose cost is linear in the computation.
Thaler~\cite{thaler13} gave linear-time provers for matrix products and regular circuits; Libra~\cite{libra} made GKR linear-time for general circuits and added zero knowledge; Spartan~\cite{spartan} and HyperPlonk~\cite{hyperplonk} built SNARKs for constraint systems around zerochecks over the Boolean hypercube.
Lookup arguments started with Plookup~\cite{plookup}; LogUp~\cite{logup} replaced grand products by sums of logarithmic derivatives, and Papini and Hab\"ock~\cite{logupgkr} proved those sums with GKR, which removes any commitment that depends on the lookup challenge.
Lasso and Jolt~\cite{lasso,jolt} use structured tables and sumcheck-based lookups for virtual machines.
Our domains combine the last GKR layer of all lookups with the zerocheck of the same columns in one sumcheck, and our chunking follows from the observation that LogUp only proves inclusion modulo the characteristic.

\paragraph{Small fields and sumcheck prover algorithms.}
STARK provers moved from 64-bit fields to 31-bit fields (Mersenne-31, BabyBear, KoalaBear) for speed on CPUs and GPUs~\cite{plonky3,circlestark}; Binius~\cite{binius} works over binary towers.
In all of them, soundness needs challenges from an extension field.
Gruen~\cite{gruen} showed how to remove the equality factor from zerocheck round messages and how to skip the first round's extension arithmetic.
Bagad, Domb and Thaler~\cite{bdt24} and Dao et al.~\cite{ddbdt26} gave sumcheck provers that keep most work in the base field when the summed polynomials take small values.
Our prover evaluates the first round of every sumcheck on base-field tables and folds them into extension tables in the second round; the more aggressive small-value algorithms of~\cite{bdt24,ddbdt26} would apply to our constraint programs only with extension-field coefficients, which \cref{sec:gpu-small} discusses.

\paragraph{Hash-based polynomial commitments.}
FRI~\cite{fri} tests proximity to Reed--Solomon codes and underlies STARKs~\cite{stark}; its soundness up to the Johnson bound follows from the proximity gaps of Ben-Sasson et al.~\cite{bciks}, and Hab\"ock~\cite{habock24basefold} analyzes BaseFold over Reed--Solomon codes in the same list-decoding regime.
Multilinear commitments from linear codes include Ligero~\cite{ligero} and Brakedown~\cite{brakedown}; BaseFold~\cite{basefold} folds a foldable code alongside a sumcheck, and WHIR~\cite{whir} reduces the verifier's queries.
Our commitment uses BaseFold over Reed--Solomon codes on multiplicative subgroups, batches columns of different lengths by joining them at the matching fold level, as in batched FRI for mixed degrees, and arranges all columns of a layer under one Merkle tree whose height differs per column size.

\paragraph{GPU provers.}
GPU acceleration of proof systems has concentrated on multi-scalar multiplication and number-theoretic transforms for pairing-based SNARKs~\cite{gzkp,cuzk} and on STARK provers over small fields (for example sppark and ICICLE).
Sable's kernels are generated per constraint program and per field, and its transcript runs on the GPU, so that the whole prove pass of a layer can be replayed as one CUDA graph.

\paragraph{Integer inference and quantization.}
Integer-only inference fixes the arithmetic of a network so that it runs on integer units~\cite{jacob2018,ibert}.
LLM.int8()~\cite{llmint8}, SmoothQuant~\cite{smoothquant}, QuaRot~\cite{quarot} and SpinQuant~\cite{spinquant} quantize transformers to 8 bits or fewer by handling activation outliers, through mixed precision, scale migration or rotations.
\(\FO\) uses a random orthogonal rotation fused into the weights in the manner of QuaRot, per-row dynamic exponents for activations, and per-column exponents for weights, and it replaces every floating-point non-linearity with a table.

\paragraph{Recursion.}
Incrementally verifiable computation~\cite{valiant} and recursive SNARKs compress long proofs by proving their verifier.
Hash-based recursion that evaluates the hash inside the circuit is standard for STARKs and is usually paired with an algebraic hash; we keep BLAKE3 everywhere, which makes the recursion circuit larger and avoids a further hash assumption.

\section{Preliminaries}
\label{sec:prelim}

\subsection{Notation and fields}

For an integer \(n\), \([n] = \{0, \ldots, n-1\}\).
An index \(i \in [2^n]\) is identified with its bit vector \((i_0, \ldots, i_{n-1}) \in \bits^n\), \(i = \sum_j 2^j i_j\), so coordinate \(j\) of a point is bit \(j\) of the index (little-endian).
\(A = [-2^{23}, 2^{23})\) is the activation range.
\(\lfloor \cdot \rceil\) denotes rounding half up, \(\lfloor z + 1/2 \rfloor\).

Throughout, \(p = 2^{31} - 2^{24} + 1\) (KoalaBear), so \(p - 1 = 2^{24} \cdot 127\) and \(\Fp^\times\) contains a cyclic subgroup of order \(2^k\) for every \(k \le 24\).
We use the tower
\[
\F_{p^2} = \Fp[x]/(x^2 - 3), \qquad \Fq = \F_{p^2}[y]/(y^2 - x), \qquad \Fo = \Fq[z]/(z^2 - y),
\]
so that \(\Fq \cong \Fp[Y]/(Y^4 - 3)\) and \(\Fo \cong \Fp[X]/(X^8 - 3)\).
These are fields: \(p \equiv 2 \pmod 3\) and \(p \equiv 1 \pmod 4\), so by quadratic reciprocity \(3\) is a non-residue modulo \(p\), and a binomial \(X^{2^k} - a\) with \(a\) a non-residue is irreducible over \(\Fp\) when \(p \equiv 1 \pmod 4\)~\cite[Thm.~3.75]{lidl}.
\(|\Fq| \approx 2^{124}\) and \(|\Fo| \approx 2^{248}\).
Witness values lie in \(\Fp\); interactive-proof challenges lie in \(\Fq\) and commitment challenges in \(\Fo \supset \Fq\).

\subsection{Multilinear extensions}

The multilinear extension of \(f : \bits^n \to \F\) is the unique multilinear \(\mle f : \F^n \to \F\) that agrees with \(f\) on \(\bits^n\):
\[
\mle f(r) = \sum_{x \in \bits^n} \eq(r, x)\, f(x), \qquad \eq(r, x) = \prod_{j=0}^{n-1} \big(r_j x_j + (1 - r_j)(1 - x_j)\big).
\]
A vector of length \(2^n\) is a function on \(\bits^n\) through the index convention above.
We write \(\mle f(r)\) for the \emph{evaluation} of a vector at \(r\) and call a pair \((r, v)\) with claimed \(v = \mle f(r)\) a \emph{claim} on \(f\).
For a matrix \(M \in \F^{2^a \times 2^b}\) stored row-major, \(\mle M(r_c, r_r)\) takes the column variables first.

\subsection{Sumcheck and zerocheck}

\begin{lemma}[Sumcheck~\cite{lfkn}]
\label{lem:sumcheck}
Let \(g : \F^n \to \F\) have degree at most \(d\) in each variable.
In the sumcheck protocol for \(\sum_{x \in \bits^n} g(x) = c\), the prover sends in round \(j\) a univariate polynomial of degree at most \(d\), the verifier checks it against the running claim and replies with a random \(r_j \in \F\), and at the end the claim is reduced to \(g(r) = v\).
If \(\sum_x g(x) \ne c\), the verifier accepts a false final claim with probability at most \(nd/|\F|\), and at most \(d/|\F|\) per round.
\end{lemma}

All our sumchecks have the form \(\sum_x w(x) \sum_t c_t \prod_{k \in t} T_k(x)\) for multilinear tables \(T_k\), coefficients \(c_t\) and either \(w = 1\) or \(w = \eq(r, \cdot)\).
Products of tables are formed as polynomials in table values, so a constraint of degree \(d\) in the columns gives a sumcheck of degree \(d\) (plus one for \(\eq\)).
In the \(\eq\) case the prover sends the round polynomial without the \(\eq\) factor of the round variable, as in~\cite{gruen}, which lowers its degree by one.

A \emph{zerocheck} proves that a polynomial expression \(C(f_1(x), \ldots, f_m(x))\) in committed vectors vanishes on \(\bits^n\).
It runs the sumcheck for \(\sum_x \eq(\tau, x) C(x) = 0\) for a random \(\tau\).
If \(C(x^*) \ne 0\) for some \(x^*\), the multilinear extension of \(x \mapsto C(x)\) is a nonzero multilinear polynomial, which vanishes at a random \(\tau\) with probability at most \(n/|\F|\).

A \emph{product sumcheck} reduces a claim on a matrix product \(C = AB\), \(\mle C(r_x, r_y) = v\), to the sumcheck \(\sum_{k} \mle A(r_x, k)\, \mle B(k, r_y) = v\), whose prover costs \(O(mk + kn)\) field operations after folding \(A\) and \(B\) at \(r_x\) and \(r_y\)~\cite{thaler13}.

\subsection{Fractional sums and LogUp}

\begin{lemma}[LogUp~\cite{logup}, over a prime field]
\label{lem:logup}
Let \(a_1, \ldots, a_N \in \F\) with \(N < p\), let \(T \subset \F\) be a set, and let \(m : T \to \F\) be arbitrary.
If the rational functions
\[
\sum_{i=1}^{N} \frac{1}{Z - a_i} \quad\text{and}\quad \sum_{v \in T} \frac{m(v)}{Z - v}
\]
in \(\F(Z)\) are equal, then every \(a_i \in T\), and \(m(v)\) equals \(|\{i : a_i = v\}|\) modulo \(p\).
\end{lemma}

The bound \(N < p\) is needed: \(p\) copies of a value \(a \notin T\) contribute \(p/(Z - a) = 0\).  No bound on the multiplicities is needed, which is why they may be committed as arbitrary field elements.
LogUp tests the identity at a random \(Z\).
A \emph{tuple} \((u_0, \ldots, u_{k-1})\) is compressed to the single value \(u_0 + \beta_1 u_1 + \cdots + \beta_{k-1} u_{k-1}\) for independent random \(\beta_i\), which we collect as \(\beta = (\beta_1, \ldots, \beta_8)\); \cref{lem:tuple} bounds the error of the combined test.
Following~\cite{logupgkr}, the sums are evaluated by a binary tree of fractions: the leaves are \((n_i, q_i) = (1, Z - a_i)\), and a parent is
\[
(n, q) = (n_0 q_1 + n_1 q_0,\ q_0 q_1),
\]
so the root is a fraction \((P, Q)\) with \(P/Q\) equal to the sum.
GKR proves the tree layer by layer: a claim on the multilinear extensions of layer \(k\) at a point becomes, by one sumcheck of degree~3, a claim on layer \(k+1\) at a new point.
No committed value depends on \(Z\), so every column, including the multiplicities, is committed before \(Z\) is drawn.

\subsection{Codes, commitments and Fiat--Shamir}
\label{sec:prelim-fs}

Let \(H_k \subset \Fp^\times\) be the subgroup of order \(2^k\).
The Reed--Solomon code \(\mathrm{RS}[k]\) of rate \(1/2\) encodes a message \(m \in \Fp^{2^{k-1}}\), read as the coefficients of \(f(X) = \sum_j m_j X^j\), as the evaluations of \(f\) on \(H_k\).
Positions are in bit-reversed order, so the two symbols \(f(\omega), f(-\omega)\) are adjacent, and the four symbols that reduce to one symbol after two folds are contiguous.
Folding with a challenge \(r\) maps the pair \((a, b) = (f(\omega), f(-\omega))\) to
\[
\tfrac{a + b}{2} + r\Big(\tfrac{a - b}{2\omega} - \tfrac{a + b}{2}\Big),
\]
which is the encoding in \(\mathrm{RS}[k-1]\) of the folded message \((1 - r) m_{\mathrm{even}} + r\, m_{\mathrm{odd}}\).
The folded message is the restriction of the message's multilinear extension to \(x_0 = r\), so a sequence of folds with challenges \(r_0, r_1, \ldots\) ends in the constant \(\mle m(r)\).
BaseFold~\cite{basefold} interleaves these folds with a sumcheck of \(\eq(\rho, x)\, \mle m(x)\) to prove \(\mle m(\rho) = v\).
Its soundness reduces to the proximity of the committed oracles to the code, which FRI-style queries test; in the unique-decoding and Johnson regimes the error is bounded by the proximity gaps of Ben-Sasson et al.~\cite{bciks}.

Merkle trees are built with BLAKE3~\cite{blake3}; a tree's \emph{cap} is the list of its \(2^{7}\) nodes at depth \(7\), sent in place of the root, and paths stop at the cap.
The protocol is made non-interactive with the Fiat--Shamir transform~\cite{fiatshamir}.
The transcript absorbs every prover message, prefixed with its length in bytes, and derives each challenge from its state, by rejection sampling of 31-bit words into \(\Fp\) coordinates; after a challenge it absorbs a separator and the first 32 output bytes.
Its hash is the BLAKE3 compression function \(\mathsf{C}(h, m, t, \ell, f)\) (chaining value \(h\), 64-byte block \(m\), counter \(t\), block length \(\ell\), flags \(f\)) in a sequential mode of our own, which is not a standardized BLAKE3 mode.
The state is a chaining value \(h\), initially the BLAKE3 IV, and a pending block.
When a block is full and more input arrives, \(h \gets \mathsf{C}(h, m_t, t, 64, 0)\) for the \(t\)-th block.
Output block \(j\) is \(\mathsf{C}(h, m, j, \ell, \textsc{chunk\_end} \mid \textsc{root})\) on the pending block \(m\) of length \(\ell \le 64\), which is never empty unless nothing was absorbed.
This is a Merkle--Damg\aa rd iteration whose final call carries the length of the last block and flags that no absorbing call uses, so the encoding of the absorbed string is prefix-free and absorbing and output calls are separated.
With the compression function modeled as ideal, as in BLAKE3's own analysis, such an iteration is indifferentiable from a random oracle~\cite{coron05}, and the Fiat--Shamir analysis below applies to it unchanged.
Compared with BLAKE3's tree mode, which derives every output from the root of a tree over 1{,}024-byte chunks, a challenge costs one compression instead of one per level of the tree; this matters because a proof draws about a million challenges, each on the prover's critical path and each to be arithmetized by a recursive verifier.
For interactive oracle proofs~\cite{bcs16} with round-by-round soundness error \(\epsilon\)~\cite{rbr}, the resulting argument has soundness error at most \(Q\epsilon\) against provers that make \(Q\) random-oracle queries, up to the collision probability of the hash.
We report security in bits as \(\log_2(1/\epsilon)\), the convention used for deployed STARKs.

\subsection{Verifiable inference}

\begin{definition}[Statement and relation]
\label{def:relation}
A \emph{statement} is a tuple \(\mathsf{st} = (C_W, \mathsf{policy}, x, y, n_c)\): a weight commitment, a decoding policy (temperature index \(\theta\), maximum output length \(G_{\max}\), stop tokens), a prompt \(x \in V^{P}\), generated tokens \(y \in V^{G}\), and a client nonce \(n_c \in \bits^{256}\).
The seed is \(r = \Sha(\texttt{seed} \,\|\, n_c \,\|\, C_W \,\|\, \Sha(x) \,\|\, \mathsf{policy})\).
\(C_W\), the seed and the digests of the statement (\cref{sec:transcript}) use SHA-256.
The main verifier evaluates SHA-256 over the caps of the weight units (for \(C_W\)) and over the model's constants and tables (for \(h_{\mathrm{circ}}\)); a recursive verifier would take those caps and constants as public inputs, so that their SHA-256 digests are checked outside the circuit.
The relation \(\mathcal{R}\) contains \((\mathsf{st}, W)\) exactly when \(C_W\) is a commitment to \(W\), \(y\) is the autoregressive output of \(\FO(W, \cdot, r)\) from \(x\) (\(y_t\) is sampled from the logits at position \(P + t - 1\) of \(x \| y_{<t}\)), and \(y\) obeys the stopping rule: no \(y_t\) with \(t < G\) is a stop token, and either \(y_G\) is one or \(G = G_{\max}\).
\end{definition}

Because \(\FO\) is deterministic and causal, all \(S = P + G - 1\) positions can be evaluated in one teacher-forced pass over \(x\) followed by the first \(G - 1\) generated tokens (the last generated token is output but never read), and the logits at each generated position equal those of a token-by-token decode.
The prover therefore proves one forward pass over \(S\) positions together with the sampling decisions at the last \(G\) positions.
\begin{definition}[Batched statement]
\label{def:batch}
A \emph{batched statement} \((C_W, \mathsf{policy}, (x_b, y_b)_{b < B}, n_c)\) has \(B\) prompts of equal length and generated sequences of equal length \(G\), one policy and one nonce.
Its seed is \(r = \Sha(\texttt{seed} \,\|\, n_c \,\|\, C_W \,\|\, \Sha(x_0 \| \cdots \| x_{B-1}) \,\|\, \mathsf{policy})\), and sequence \(b\) samples its token \(t\) with \(u_{b,t} = \mathrm{PRF}(r, bG + t)\).
\(\mathcal{R}\) contains \((\mathsf{st}, W)\) when every \((x_b, y_b)\) satisfies the conditions of \cref{def:relation} with these sampling values.
\end{definition}

The verifier of a batched statement needs every \(x_b\) and \(y_b\), and the proof opens values of every sequence's trace, so a batch is either one client's sequences or is verified by a party entitled to see all of them.

An argument for \(\mathcal{R}\) is \emph{sound} if no efficient prover makes the verifier accept a statement outside the language of \(\mathcal{R}\), except with small probability.
The stopping rule is part of the relation because otherwise every prefix of a valid output would also be valid, and a server could truncate answers.

\section{Threat Model and Scope}
\label{sec:threat}

\paragraph{Adversary and trust anchors.}
The adversary is the server, which runs the prover and wants a verifier to accept an output that is not \(\FO(W, x, r)\) for the committed \(W\).
The verifier trusts three things.
(i) A \emph{registry} that publishes, with the model release, \(C_W\) and a digest \(h_{\mathrm{circ}}\) of every constant outside the weights (exponents, shifts, biases, tables and the commitment digests of the fixed tables).
For an open-weight model anyone recomputes both from the checkpoint and the published quantization program, so the registry needs no trust beyond that recomputation.
The server must never be the source of either, since a server that supplies \(h_{\mathrm{circ}}\) chooses the tables.
(ii) Its own tokenization: the statement is over token ids, so the client tokenizes and must receive the full sequence, including any system prompt the server prepends; a proof over a sequence with a hidden prefix proves a different statement and is rejected.
(iii) Its nonce \(n_c\), which fixes the sampling seed.

\paragraph{Who verifies.}
The proof is not zero-knowledge, it is 106 to 477\,MiB long for the three models we evaluate, and our verifier takes 19 to 56\,s on one CPU core (\cref{sec:eval-verify}).
As built, it is therefore an auditor's proof: the verifier is a client or an auditor that can afford this, for a model whose weights are public.
A recursive proof of the verifier with zero-knowledge outer layers is the intended form for every client, and \cref{sec:recursion} reports its first components.

\paragraph{Open weights.}
The proof reveals the trace at the queried positions, and the weights are committed once and reopened by every proof.
Each proof opens every weight column at 153 random leaves of four Reed--Solomon symbols, and a column of \(2^{23}\) values is determined by any \(2^{23}\) of its \(2^{24}\) codeword symbols.
After \(T\) proofs, a client holds an expected fraction \(1 - e^{-153 T / 2^{22}}\) of the leaves, so about \(2^{22} \ln 2 / 153 \approx 1.9 \times 10^4\) proofs suffice to recover every weight by linear algebra.
A provider that served a closed model with this prover would therefore publish it.
The system as built is for open-weight models; closed weights need a hiding weight commitment, which the planned zero-knowledge layers provide.
The soundness theorem also assumes that \(C_W\) commits to in-range weights (\(|w| \le 127\), \cref{sec:security}).
For open weights this is checked by recomputing \(C_W\); for closed weights, a one-time range proof per model release would replace the assumption.

\paragraph{Batches.}
A batched proof, such as our \(4 \times 4\)K shape, proves that each of several sequences is an output of \(\FO\) under one \(C_W\), one policy and one nonce, with the seed derived from all prompts (\cref{def:batch}).
Its verifier needs all prompts and outputs and sees opened values of every sequence's residual stream.
A batch is therefore either one client's sequences or is verified by an auditor entitled to see all of them; batching the requests of different clients would leak their prompts to each other.

\paragraph{Refusals.}
A server can always decline to answer, and two kinds of refusal deserve mention.
First, \(\FO\) is a partial function: on inputs where an intermediate value leaves its range, \(\FO\) is undefined and the honest prover aborts (\cref{sec:fo-undefined}).
For open weights the client can evaluate \(\FO\) itself and confirm such an abort; \cref{sec:eval-abort} measures how often one occurs.
Second, at temperature \(\theta > 0\) a server could refuse requests until a client's fresh nonce yields a seed it likes.
Clients can defend against this by counting refusals per prompt, or by sending nonces only under a commitment that obliges the server to answer.
At \(\theta = 0\) the proof is a deterministic function of the statement and the nonce plays no role, so a server may cache and replay the proof of a repeated prompt, which is harmless.

\paragraph{Verifier cost under attack.}
The verifier checks the statement, the policy, and the weight and fixed-table commitments against the registry before it replays the transcript, and it checks the 11-bit and 30-bit proofs of work when it reaches them, so a malformed proof fails early.
A proof that fails only at its last query still costs a full verification.

\section{The integer function \texorpdfstring{\(\FO\)}{F\_O}}
\label{sec:function}

\(\FO\) is defined for decoder-only transformers in the style of Qwen2: pre-normalization with RMSNorm, grouped-query attention with biases on the query, key and value projections, rotary position embeddings (RoPE)~\cite{rope}, and a gated MLP with SiLU.
Our running example is Qwen2.5-32B-Instruct~\cite{qwen25}: \(L = 64\) layers, width \(d = 5120\), MLP width \(d_{\mathrm{ff}} = 27648\), \(40\) query heads and \(8\) key-value heads of dimension \(d_h = 128\), and a vocabulary of \(152{,}064\) tokens.
\Cref{sec:qwen3} extends \(\FO\) to Qwen3~\cite{qwen3}, which normalizes queries and keys per head and, in its mixture-of-experts models, routes each token to a few of many MLPs.
\(\FO\) has no floating-point step and no data-dependent loop.
Every operation below is an exact integer definition.
The weights and embeddings are fixed by \(C_W\), and every other constant (exponents, shifts, biases and the contents of the tables) by the circuit digest \(h_{\mathrm{circ}}\) of the statement (\cref{sec:transcript}).

\subsection{Number formats}
\label{sec:formats}

\begin{itemize}
\item \emph{Residual stream.} Integers in \(A = [-2^{23}, 2^{23})\) with \(f_h = 10\) fractional bits.
\item \emph{Matrix-product inputs.} A row of activations is quantized to integers in \([-127, 127]\) with one power-of-two exponent per row, chosen as the least exponent at which the row fits (\(\gad{Requant8}\) below).
\item \emph{Weights.} Integers in \([-127, 127]\) with one power-of-two exponent per output column (or per tensor).
\item \emph{Scales off the reduction axis.} In every product \(C = AB\), the scale of \(A\) depends only on the row and the scale of \(B\) only on the column, so the accumulator \(\sum_t a_{mt} b_{tn}\) is an exact integer and the combined scale is applied once, by a shift, after the sum.
\item \emph{Queries, keys and values.} Static per-head exponents \(e^Q_h, e^K_g, e^V_g\) with \(e^Q_h + e^K_{g(h)} = E_{\mathrm{attn}}\) for one global constant, so that one exponential table serves every head and layer.
\end{itemize}
Every exact sum in a linear layer has at most \(2^{15}\) terms of magnitude at most \(127^2\), so \(|\mathrm{acc}| < 2^{29}\).

\subsection{Primitive operations}
\label{sec:primitives}

Let \(a\) be a row of integers.
\begin{itemize}
\item \(\gad{Requant8}(a)\): \(e\) is the least integer in \([e_0, e_{\max}]\) such that \(2a_i \in [-255 \cdot 2^e, 255 \cdot 2^e)\) for all \(i\); \(q_i = \lfloor a_i / 2^e \rceil\).
Since the interval is half-open, \(|q_i| \le 127\): the rule never clips.
\(\gad{ShiftQ8}\) is the same rule applied to a row of accumulators, with \(e \in [2, 22]\).
\item \(\gad{NormScale}(q)\): \(\Sigma = \sum_i q_i^2 + 1\); write \(\Sigma = 2^{2E} \delta + \rho\) with \(\delta \in [2^{14}, 2^{16})\), \(0 \le \rho < 2^{2E}\) (for \(E \ge 0\); for \(E < 0\), \(\delta = 2^{-2E}\Sigma\));
\(\mu = \tau_{\mathrm{rsqrt}}(\delta) = \lfloor 127 \cdot 2^{15} / \sqrt{\delta} \rfloor\); \(q'_i = \lfloor (q_i \mu + 2^{14}) / 2^{15} \rfloor\).
The exponent \(E\) enters the next shift.
This is RMSNorm with the gain vector fused into the next weight matrix.
\item \(\gad{ShiftRound}(\mathrm{acc}, k) = \lfloor \mathrm{acc} / 2^k \rceil\) for \(k > 0\), and \(\mathrm{acc} \cdot 2^{-k}\) for \(k \le 0\), with \(k \in [-16, 24]\) and the output required to lie in \(A\).
\item \(\gad{ShiftQ8S}(\mathrm{acc}, k, b) = \clip(\lfloor \mathrm{acc} / 2^k \rceil + b, -127, 127)\) for a shift \(1 \le k \le k_{\max}(b)\) and an integer bias \(b\): one rounding from the accumulator to a static, saturating 8-bit value.
Here \(k_{\max}(b)\) is the largest \(k \le 22\) with \(2^k(|b| + 128) < p - 2^{30} - d \cdot 127^2\), which keeps the gadget's relations wrap-free (\cref{tab:bounds}).
\item \(\gad{RoPE8}(q_1, q_2; \cos, \sin) = (\lfloor (q_1 \cos - q_2 \sin + 2^{14}) / 2^{15} \rfloor, \lfloor (q_2 \cos + q_1 \sin + 2^{14}) / 2^{15} \rfloor)\) on 8-bit inputs and public 16-bit integers \(\cos = \lfloor 2^{15} \cos\phi \rceil\), \(\sin = \lfloor 2^{15}\sin\phi \rceil\); outputs lie in \([-180, 180]\).
\item \(\gad{MulRound}(x, y, k) = \lfloor x y / 2^k \rceil\) for \(x \in A\), \(|y| \le 2^{17}\), \(k \in [13, 30]\), evaluated through the split \(x = 2^{12} x_h + x_l\) as in \cref{sec:g-mulround}.
\item \(\gad{SatAdd}(h, c) = \clip(h + c, -2^{23}, 2^{23} - 1)\).
\item Tables: \(\tau_{\exp}(x) = \lfloor 2^{10} \exp(-x \cdot 2^{E_{\mathrm{attn}}} / \sqrt{d_h}) \rceil\) for \(x \in [0, 2^{23})\); \(\tau_{\mathrm{SiLU}}(c)\) the SiLU of \(c \cdot 2^{e_G}\) in units \(2^{e_A}\), clipped to 16 bits, for \(|c| \le 2^{14}\); \(\tau_{\mathrm{recip}}(\delta) = \lfloor 2^{32} / \delta \rfloor\) for \(\delta \in [2^{15}, 2^{16})\).
\(\tau_{\exp}\), \(\tau_{\mathrm{SiLU}}\), the RoPE values \(\cos, \sin\) and the sampling table \(\tau_{\exp,\theta}\) are correctly rounded: each entry is the integer nearest to the exact real value (ties to even).
They are computed in double precision, and every entry within \(10^{-4}\) of a rounding boundary is recomputed in 60-digit arithmetic, so the tables do not depend on the platform's floating point; \(h_{\mathrm{circ}}\) pins them.
\end{itemize}

\subsection{One layer}
\label{sec:layer}

Given the residual rows \(h \in A^{S \times d}\), one layer computes:
\begin{enumerate}
\item \(q = \gad{Requant8}(h)\), \(q' = \gad{NormScale}(q)\) with exponent \(E_h\) per row.
\item \(\mathrm{acc} = q' W_{qkv}\); \((Q, K, V) = \gad{ShiftQ8S}(\mathrm{acc}, k_{mn}, b_n)\) with \(k_{mn} = \kappa_n + E_{h,m}\), defined when \(E_{h,m} \le K_{qs} = \min_n (k_{\max}(b_n) - \kappa_n)\).
\item \(\hat Q = \gad{RoPE8}(Q)\), \(\hat K = \gad{RoPE8}(K)\) per head and position.
\item Attention per head (\cref{sec:attention}) gives \(o \in A^{S \times 5120}\).
\item \(q_o = \gad{Requant8}(o)\) with exponent \(e_o\); \(h \leftarrow \gad{SatAdd}(h, \gad{ShiftRound}(q_o W_o, K_o - e_o))\).
\item \(q = \gad{Requant8}(h)\), \(q' = \gad{NormScale}(q)\).
\item Gate: \(c = \gad{ShiftRound}(q' W_g, k_{g,mn})\) with \(k_{g,mn} = \kappa^g_n + \max(E_{h,m}, K_{gl}) \le 15\) and \(|c| \le 2^{14}\), \(a = \tau_{\mathrm{SiLU}}(c)\).
The layer constant \(K_{gl} = 1 - \min_n \kappa^g_n\) is the least row exponent at which every gate shift is positive; a row of smaller norm is shifted as if its exponent were \(K_{gl}\).
Up: \(q_u = \gad{ShiftQ8}(q' W_u)\) with exponent \(e_u\).
Product: \(m = a \cdot q_u\), \(|m| < 2^{22}\).
\item \(q_m = \gad{Requant8}(m)\) with exponent \(e_m\); \(h \leftarrow \gad{SatAdd}(h, \gad{ShiftRound}(q_m W_d, K_d - e_m - e_u + E_h))\).
\end{enumerate}
The embedding of token \(x_t\) is row \(x_t\) of the committed embedding table.
After the last layer, \(q' = \gad{NormScale}(\gad{Requant8}(h))\) and the logits are \(z = \gad{ShiftRound}(q' W_{\mathrm{lm}}, K_z + E_h)\) in units of \(2^{-8}\).

\subsection{Attention}
\label{sec:attention}

For query row \(m\) and key \(j \le m\) of the same sequence, per query head \(h\) with key-value group \(g\):
\begin{enumerate}
\item \(\mathrm{acc}_{mj} = \sum_c \hat Q_{mc} \hat K_{jc}\), \(|\mathrm{acc}| \le 128 \cdot 180^2 < 2^{22}\).
\item \(M_m = \max_{j \le m} \mathrm{acc}_{mj}\).
\item \(P_{mj} = \tau_{\exp}(M_m - \mathrm{acc}_{mj}) \in [0, 2^{10}]\) for \(j \le m\) and \(P_{mj} = 0\) otherwise.
\item \(D_m = \sum_j P_{mj} \in [2^{10}, 2^{30}]\) for \(S \le 2^{20}\).
\(D_m = 2^{E_m} \delta_m + \rho_m\) with \(\delta_m \in [2^{15}, 2^{16})\) and \(E_m \in [-5, 14]\) (a left shift when \(E_m < 0\)).
\item For key blocks \(b\) of \(2^{12}\) consecutive positions, \(N^{(b)}_{mc} = \sum_{j \in b} P_{mj} V_{jc}\), \(|N^{(b)}| < 2^{12} \cdot 2^{10} \cdot 127 < 2^{29}\); with \(k_m = \max(7, E_m)\),
\(N_{mc} = \sum_b \gad{ShiftRound}(N^{(b)}_{mc}, k_m)\), and \(|N_{mc}| < 2^{23}\).
\item \(o_{mc} = \gad{MulRound}(N_{mc}, \tau_{\mathrm{recip}}(\delta_m), 25 + E_m - k_m + s_h)\), where \(s_h\) aligns the head's value exponent with the group's.
\end{enumerate}
Blocking the \(PV\) product keeps every partial sum below \(2^{29}\) for any context up to \(2^{20}\) tokens, and the per-row shift \(k_m\) keeps the sum of rounded blocks in \(A\).
Causal masking is part of the definition: an entry above the diagonal contributes exactly zero.

\subsection{Sampling and stopping}
\label{sec:sampling}

At temperature index \(\theta = 0\), \(y_t\) is the smallest index that attains \(\max_j z_{tj}\).
For \(\theta > 0\) with temperature \(T_\theta\), \(p_{tj} = \lfloor 2^{12} \exp(-(\max_j z_{tj} - z_{tj}) / (2^8 T_\theta)) \rceil\), \(Z_t = \sum_j p_{tj}\), \(u_t\) the low 30 bits of \(\mathrm{BLAKE3}(r \,\|\, t)\), \(w_t = \lfloor u_t Z_t / 2^{30} \rfloor\), and \(y_t\) is the unique index with
\[
\textstyle\sum_{j < y_t} p_{tj} \;\le\; w_t \;<\; \sum_{j \le y_t} p_{tj}.
\]
Tokens whose probability is below about \(2^{-13}\) of the mode's cannot be sampled; this is part of \(\FO\).

\subsection{Qwen3: QK-norm and mixture of experts}
\label{sec:qwen3}

We also define \(\FO\) for two Qwen3 models~\cite{qwen3}.
Qwen3-32B is dense: \(L = 64\), \(d = 5120\), \(d_{\mathrm{ff}} = 25600\), \(64\) query heads and \(8\) key-value heads of dimension \(128\), and \(151{,}936\) vocabulary rows.
Qwen3-30B-A3B is a mixture of experts: \(L = 48\), \(d = 2048\), \(32\) query heads and \(4\) key-value heads, and in every layer \(128\) expert MLPs of width \(d_e = 768\), of which each token uses \(k = 8\).
Both drop the query, key and value biases (\(b = 0\) in \(\gad{ShiftQ8S}\)) and add two operations.

\paragraph{QK-norm.}
Qwen3 applies RMSNorm with a learned gain to each query and key head before RoPE.
In \(\FO\), for each query or key head row \(a\) of the accumulator \(q' W_{qkv}\) (\(d_h\) values):
\(q_8 = \gad{ShiftQ8}(a)\); \(q'' = \gad{NormScale}(q_8)\) with exponent \(E_2\); and \(Q = \gad{ShiftQ8S}(q'' G, k_n, 0)\), where \(G\) is the head's gain, with \(\sqrt{d_h}/127\) folded in, quantized to 8 bits with one exponent, and \(k_n = \max(\kappa^n_h + E_2, 1) \le 22\) with \(\kappa^n_h\) a per-head constant.
Values keep the path of step 2.
The QKV weight columns of one head share one exponent, so that the head row of the accumulator has a single scale.

\paragraph{Mixture of experts.}
The MLP of a Qwen3-30B-A3B layer reads the normalized rows \(q'\) (exponent \(E_h\)) of step 6.
\begin{enumerate}
\item \emph{Router.} \(z = \gad{SatShift}(q' W_r, k_r)\), whose accumulator is below \(d \cdot 127^2 < 2^{25}\) for \(d \le 2048\), the rounded quotient \(\lfloor \mathrm{acc} / 2^{k_r} \rceil\) saturated to \([-(2^{15} - 1), 2^{15} - 1]\), with \(k_r = \max(E_h - e_r - 8, 1) \le 14\): logits in units of \(2^{-8}\), one weight exponent \(e_r\) per layer.
\item \emph{Selection.} The \(k\) largest logits of the row, ties to the smaller expert index, give slots \(s = 1, \ldots, k\) with experts \(e_s\).
\item \emph{Weights.} \(p_s = \tau_r(z_{\max} - z_s)\) with \(\tau_r(x) = \lfloor 2^{12} \exp(-x / 2^8) \rceil\) for \(x \in [0, 2^{16})\), correctly rounded, and \(w_s = \lfloor (2^{15} p_s + \Sigma_p) / (2 \Sigma_p) \rfloor\) with \(\Sigma_p = \sum_s p_s\): the softmax over the selected experts, renormalized as in Qwen3, in units of \(2^{-14}\).
\item \emph{Experts.} Slot \(s\) runs the dense MLP of step 7 with the weights of expert \(e_s\) (one exponent per expert and matrix): \(c = \gad{ShiftRound}(q' W_g^{e_s}, k_g)\) with \(k_g = \max(e_G - e^g_{e_s} + E_h, 1) \le 15\) and \(a = \tau_{\mathrm{SiLU}}(c)\); \(q_u = \gad{ShiftQ8}(q' W_u^{e_s})\) with exponent \(e_u\); \(q_m = \gad{Requant8}(a q_u)\) with exponent \(e_m\).
\item \emph{Merge.} The slots have different scales \(\epsilon_s = e_m + e_u + e^u_{e_s} + e^d_{e_s}\); with \(\epsilon = \max_s \epsilon_s\), each weight is aligned to the largest, \(\omega_s = \gad{ShiftRound}(w_s, \epsilon - \epsilon_s)\) with \(\epsilon - \epsilon_s \le 24\), and the \(k d_e\) products \(q_m \omega_s\) of the token are requantized together to integers in \([-1023, 1023]\) with one exponent \(e_2 \in [0, 17]\) (inputs below \(127 \cdot 2^{14}\) need at most \(e_2 = 11\)).
\item \emph{Down.} One accumulator over the slots, \(\mathrm{acc} = \sum_s q_{2,s} W_d^{e_s}\), with \(|\mathrm{acc}| \le k d_e \cdot 127 \cdot 1023 < 2^{29.6}\), and \(h \leftarrow \gad{SatAdd}(h, \gad{ShiftRound}(\mathrm{acc}, k_d))\) with \(k_d = \min(K - e_2 - \epsilon + E_h, 24)\) for a model constant \(K\); as in the dense down projection, \(k_d \le 0\) is a left shift, and \(k_d\) must lie in \([-16, 24]\).
\end{enumerate}
Folding the weights into one requantized row lets a single down accumulator and one rounding serve all \(k\) slots, as one product sums over the experts.

\paragraph{Row clamps.}
The \(\max(\cdot, 1)\) and \(\min(\cdot, 24)\) above, and \(K_{gl}\) in the dense gate, define rows whose exponent \(E_h\) is unusually small.
Such rows arise for a handful of under-trained tokens at the first position, whose residual has a tiny norm; without the clamps their shifts leave the ranges of the gadgets and \(\FO\) is undefined on any prompt that starts with them.
A clamp changes no row on which the unclamped shift is in range.

\subsection{From a checkpoint to \texorpdfstring{\(\FO\)}{F\_O}}
\label{sec:quantize}

A deterministic program maps a BF16 checkpoint to the integer model.
It multiplies the residual basis by a fixed random orthogonal matrix \(R\) (generated from a public seed), as in QuaRot~\cite{quarot}, which spreads activation outliers across channels without changing the floating-point function; fuses each RMSNorm gain and \(R\) into the adjacent weight matrices; and quantizes each weight matrix to 8 bits with per-column power-of-two exponents.
It then fixes the static exponents (\(e^Q_h, e^K_g, e^V_g, E_{\mathrm{attn}}\), the SiLU input and output units \(e_G, e_A\), and the per-layer shift constants) by a two-pass calibration on 1{,}024 tokens of English technical text (the beginning of a LaTeX document, an early specification of this system), choosing each exponent from the largest magnitude it must represent.
For models with QK-norm, one more choice matters.
The attention exponent \(E_{\mathrm{attn}}\) is set by the head that needs the most range, so in most key-value groups \(e^Q_h + e^K_g = E_{\mathrm{attn}}\) leaves slack: \(e^K_g\) may lie anywhere between the least exponent that fits \(1.25 \max |K|\) and \(E_{\mathrm{attn}} - \max_h e^Q_h\), the largest that still lets every query head of the group fit.
Giving all of the slack to \(K\) leaves the normalized keys with about 2 of their 8 bits in most Qwen3 layers.
The program instead picks, per group, the \(e^K_g\) in this range that minimizes the rounding noise of the scores, \(\sum_{h \in g} \big(4^{e^Q_h} \|K_g\|^2 + 4^{e^K_g} \|Q_h\|^2\big)\) where \(e^Q_h = E_{\mathrm{attn}} - e^K_g\) and \(\|\cdot\|^2\) are mean squares over the calibration text.  We call this the \emph{Q/K split}.
This moves only calibration constants: \(E_{\mathrm{attn}}\), the exponential table and every bound of \cref{sec:constraints} stay the same, and so do the proof's size and cost.
We did not apply the split to Qwen2.5, which has no QK-norm; its model gives all of the slack to \(K\).
The first position of a prompt attends only to itself, so its activations depend on the token alone, and some tokens there carry far larger activations than any row of the text.
The program therefore runs every vocabulary token at position 0, widens \(e_G\) and \(e_A\) to the largest gate input and SiLU output it sees, lowers the residual's fractional bits \(f_h\) until every output of the attention and down projections lies in \(A\), and repeats until nothing changes.
Finally it raises the weight exponent of every \(W_{qkv}\) column \(n\) with \(k_{\max}(b_n) - \kappa_n\) below the largest row exponent \(E_h\) that \(\gad{NormScale}\) can produce (\(6\) for \(d = 5120\)), by the missing amount, and requantizes those columns, so that \(K_{qs}\) never binds and the shift condition of \(\gad{ShiftQ8S}\) holds on every input; it does the same for every gate column (or expert) whose shift could exceed 15.
For Qwen2.5-32B-Instruct the sweep changes nothing, and the refit changes 134 of the 458{,}752 QKV columns, in layers 1 and 2, by one or two bits.
For Qwen3-32B the sweep lowers \(f_h\) from 10 to 8 and raises \(e_G, e_A\) by one, and the refit changes 12{,}936 of the 1{,}638{,}400 gate columns; for Qwen3-30B-A3B the sweep and the refits change nothing.

\paragraph{Where \(\FO\) is undefined.}
\label{sec:fo-undefined}
Three conditions of the definition are not implied by the formats, and \(\FO\) is undefined on inputs that violate one: the SiLU input \(|c| \le 2^{14}\) with gate shift \(k_g \le 15\); the shifts of the attention output and down projection in \([-16, 24]\); and, for a left shift, an output in \(A\).
QK-norm adds one: the unclamped head shift must satisfy \(\kappa^n_h + E_2 \ge -31\), the range of its clamp in the proof (real rows have \(k_n \ge 4\)).  A mixture-of-experts layer adds two: the router shift \(k_r \le 14\) and the slot alignment \(\epsilon - \epsilon_s \le 24\).
The upper bounds on the gate and QK-norm shifts hold on every input by the refits above.
An honest prover detects the violation and aborts.
The calibration sets the static exponents with margin, and \cref{sec:eval-abort} measures how often an input reaches these limits on real and adversarial text.
The program, its seed and its calibration data are published with \(C_W\), so anyone can recompute \(C_W\) from an open checkpoint.
The program uses floating point (32-bit matrix products for the rotation and the weight scales, rounded half to even), so a recomputation on other hardware can differ in rare ties; the integer model it produced, fixed by \(C_W\) and \(h_{\mathrm{circ}}\), is what defines \(\FO\), and quantization is outside \(\FO\).
For Qwen2.5-32B-Instruct the program takes under four minutes on one GPU: 17\,s to quantize the weights, 13\,s to calibrate and refit, and the rest to load the checkpoint and save the result; two runs give bit-identical models.

\begin{lemma}[Determinism]
\label{lem:det-fo}
For every prompt, seed and policy on which \(\FO\) is defined, \(\FO\) has exactly one output, and every intermediate value it defines lies in the range stated for it above.
\end{lemma}

\begin{proof}[Proof sketch]
Each operation is a composition of integer additions, products, floor divisions by powers of two, table lookups and clipping, with no choice.
The ranges follow from the bounds stated with each operation: \(\gad{Requant8}\) never clips by its half-open interval, the attention bounds follow from \(\tau_{\exp} \le 2^{10}\) and the block size, and \(\gad{SatAdd}\) clips the residual into \(A\).
\cref{app:nowrap} lists the bounds.
\end{proof}

\section{A wrap-free constraint system}
\label{sec:constraints}

Each operation of \(\FO\) is proved by a \emph{gadget}: committed witness columns, polynomial constraints that must vanish row by row, and lookups into a fixed table.
The design rule is that every constraint, read as an identity over \(\Z\) on values that satisfy the gadget's lookups, has both sides below \(p\) in absolute value, and below \(2^{30.9}\) except where a bias enters (\(\gad{ShiftQ8S}\)).
The identity then holds in \(\Fp\) exactly when it holds in \(\Z\) (\cref{lem:nowrap}).

\subsection{The tagged table}
\label{sec:table}

All static lookups go into one table \(\Tab\) of \(2^{26}\) rows and seven columns \((\mathrm{tag}, v_0, \ldots, v_5)\), compressed as \(\mathrm{tag} + \beta_1 v_0 + \beta_2 v_1 + \cdots\).
\cref{tab:tags} lists its tags.
A range check \(u \in [0, 2^k)\) is the single lookup \((k, u)\), so every bound costs one lookup whatever its width, and the shift-dependent bounds of the rounding gadgets come from one row per shift value.
The rows of \(\Tab\) are distinct tuples, as \cref{lem:logup} requires: within a tag every row has a different key (the value, shift or table input), and the tag separates rows of different tags, so for instance the range rows \((k, v)\) for different \(k\) and the same \(v\) are different tuples.

\begin{table}[t]
\centering\small
\begin{tabular}{@{}p{0.13\linewidth}p{0.40\linewidth}p{0.41\linewidth}@{}}
\toprule
Tag & Row & Meaning \\
\midrule
\(k \in [0, 24]\) & \((k, v)\), \(v < 2^k\) & range \\
\(\star_8\), \(\star_9\) & \((v)\), \(v < 255\) or \(v < 361\) & \(q + 127\) of an 8-bit value; \(r + 180\) after \(\gad{RoPE8}\) \\
\(\mathsf{POW}\) & \((e + 16, 2^e)\), \(e \in [-16, 30]\) & power of two in \(\Fp\) \\
\(\mathsf{SR}\) & \((k + 16, s, R, t_R, t_B, o_B)\), \(k \in [-16, 24]\) & regime of a shift by \(k\) (\cref{sec:g-shiftround}) \\
\(\mathsf{SRA}\) & \((k + 16, t_A, o_A, f)\), \(k \in [-16, 24]\) & input bound of a possible left shift \\
\(\mathsf{NPOW}\) & \((E + 15, E^+, 2^{E^+}, 2^{E^-}, [E \text{ even}], [E < 0])\) & signed normalization exponent, \(|E| \le 15\) \\
\(\mathsf{RSQ}\), \(\mathsf{RECIP}\) & \((\delta - \delta_0, \tau(\delta))\) & \(\lfloor 127 \cdot 2^{15}/\sqrt{\delta} \rfloor\), \(\lfloor 2^{32}/\delta \rfloor\) on 16-bit mantissas \\
\(\mathsf{SILU}\) & \((c + 2^{14}, \tau_{\mathrm{SiLU}}(c))\), \(|c| \le 2^{14}\) & MLP gate \\
\(\mathsf{EXP}\) & \((x, \tau_{\exp}(x))\), \(x < 2^{23}\) & softmax \\
\(\mathsf{EXPT}\) & \((d, \tau_{\exp,\theta}(d))\), \(d < 2^{24}\) & sampling at \(\theta > 0\) \\
\bottomrule
\end{tabular}
\caption{Tags of the static lookup table \(\Tab\).  Dynamic lookups (gathers) use tags of their own, outside this table.}
\label{tab:tags}
\end{table}

\subsection{Rounding gadgets}

\paragraph{\(\gad{ShiftRound}\).}
\label{sec:g-shiftround}
Inputs: an accumulator \(\mathrm{acc}\) with \(|\mathrm{acc}| + 2^{30} + 2^{23} < p\) (in our models \(|\mathrm{acc}| < 2^{29}\), and \(2^{29.6}\) in a mixture-of-experts down projection) and a per-row shift \(k\).
Witnesses: the output \(c\) and, per row, the regime \((s, R, t_R, t_B, o_B)\), looked up under tag \(\mathsf{SR}\) together with \(k\).
The prover does not choose \(k\): the first coordinate of the \(\mathsf{SR}\) lookup is the expression that defines the shift in \(\FO\), a public layer constant plus or minus committed row exponents (for the attention output, \(k = K_{co} - e_o\) with \(e_o\) the exponent of the preceding \(\gad{Requant8}\); for the logits, \(E_h + K_z\) with \(E_h\) from \(\gad{NormScale}\)), so the lookup binds the whole regime to the honest shift.
For \(k \ge 1\), \(s = 2^k\), \(R = 2^{k-1}\), \(t_R = k\); for \(k \ge 7\), \((t_B, o_B) = (31 - k, 2^{30 - k})\), and for \(1 \le k < 7\), \((t_B, o_B) = (24, 2^{23})\).
Constraints:
\[
(t_R,\ \mathrm{acc} + R - s\,c) \in \Tab, \qquad (t_B,\ c + o_B) \in \Tab.
\]
The first lookup states \(0 \le \mathrm{acc} + 2^{k-1} - 2^k c < 2^k\), so \(c = \lfloor \mathrm{acc}/2^k \rceil\).
The second bounds \(|2^k c| \le 2^{30}\) for \(k \ge 7\), and \(c \in A\) otherwise; without it, \(c + 2^{-k} p\) would satisfy the first relation in \(\Fp\).
For \(k \le 0\) the regime is \(s = 2^{k} \in \Fp\) (an inverse power of two), \(R = t_R = 0\), so the first lookup forces \(c = 2^{-k} \mathrm{acc}\) in \(\Fp\), and \(c \in A\).
That relation alone can wrap: with \(|\mathrm{acc}|\) near \(2^{29}\) and \(-k\) up to \(16\), the integer \(2^{-k}\mathrm{acc}\) may exceed \(p\), and its residue may land in \(A\).
Gadgets whose shift can be negative (the attention output, the down projection and the logits) therefore also look up \((\mathsf{SRA}, k + 16, t_A, o_A, f)\) per row, with \(f = [k \le 0]\) and \((t_A, o_A) = (24 + k, 2^{23 + k})\) if \(f = 1\), else \((0, 0)\), and per element
\[
(t_A,\ f\,\mathrm{acc} + o_A) \in \Tab.
\]
For \(k \ge 1\) the input is \(0\) and the lookup is vacuous; for \(k \le 0\) it states \(|\mathrm{acc}| \le 2^{23 + k}\), so \(|2^{-k}\mathrm{acc}| \le 2^{23}\) and the relation is wrap-free.
Cost per output: 1 committed element and 2 lookups, plus 1 lookup where the shift can be negative.

\paragraph{\(\gad{Requant8}\) and \(\gad{ShiftQ8}\).}
\label{sec:g-requant}
Per row: the exponent \(e\) with \(s = 2^e\) from \((\mathsf{POW}, e + 16, s)\); per element: \(q_i\).
For \(\gad{Requant8}\) (inputs \(|a_i| < 2^{23}\)):
\[
(e + 1,\ 2 a_i + s - 2 s q_i) \in \Tab, \qquad (\star_8,\ q_i + 127) \in \Tab.
\]
For \(\gad{ShiftQ8}\) (inputs \(|\mathrm{acc}_i| < 2^{29}\), \(e \ge 2\)) the first lookup is \((e, \mathrm{acc}_i + s/2 - s q_i)\).
These force \(q_i = \lfloor a_i / s \rceil\) and \(|q_i| \le 127\), hence every \(a_i\) fits at exponent \(e\).
\emph{Minimality}: a flag \(f = [e > e_0]\) with \(f(1 - f) = 0\) and \((1 - f)(e - e_0) = 0\); a selected index \(j^\ast\) (\cref{sec:selected}) and a sign \(\sigma \in \{0, 1\}\); and, when \(f = 1\),
\[
w = \begin{cases} 4 a_{j^\ast} - 255 s & \sigma = 1\\ -4 a_{j^\ast} - 255 s - 1 & \sigma = 0 \end{cases} \;\ge\; 0,
\]
proved by splitting \(w\) into two 15-bit limbs.
For \(\gad{ShiftQ8}\), whose inputs reach \(2^{29}\), \(4 a_{j^\ast}\) would wrap modulo \(p\); since \(s \ge 4\) there, the inequality is divided by 4,
\(w = \pm a_{j^\ast} - 255 s/4 - [\sigma = 0] \ge 0\), with \(s/4\) the field constant \(4^{-1} s\), and \(w\) is split into two 14-bit limbs (\(w < 2^{28}\)).
\(w \ge 0\) states that \(a_{j^\ast}\) does not fit at exponent \(e - 1\), so \(e\) is the least admissible exponent.
Cost per element: 1 committed element and 2 lookups, plus a few per row.

\paragraph{\(\gad{NormScale}\).}
\(\Sigma = \sum_i q_i^2 + 1\) is virtual (a sumcheck over the row).
Signed normalization: \(2^{E^-} \Sigma = 2^{E^+} \delta + \rho\) with the exponent tuple under tag \(\mathsf{NPOW}\), \([E \text{ even}] = 1\), \((E^+, \rho) \in \Tab\), \((16, [E < 0]\,\Sigma) \in \Tab\), and \(\delta\) bounded by the domain of the next table.
The third lookup bounds the input of a left shift: with \(\Sigma < 2^{16}\), \(2^{E^-}\Sigma < 2^{30} < p\), so the relation holds over \(\Z\).
Without it, a \(\Sigma\) with \(2^{E^-}\Sigma \ge p\) whose residue lies in \([2^{14}, 2^{16})\) would have a second witness; for example \(130050 \cdot 2^{14} = p + 32767\), so \(\Sigma = 130050\) would pass at \(E = -14\) with \(\delta = 32767\).
An honest left shift has \(\delta = 2^{E^-}\Sigma < 2^{16}\), so the bound never excludes it.
Then \((\mathsf{RSQ}, \delta - 2^{14}, \mu) \in \Tab\), and per element
\((15,\ q_i \mu + 2^{14} - 2^{15} q'_i) \in \Tab\) and \((\star_8, q'_i + 127) \in \Tab\).
The left shift (\(E < 0\)) handles rows whose sum of squares is small, such as a row dominated by one channel; with a right shift only, those rows would have no valid witness.

\paragraph{\(\gad{ShiftQ8S}\).}
Witnesses \(q\) and \(o \in \{-1, 0, 1\}\) with \(o^3 = o\).
If \(o = 0\): \((k, \mathrm{acc} + 2^{k-1} - 2^k (q - b)) \in \Tab\) and \((\star_8, q + 127) \in \Tab\).
If \(o = \pm 1\): \(q = 127 o\) and \(w = o(\mathrm{acc} + 2^k b) - 255 \cdot 2^{k-1} - [o = -1] \ge 0\) by two 15-bit limbs, which states that the rounded value lies beyond \(\pm 127\).
The lookups are shared between the cases by multiplying their inputs by \(1 - o^2\) and \(o^2\).
Both cases are wrap-free only for \(k \le k_{\max}(b)\) (\cref{sec:primitives}); at \(k = 24\), for instance, \(2^{24} \cdot 127 = p - 1\) and a shifted quotient would pass the first lookup.
The per-row domain therefore checks \((5,\ K_{qs} - E_h) \in \Tab\), where the layer constant \(K_{qs} = \min_n(k_{\max}(b_n) - \kappa_n)\) is a public column, so every shift \(\kappa_n + E_h\) of the row is admissible.
Cost per output: 3 committed elements, 2 lookups, constraint degree 3.

\paragraph{\(\gad{RoPE8}\).}
\((15,\ q_1 \cos - q_2 \sin + 2^{14} - 2^{15} r_1) \in \Tab\) and \((\star_9, r_1 + 180) \in \Tab\), and the same for \(r_2\); all terms are below \(2^{23}\).

\paragraph{MLP gate.}
\label{sec:g-gate}
\((k, \mathrm{acc} + 2^{k-1} - 2^k c) \in \Tab\) and \((\mathsf{SILU}, c + 2^{14}, a) \in \Tab\).
The table's domain bounds \(|c| \le 2^{14}\), which with \(k \le 15\) keeps \(|2^k c| \le 2^{29}\), so no separate quotient bound is needed.
The bound \(k \le 15\) is enforced per row by \((5,\ K_{gm} - E_h) \in \Tab\) with the public layer constant \(K_{gm} = 15 - \max_n \kappa^g_n\).

\emph{Row clamps.} The gate's row exponent \(E_g = \max(E_h, K_{gl})\) is committed per row with a flag \(f\):
\[
f(1 - f) = 0, \qquad E_g = f K_{gl} + (1 - f) E_h, \qquad (24,\ (1 - f)(E_h - K_{gl})),\ (24,\ f\,(K_{gl} - 1 - E_h)) \in \Tab.
\]
The lookups state \(E_h \ge K_{gl}\) if \(f = 0\) and \(E_h < K_{gl}\) if \(f = 1\), so the flag, and with it \(E_g\), is unique; all inputs are far below \(2^{24}\) in absolute value, so neither lookup wraps.
The other clamps of \(\FO\) (\cref{sec:qwen3}) have the same form.
For \(k = \max(v, 1)\): \(k = f + (1 - f) v\), \((24, f\,(0 - v))\) and \((5, k - 1)\), so \(f = 1\) only if \(v \le 0\) and \(f = 0\) only if \(v \ge 1\).
For \(k = \min(v, 24)\): \(k = 24 f + (1 - f) v\) and \((24, f\,(v - 25))\), with \(k \le 24\) from the shift table.
\(K_{gl} \le K_{gm}\) is checked when the model is loaded, so the gate shifts of a clamped row also stay at most 15.
The product \(m = a \cdot q_u\) is virtual: the next \(\gad{Requant8}\) reads it as a degree-2 expression.

\paragraph{\(\gad{MulRound}\).}
\label{sec:g-mulround}
Witnesses \(x_h, t, c\), with \(x_l = x - 2^{12} x_h\) virtual:
\[
(12, x_l),\ (12, x_h + 2^{11}),\ (12,\ x_l y - 2^{12} t),\ (18,\ t + 2^{17}) \in \Tab,
\]
followed by the \(\gad{ShiftRound}\) relation on \(H = x_h y + t\) with shift \(k - 12\).
Dropping \(v = x_l y - 2^{12} t \in [0, 2^{12})\) does not change the rounded quotient, because \(H + 2^{k - 13}\) is an integer and \(2^{k-12} \ge 2\).
The bound on \(t\) is necessary: \(127 \cdot 2^{24} = p - 1\), so without it \((t + 127 \cdot 2^{12}, v + 1)\) also satisfies \(x_l y = 2^{12} t + v\) in \(\Fp\).

\paragraph{\(\gad{SatAdd}\).}
With \(s = h + c \in [-2^{24}, 2^{24})\), witnesses \(h'\) and \(o \in \{-1,0,1\}\):
\(h' = (1 - o^2) s + \tfrac{o^2 + o}{2}(2^{23} - 1) - \tfrac{o^2 - o}{2} 2^{23}\), and one lookup
\((24,\ (1 - o^2)(h' + 2^{23}) + o^2 (o s - 2^{23} - [o = -1])) \in \Tab\).

\subsection{Selected indices and gathers}
\label{sec:selected}

A gadget that selects one element of a row, such as the minimality witness of \(\gad{Requant8}\), commits the index as Boolean bits \(b_1, \ldots, b_\ell\) with \(b_k(1 - b_k) = 0\) and reads the selected value as the virtual \(a_{j^\ast} = \sum_i \eq(b, i)\, a_i\).
The Boolean constraints are necessary.
For \(b \notin \bits^\ell\), the sum is the row's multilinear extension at \(b\), which a prover can steer to any value by solving for \(b\).

The attention row maximum uses a \emph{gather} instead.
Per query row \(m\) and head, the prover commits a key index \(j^\ast_m\) and the gathered key row \(K^g_m = \hat K_{j^\ast_m, \cdot}\), and per key position \(j\) the number \(\mu_j\) of (row, head) pairs that select it.
The gather is a dynamic lookup set with its own tag:
\[
\sum_{m, c} \frac{1}{Z - (\mathrm{tag} + \beta_1 j^\ast_m + \beta_2 c + \beta_3 K^g_{mc})} \;=\; \sum_{j, c} \frac{\mu_j}{Z - (\mathrm{tag} + \beta_1 j + \beta_2 c + \beta_3 \hat K_{jc})}.
\]
Two range lookups place \(j^\ast_m\) between the first position of \(m\)'s sequence and \(m\).
The maximum is then \(M_m = \sum_c \hat Q_{mc} K^g_{mc}\), proved by a product sumcheck, and the softmax lookup below forces \(M_m \ge \mathrm{acc}_{mj}\) for every valid \(j\).
Hence \(M_m\) is the maximum (\cref{lem:rowmax}).
The gather adds \(d_h + 2\) committed elements per (row, head) and \(d_h\) per key, in place of the \(\ell\) index bits and an \(\eq\)-weighted sum over a whole row of scores.

\subsection{Attention}
\label{sec:attn-proof}

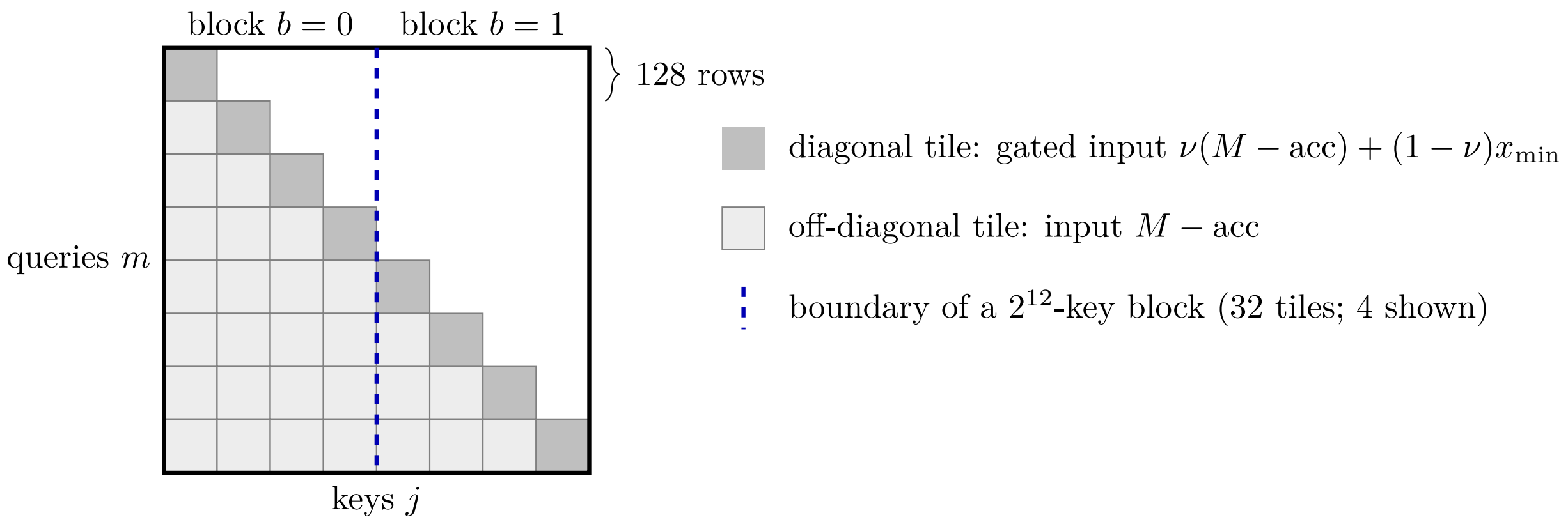
\begin{figure}[t]
\centering
\begin{tikzpicture}[scale=0.55]
  \foreach \i in {0,...,7} {
    \foreach \j in {0,...,\i} {
      \pgfmathsetmacro{\fillc}{ifthenelse(\i==\j,"black!25","black!7")}
      \fill[\fillc] (\j, -\i) rectangle ++(1,-1);
      \draw[black!50] (\j, -\i) rectangle ++(1,-1);
    }
  }
  \draw[very thick] (0,0) -- (8,0) -- (8,-8) -- (0,-8) -- cycle;
  \draw[very thick, dashed, blue!70!black] (4,0) -- (4,-8);
  \node[above] at (2,0) {\small block \(b = 0\)};
  \node[above] at (6,0) {\small block \(b = 1\)};
  \node[left] at (0,-4) {\small queries \(m\)};
  \node[below] at (4,-8) {\small keys \(j\)};
  \draw[decorate,decoration={brace,amplitude=4pt}] (8.3,0) -- (8.3,-1) node[midway,right=5pt] {\small 128 rows};
  \begin{scope}[shift={(10.5,-1.5)}]
    \fill[black!25] (0,0) rectangle ++(0.8,-0.8); \node[right] at (1,-0.4) {\small diagonal tile: gated input \(\nu(M - \mathrm{acc}) + (1 - \nu)x_{\min}\)};
    \fill[black!7] (0,-1.5) rectangle ++(0.8,-0.8); \draw[black!50] (0,-1.5) rectangle ++(0.8,-0.8); \node[right] at (1,-1.9) {\small off-diagonal tile: input \(M - \mathrm{acc}\)};
    \draw[very thick, dashed, blue!70!black] (0.4,-3) -- (0.4,-3.8); \node[right] at (1,-3.4) {\small boundary of a \(2^{12}\)-key block (32 tiles; 4 shown)};
  \end{scope}
\end{tikzpicture}
\caption{Tile list of one head.  Each kept tile holds \(2^{14}\) entries of \(P\).  Only diagonal tiles need the causal mask in the softmax lookup.}
\label{fig:tiles}
\end{figure}

The causal score matrix of a head is split into \(128 \times 128\) tiles, and only the tiles on or below the diagonal are kept (\cref{fig:tiles}): \(\frac{1}{2}(S/128)(S/128 + 1)\) tiles per sequence and head.
The probabilities \(P\) are committed per tile.
The scores \(\mathrm{acc}\) are virtual, and so are the block partials \(N^{(b)}\) and the denominators \(D\).

\begin{enumerate}
\item \emph{Softmax lookups.} For every entry of every kept tile, \((\mathsf{EXP}, x_{mj}, P_{mj}) \in \Tab\), with \(x_{mj} = M_m - \mathrm{acc}_{mj}\) on off-diagonal tiles and \(x_{mj} = \nu_{mj}(M_m - \mathrm{acc}_{mj}) + (1 - \nu_{mj}) x_{\min}\) on diagonal tiles, where \(\nu\) is the public causal mask of a tile and \(\tau_{\exp}(x_{\min}) = 0\).
The table's domain \(x < 2^{23}\) enforces \(M_m \ge \mathrm{acc}_{mj}\), and masked entries read exactly \(0\).
These lookups and nothing else form the tile domain (\cref{sec:domains}), with \(2^{14}\) entries per block and one block per tile.
\item \emph{Scores.} The domain ends in claims on \(\mathrm{acc}\) at points inside the tiles.
They are reduced to claims on \(\hat Q\) and \(\hat K\) by one product sumcheck per key-value group over (query row, channel), with the tile structure entering through public weights.
\item \emph{\(N\) and \(D\).} The rounding of the block partials, \(N_{mc} = \sum_b \gad{ShiftRound}(N^{(b)}_{mc}, k_m)\), yields a claim on the virtual \(N^{(b)}\), and the normalization of \(D\) yields a claim on \(D\).
Both are proved by one batched sumcheck over the keys \(j\):
\[
\textstyle \gamma_1 \sum_j \mle P(r_m, j)\, \eq\big(r_b, b(j)\big)\, \mle V(j, r_c) + \gamma_2 \sum_j \mle P(r'_m, j) \;=\; \gamma_1 \mle N^{(\cdot)}(r_m, r_b, r_c) + \gamma_2 \mle D(r'_m).
\]
Since blocks are aligned to \(2^{12}\) keys, \(b(j)\) is the high bits of \(j\), and \(\eq(r_b, b(j))\) is a public table.
\item \emph{Normalization.} \(D_m = 2^{E_m}\delta_m + \rho_m\) by the signed normalization gadget, with \(E_m \in [-5, 14]\) by two range lookups; \(k_m = \max(7, E_m)\) by a flag \(f_m = [E_m \ge 7]\), fixed by \((4, f_m(E_m - 7) + (1 - f_m)(6 - E_m)) \in \Tab\); \((22, (1 - f_m) D_m) \in \Tab\), which plays the role of \(\gad{NormScale}\)'s input bound: every left shift has \(D_m < 2^{22}\), so \(2^{E_m^-} D_m < 2^{27} < p\) at every \(S\), and an honest \(D_m < 2^{22}\) is exactly the case \(E_m < 7\); \((\mathsf{RECIP}, \delta_m - 2^{15}, \mathrm{recip}_m) \in \Tab\); and \(o = \gad{MulRound}(N, \mathrm{recip}, 25 + E_m - k_m + s_h)\).
\item \emph{Collecting claims on \(P\).} Claims on \(P\) arrive at points of the full \(S \times S\) square and at points inside the tile domain.
One sumcheck over the tile list, with public tile-to-row and tile-to-column weights, reduces all of them to claims on the committed tiles.
\end{enumerate}

\subsection{Linear layers, embedding and output}

A linear layer with weight \(W\) and input \(q\) reduces a claim \(\mle{\mathrm{acc}}(r_o, r_m) = v\) by the product sumcheck \(\sum_t \mle q(t, r_m)\, \mle W(r_o, t) = v\).
Weights are committed in blocks of \(2^{10} \times 2^{10}\); at the end the prover sends the value of every weight block at the final point, and the verifier checks their combination against the sumcheck's final claim.
Bias terms enter \(\gad{ShiftQ8S}\) as constants of the layer.

The embedding rows are selected by public token ids: a claim \(\mle h_0(r_d, r_t)\) on the first residual becomes, after a sumcheck over the vocabulary with the public weights \(w(v) = \sum_{t : x_t = v} \eq(r_t, t)\), a claim on the committed embedding table.

At \(\theta = 0\) the output constraints are, for every generated position \(t\) and every \(j\) in the vocabulary,
\[
\big(24,\ z_{t, y_t} - z_{tj} - [j < y_t]\big) \in \Tab,
\]
where \(z_{t, y_t}\) is read through the public one-hot of \(y_t\) and \([j < y_t]\) is public; the verifier evaluates both in closed form.
These state \(z_{t, y_t} > z_{tj}\) for \(j < y_t\) and \(z_{t, y_t} \ge z_{tj}\) for \(j > y_t\), so \(y_t\) is the first maximum.
The vocabulary here is the \(152{,}064\) rows of the checkpoint's output matrix, which the commitment layout holds without padding (\(152{,}064 = 297 \cdot 2^9\)); \(\FO\)'s argmax ranges over the same rows.
Qwen's checkpoint pads its \(151{,}665\)-token tokenizer vocabulary to this size with rows of its own; those rows are real, generally nonzero rows of \(W_{\mathrm{lm}}\), and they take part in both the definition and the constraints.
The logits are signed, and the constraints compare logits with each other only, so they hold for the honest assignment whatever the signs.

\paragraph{Sampling at \(\theta > 0\).}
\label{sec:g-sampling}
The probabilities come from \((\mathsf{EXPT}, z^{\max}_t - z_{tj}, p_{tj}) \in \Tab\), where \(z^{\max}_t\) is the logit at a committed index (Boolean bits, bounded by a range lookup when the vocabulary is not a power of two) and the table's domain forces \(z^{\max}_t \ge z_{tj}\).
One sumcheck over the vocabulary, with the public indicator tables \([j < y_t]\) and \([j = y_t]\), gives \(Z_t = \sum_j p_{tj}\), \(A_t = \sum_{j < y_t} p_{tj}\) and \(P_t = p_{t,y_t}\); all three are below \(2^{12} \cdot 152{,}064 < 2^{29.3}\).
The value \(u_t \in [0, 2^{30})\) is public, derived from the seed, and given to the domain as three public 10-bit limbs \(u_{t,i}\).
The prover commits 10-bit limbs \(z_{t,i}\) of \(Z_t\), six 10-bit limbs \(\tau_{t,0}, \ldots, \tau_{t,5}\) of the product \(X_t = u_t Z_t < 2^{60}\), and carries \(\kappa_{t,0}, \ldots, \kappa_{t,4}\), with
\[
Z_t = \sum_{i<3} 2^{10 i} z_{t,i}, \qquad
\sum_{i + i' = l} u_{t,i}\, z_{t,i'} + \kappa_{t,l-1} = \tau_{t,l} + 2^{10} \kappa_{t,l} \quad (l = 0, \ldots, 4), \qquad \kappa_{t,4} = \tau_{t,5},
\]
(\(\kappa_{t,-1} = 0\)), range lookups \((10, z_{t,i})\), \((10, \tau_{t,l})\) and \((12, \kappa_{t,l})\), and \(w_t = \tau_{t,3} + 2^{10}\tau_{t,4} + 2^{20}\tau_{t,5}\).
The terms of each limb relation sum to less than \(3 \cdot 2^{20} + 2^{12} + 2^{10} + 2^{22} < 2^{23}\) in absolute value, so the relations hold over \(\Z\), and together they state \(X_t = \sum_l 2^{10 l} \tau_{t,l}\) over \(\Z\); hence \(w_t = \lfloor u_t Z_t / 2^{30} \rfloor\) exactly, and \(w_t < Z_t < 2^{29.3}\) because \(u_t < 2^{30}\).
The inverse-CDF condition \(A_t \le w_t < A_t + P_t\) is two linear relations with witnesses split into 15-bit limbs:
\[
w_t - A_t = d_{t,1}, \qquad A_t + P_t - 1 - w_t = d_{t,2}, \qquad d_{t,i} = d^{\mathrm{lo}}_{t,i} + 2^{15} d^{\mathrm{hi}}_{t,i},\ (15, d^{\mathrm{lo}}_{t,i}), (15, d^{\mathrm{hi}}_{t,i}) \in \Tab.
\]
Their terms are bounded by \(w_t, A_t + P_t < 2^{29.3}\) and \(d_{t,i} < 2^{30}\), so each side stays below \(2^{30} + 2^{29.3} < 2^{30.6} < p\), and a violated inequality would need \(d_{t,i}\) near \(p\), above its \(2^{30}\) range.
Cost per generated token: 21 committed elements in the per-row domain and 18 range lookups, besides the \(|V|\) probabilities and their table lookups.

\subsection{Qwen3: QK-norm, router and experts}
\label{sec:qwen3-proof}

\paragraph{QK-norm.}
A query or key head row is \(d_h = 128\) consecutive columns of the QKV accumulator, inside one column block.
Per element, the domain holds the \(\gad{ShiftQ8}\) relation \((e, \mathrm{acc} + s/2 - s q_8)\) with \((\star_8, q_8 + 127)\), the output relation of \(\gad{NormScale}\), and \(\gad{ShiftQ8S}\) of \(q'' G\), with \(G\) a public column and bias 0.
Per head row, a second domain holds the \(\gad{ShiftQ8}\) minimality witness, the normalization of \(\gad{NormScale}\), and the shift \(k_n = \max(\kappa^n_h + E_2, 1) \le 22\) as a clamp, with \(\kappa^n_h\) a public per-head column.
The head-row accumulator reaches \(2^{29}\) and its exponent \(22\), so the minimality relation takes the quartered form (\(e \ge 2\)); with the unquartered form, \(4 a_{j^\ast} - 255 \cdot 2^{22}\) exceeds \(p\), and a prover could claim exponent 22 for any head row (\cref{sec:eval-tests}).
The selected element of every head row and its sum of squares are read by one gather and row-sum sumcheck.

\paragraph{Router.}
\(\gad{SatShift}\) is \(\gad{ShiftQ8S}\) with bias 0 and saturation bound \(2^{15} - 1\) in place of 127, with \((16, z + 2^{15} - 1)\) and \((16, 2^{15} - 1 - z)\) for the output; the shift \(k_r = \max(E_h - e_r - 8, 1) \le 14\) is a clamp.
The bound 14 is what keeps the saturated case wrap-free: its relation reaches \(|\mathrm{acc}| + 2^{k_r - 1}(2^{16} - 1) + 2^{30}\), above \(p\) at \(k_r = 15\).

\paragraph{Selection.}
Per token \(t\), the prover commits a threshold index \(j_t\) (Boolean bits, \(j_t < E\)) whose logit is the threshold \(\vartheta_t\), and per (token, expert \(j\)) a selection bit \(\sigma_{tj}\) and a tie flag \(\phi_{tj} = [z_{tj} = \vartheta_t]\), enforced by \(\phi(z - \vartheta) = 0\) and
\[
\big(24,\ (1 - \phi)\big(\sigma (z - \vartheta - 1) + (1 - \sigma)(\vartheta - z - 1)\big) + \phi\big(\sigma (j_t - j) + (1 - \sigma)(j - j_t - 1)\big)\big) \in \Tab,
\qquad \textstyle\sum_j \sigma_{tj} = k.
\]
An expert is selected exactly if its logit exceeds the threshold, or equals it at an index up to \(j_t\); with \(k\) selected, these are the \(k\) largest logits with ties to the smaller index, and \(\vartheta_t\), \(j_t\) are those of the last one.
The \(k\) slots of a token carry \((t, e_s, z_s, e^g_{e_s}, e^u_{e_s}, e^d_{e_s})\), and a dynamic multiset, like the attention gather, matches them with multiplicity 1 against the candidates \((t, j, z_{tj}, e^g_j, e^u_j, e^d_j)\) with multiplicity \(\sigma_{tj}\), where the exponents of candidate \(j\) are public columns.
So the slots are the selected experts, each with its own exponents.

\paragraph{Weights and merge.}
A one-hot slot flag attains \(z_{\max}\), and the table \((\mathsf{RTEXP}, z_{\max} - z_s, p_s)\) forces \(z_{\max} \ge z_s\).
The division \(w_s = \lfloor (2^{15} p_s + \Sigma_p)/(2\Sigma_p) \rfloor\) is \(\rho = 2^{15} p_s + \Sigma_p - 2 \Sigma_p w_s\) with \((24, \rho)\) and \((24, 2\Sigma_p - 1 - \rho)\), and \(w_s \in [0, 2^{14}]\) by two range lookups; with \(\Sigma_p \le 2^{15}\), every term is below \(2^{30}\).
The alignment \(\omega_s = \gad{ShiftRound}(w_s, \epsilon - \epsilon_s)\) uses a one-hot slot flag for the maximum \(\epsilon\) and \((24, \epsilon - \epsilon_s)\).
The requantization of the token's \(k d_e\) products to \([-1023, 1023]\) is \(\gad{Requant8}\) with 1023 in place of 127 (minimality \(\pm 4 a_{j^\ast} - 2047 s\)), for inputs \(|q_m \omega_s| \le 127 \cdot 2^{14}\).

\paragraph{Experts.}
Each expert product is a gathered sumcheck.
A claim on a slot's accumulator \(q' W^{e_s}\) is reduced by a sumcheck over the input dimension to claims on \(q'\) and on the slot's weight row, and a second sumcheck over the expert index, against the one-hot of \(e_s\) built from the slot's committed index bits, turns the latter into a claim on the stacked expert weights.
The down projection sums over the slots in the same way, so the token's one accumulator, below \(2^{29.6}\), enters a single \(\gad{ShiftRound}\).

\subsection{Determinism and no wrap}

\begin{lemma}[No wrap]
\label{lem:nowrap}
Every constraint of this section, read as a polynomial identity over \(\Z\) in variables that satisfy the gadget's lookups, has absolute value below \(p\): below \(2^{30.9}\) for every gadget (\cref{tab:bounds}) except \(\gad{ShiftQ8S}\), whose bound depends on the bias through \(k_{\max}\).
Hence the identity holds in \(\Fp\) if and only if it holds in \(\Z\).
\end{lemma}

\begin{lemma}[Determinism]
\label{lem:det}
Fix the public inputs and committed weights of a valid model: weights in \([-127, 127]\) and embedding rows in \(A\).
For inputs in range, every gadget has exactly one assignment of its output columns that satisfies its constraints, and it equals the definition of \cref{sec:function}.
Consequently, a full assignment that satisfies all constraints, lookups and gathers is the execution of \(\FO\).
\end{lemma}

\cref{app:nowrap} gives the bounds and the proofs.
Both lemmas are statements about the constraint system alone; \cref{sec:security} combines them with the soundness of the protocol.

\begin{remark}[Left shifts]
\label{rem:leftshift}
Without the \(\mathsf{SRA}\) bound, \(c\) would be the unique element of \(A\) congruent to \(2^{-k}\mathrm{acc}\), which equals the left shift of \(\FO\) whenever the latter lies in \(A\).
On inputs where \(2^{-k}\mathrm{acc} \notin A\), \(\FO\) is undefined and an honest prover aborts, yet the two \(\mathsf{SR}\) lookups could accept such a residue.
A cheating prover could exploit this: during development, before the bound was added, a randomized search over accumulators constructed such residues. The bound closes the gap, and a regression test (\cref{sec:eval-tests}) checks that the forged rows are rejected.
The regime \(k = 0\) is safe even without it, since \(|\mathrm{acc} - c| < 2^{29.6} + 2^{23} < p\).
For Qwen2.5-32B, the static parts of the shifts are at least \(8\) for the attention output, \(14\) for the down projection and \(1\) for the logits, so a negative shift needs a row exponent above these constants.
\end{remark}

\section{The protocol}
\label{sec:protocol}

\subsection{Overview}

\begin{figure}[t]
\centering
\begin{tikzpicture}[
  box/.style={draw, rounded corners=2pt, minimum height=0.75cm, minimum width=1.25cm, font=\small, align=center},
  arr/.style={-{Latex[length=2mm]}, thick},
  lbl/.style={font=\footnotesize, align=center}]
  \node[box, fill=black!6] (emb) at (0,0) {embed};
  \node[box] (l0) at (2.0,0) {layer 0};
  \node[box] (l1) at (4.0,0) {layer 1};
  \node (dots) at (5.6,0) {\(\cdots\)};
  \node[box] (lL) at (7.2,0) {layer \(L{-}1\)};
  \node[box, fill=black!6] (fin) at (9.6,0) {logits,\\sampling};
  \draw[arr, blue!60!black] (emb) -- (l0);
  \draw[arr, blue!60!black] (l0) -- (l1);
  \draw[arr, blue!60!black] (l1) -- (dots);
  \draw[arr, blue!60!black] (dots) -- (lL);
  \draw[arr, blue!60!black] (lL) -- (fin);
  \node[lbl, blue!60!black, above=0.55cm of l1] (cp) {commit pass: witness \(\to\) commitments, multiplicities;\\residual stream saved between layers};
  \draw[arr, red!60!black, bend left=28] (fin.south) to node[lbl, below, red!60!black] {prove pass: claims on outputs \(\to\) sumchecks \(\to\) claims on inputs, weights, committed tensors} (emb.south);
  \node[box, fill=black!6, minimum width=2.3cm] (pcs) at (12.6, 0) {open all\\commitments};
  \draw[arr, dashed] (fin.east) -- node[lbl, above] {claims} (pcs.west);
\end{tikzpicture}
\caption{The two passes.  Between the passes the transcript absorbs every commitment and the multiplicities and draws \((Z, \beta)\) after an 11-bit proof of work.  Claims on committed tensors accumulate until the tensor's commitment group is opened.}
\label{fig:passes}
\end{figure}
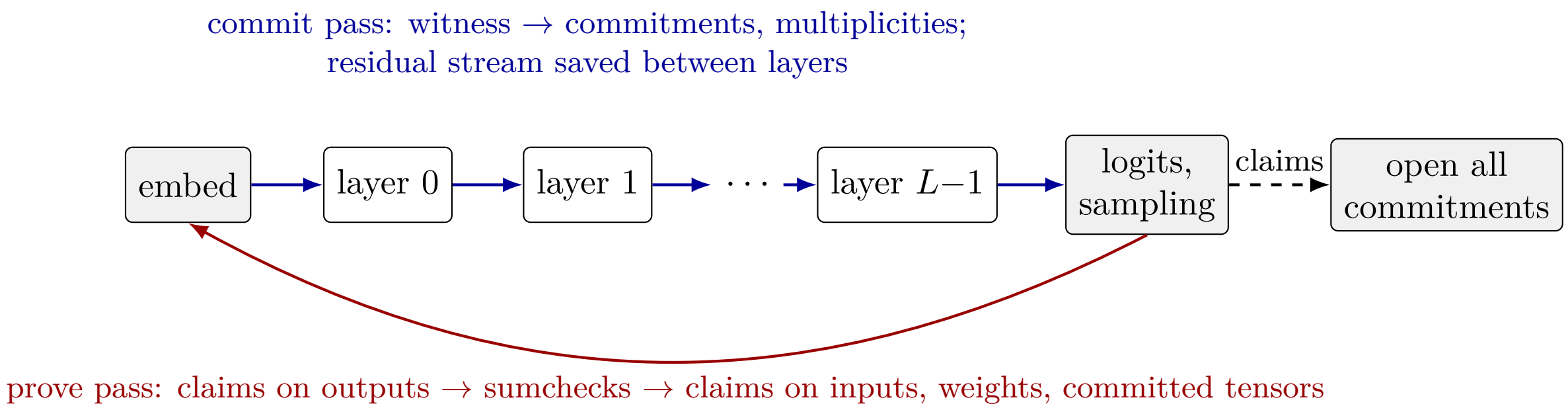

The prover makes two passes over the model (\cref{fig:passes}).
The \emph{commit pass} runs \(\FO\) forward, one layer at a time.
For each layer it builds the witness from the layer's input, commits it, and counts the table multiplicities of its lookups.
It keeps only the residual stream between layers, in host memory.
After the last layer it commits the output gadgets and the multiplicity columns, and the transcript then yields the lookup challenges \((Z, \beta)\).
The \emph{prove pass} visits the model backward, from the sampling constraints to the embedding.
Each layer's witness is recomputed from its stored input.
The layer's sumchecks consume the claims that later layers made on its output and produce claims on its input, on its weights and on its own committed tensors.
A final phase opens every committed tensor at the points the claims name.

The protocol has three building blocks: domains (\cref{sec:domains}), which prove constraints and lookups on a set of columns; the LogUp table side with chunked multiplicities (\cref{sec:chunks}); and a multilinear commitment for all committed tensors (\cref{sec:pcs}).
Product sumchecks and the attention reductions of \cref{sec:constraints} connect them.

\subsection{Domains}
\label{sec:domains}

A \emph{domain} has \(K\) blocks of \(2^n\) rows and a set of columns.
A column is \emph{full} (one value per row and block), a \emph{row} column (constant along its low \(c\) variables), or a \emph{column} column (depending only on the low \(c\) variables, possibly shared by all blocks).
Each column has a role: \emph{committed} (its claims go to the commitment), \emph{virtual} (its claims go to the protocol that defines it, such as a product sumcheck), or \emph{public} (the verifier evaluates it, in closed form where possible).
A domain carries constraints \(C_1, \ldots, C_s\), polynomials in the columns, and lookups \(\ell = 1, \ldots, L\), each with a numerator \(N_\ell\) (default \(1\)) and a tuple \((t_\ell, v_{\ell,0}, \ldots)\) of polynomials in the columns, whose compressed denominator is
\[
Q_\ell(x) = Z - \Big(t_\ell(x) + \sum_{i} \beta_{i+1} v_{\ell, i}(x)\Big).
\]
The domain asserts that every \(C_k\) vanishes on every row and that the root fractions \((P_\ell, Q_\ell)\) it sends equal \(\sum_x N_\ell(x)/Q_\ell(x)\).

For each lookup, the fraction tree pairs rows along the top in-block variable \(x_{n-1}\).
With \(y \in \bits^{n-1}\) and \(N^h(y) = N(y, h)\), the layer just above the leaves is
\[
p_\ell(y) = N^0_\ell(y) Q^1_\ell(y) + N^1_\ell(y) Q^0_\ell(y), \qquad q_\ell(y) = Q^0_\ell(y)\, Q^1_\ell(y).
\]
The domain protocol (for \(K = 1\); blocks add one phase of block variables) is:

\begin{enumerate}
\item The prover sends the root fractions.
Batched GKR over all \(L\) trees reduces them, layer by layer, to claims \(p_\ell(r) = \pi_\ell\) and \(q_\ell(r) = \kappa_\ell\) at one random point \(r \in \Fq^{n-1}\).
\item The verifier draws \(a_1, \ldots, a_L, \lambda, \gamma_1, \ldots, \gamma_s, \eta \in \Fq\).
One sumcheck over \(y \in \bits^{n-1}\) proves
\begin{equation}
\label{eq:domain}
\sum_{y} \eq(r, y) \Big[ \sum_\ell a_\ell \big(p_\ell(y) + \lambda\, q_\ell(y)\big) + \sum_k \gamma_k \big(C^0_k(y) + \eta\, C^1_k(y)\big) \Big] = \sum_\ell a_\ell (\pi_\ell + \lambda \kappa_\ell),
\end{equation}
where \(p_\ell\), \(q_\ell\) and \(C^h_k\) are expanded as polynomials in the half-columns \(f^h_j(y) = f_j(y, h)\).
\item The sumcheck ends at a point \(\rho\); the prover sends \(f^0_j(\rho)\) and \(f^1_j(\rho)\) for every column, and the verifier evaluates the bracket of \eqref{eq:domain} on them.
\item The verifier draws \(t\), and each column receives the claim \(\mle f_j(\rho, t) = (1 - t) f^0_j(\rho) + t f^1_j(\rho)\), routed according to its role.
\end{enumerate}

\begin{lemma}[Domain]
\label{lem:domain}
Suppose the columns are fixed before \(r\) is drawn, and that some constraint is nonzero on some row or some root fraction differs from the fraction sum of its leaves.
Then, except with probability at most \(\epsilon_{\mathrm{GKR}} + (n + 1)/|\Fq| + \epsilon_{\mathrm{sc}} + 1/|\Fq|\), the verifier rejects or one of the column claims of step 4 is false.
Here \(\epsilon_{\mathrm{GKR}}\) is the error of the GKR layers and \(\epsilon_{\mathrm{sc}} \le (n-1)(d+1)/|\Fq|\) that of the final sumcheck, whose bracket has degree \(d\).
\end{lemma}

The proof is in \cref{app:domain}.
Merging the GKR leaf layer into the zerocheck saves one sumcheck per domain and lets the leaf expressions be arbitrary polynomials in the columns: a remainder such as \(\mathrm{acc} + 2^{k-1} - 2^k c\) never needs its own column.
The cost is a higher degree: a constraint of degree 3 and a lookup tuple of degree 2 give a bracket of degree up to 6 in the half-columns.
Our largest domains have degree 8 to 10.

Domains with the same number of variables are proved together as blocks of one domain where possible.
For attention, each kept tile is a block of \(2^{14}\) rows, and the block index is bound after the in-block variables, so that every column ends in a single claim at a point (in-tile point, tile point).

\subsection{Chunked LogUp and the table side}
\label{sec:chunks}

Every static lookup (a lookup into \(\Tab\)) is assigned, in commit order, to a \emph{chunk} of at most \(2^{30}\) lookups; a domain's lookups are never split across chunks.
Chunk \(i\) has its own multiplicity column \(m_i \in [0, 2^{30})^{2^{26}}\), committed with the trace.
The prover sends the assignment before \(Z\) is drawn.
At the end of the prove pass the verifier sums the root fractions of the lookups of chunk \(i\) into \(S_i\), draws \(\gamma_1, \gamma_2, \ldots\), and one GKR over the table proves
\[
\sum_{v \in \Tab} \frac{\sum_i \gamma_i\, m_i(v)}{Z - \Tab(v)} \;=\; \sum_i \gamma_i S_i,
\]
ending in claims on every \(m_i\) and on the seven table columns at one point.

\begin{lemma}[Chunks]
\label{lem:chunks}
Suppose every root fraction equals the fraction sum of its lookups, and some static lookup is outside \(\Tab\).
Then, except with probability at most \(\epsilon_{\mathrm{LU}} + 1/|\Fq| + \epsilon_{\mathrm{GKR}}\), the verifier rejects or one of the claims on the multiplicity or table columns is false.
Here \(\epsilon_{\mathrm{LU}}\) bounds the probability that the fraction identity of one chunk holds at a random \((Z, \beta)\) although its multisets differ (\cref{lem:tuple}).
\end{lemma}

The proof is in \cref{app:chunks}.
The chunk bound is a soundness condition: a proof of a 16{,}384-token sequence through Qwen2.5-32B makes about \(7 \times 10^{11}\) lookups, more than \(300 p\), so a single multiplicity column would allow \(p\) copies of an out-of-table value to cancel.
The table side costs one GKR over \(2^{26}\) leaves for all chunks together, and one committed column per chunk.

Gathers and other dynamic lookups form separate sets, identified by their tags, whose fraction sums over all domains must be zero.
Their counts are below \(p\) by construction (a gather of \(d_h\) values for every (row, head) pair of one key-value group).

\subsection{Work units and the routing of claims}
\label{sec:units}

The committed tensors of a layer are grouped into \emph{units}, each owned by the step of the protocol that consumes it first in forward order:
\(\mathsf{in}\) (normalization and quantization of the residual, \(Q, K, V\), RoPE),
\(\mathsf{att}.g\) for each key-value group \(g\) (the probabilities \(P\), gathered keys, rounded block partials),
\(\mathsf{ao}\) (softmax normalization, \(\gad{Requant8}\) of the attention output, output projection, residual),
\(\mathsf{mi}\) (normalization before the MLP),
\(\mathsf{mlp}.s\) for slices \(s\) of the MLP width, and
\(\mathsf{mo}\) (down projection, residual).
Weights form one unit per layer, plus units for the embedding and the output head.
Two \emph{fixed units} hold public data that every proof for a model and shape reuses: the lookup table \(\Tab\) and the RoPE tables.

A claim is a point and a value per block of a tensor, or a point with weights over the blocks and one value.
Claims accumulate on their tensor until every protocol step that reads the unit has run; then the unit is \emph{opened} (\cref{sec:pcs-open}).
The prove pass visits steps in reverse forward order, so a unit's last claim arrives in the step that owns it.

\subsection{The commitment}
\label{sec:pcs}

All committed tensors, across all units, are committed with one multilinear polynomial commitment scheme and opened by one evaluation proof at the end.

\paragraph{Columns and packing.}
A tensor of \(K\) blocks of \(2^{c_0}\) values is packed into columns of \(2^c\) values, \(c \in [5, 23]\).
The packer picks the largest \(c\) whose zero padding stays below \(10\%\) of the tensor or below \(2^{20}\) elements (\(15\%\) for weights).
Columns of length \(2^c\) form the \emph{class} \(c\) of their unit.
Every query of the evaluation proof opens every column, so the number of columns, more than the number of committed elements, sets the proof size and the verifier's hashing.
Wide columns are therefore preferred even at some cost in padding.

\paragraph{Encoding and leaves.}
Each column of class \(c\) is encoded in \(\mathrm{RS}[c + 1]\) (rate \(1/2\), codewords of at most \(2^{24}\) symbols).
A leaf of class \(c\) holds four consecutive codeword positions of every column of the class, and is hashed with BLAKE3 in chunks of \(64\) columns whose digests are chained.
A query thus opens all columns at four positions, which also covers the first two folds.

\begin{figure}[t]
\centering
\begin{tikzpicture}[
  nd/.style={circle, draw, inner sep=0pt, minimum size=4.5mm, font=\tiny},
  lf/.style={rectangle, draw, inner sep=1pt, minimum width=6mm, minimum height=5mm, font=\scriptsize},
  lc/.style={rectangle, draw, fill=orange!25, inner sep=1pt, minimum width=6mm, minimum height=5mm, font=\scriptsize}]
  \foreach \i in {0,...,7} { \node[lf, fill=blue!12] (L\i) at (\i*1.1, 0) {\(\ell^{c_m}_{\i}\)}; }
  \foreach \i in {0,...,3} {
    \pgfmathtruncatemacro{\a}{2*\i}\pgfmathtruncatemacro{\b}{2*\i+1}
    \node[nd] (A\i) at (\i*2.2+0.55, 1.2) {};
    \draw (L\a) -- (A\i); \draw (L\b) -- (A\i);
    \node[lc] (C\i) at (\i*2.2+1.25, 1.95) {\(\ell^{c}_{\i}\)};
    \node[nd, fill=orange!10] (B\i) at (\i*2.2+0.55, 2.6) {};
    \draw (A\i) -- (B\i); \draw (C\i) -- (B\i);
  }
  \foreach \i in {0,...,1} {
    \pgfmathtruncatemacro{\a}{2*\i}\pgfmathtruncatemacro{\b}{2*\i+1}
    \node[nd] (D\i) at (\i*4.4+1.65, 3.7) {};
    \draw (B\a) -- (D\i); \draw (B\b) -- (D\i);
  }
  \node[nd] (R) at (3.85, 4.7) {};
  \draw (D0) -- (R); \draw (D1) -- (R);
  \node[right, font=\footnotesize, align=left] at (9.0, 1.2) {\(\Hash(\ell \| r)\)};
  \node[right, font=\footnotesize, align=left] at (9.0, 2.6) {\(\Hash(\Hash(\ell\|r) \,\|\, \ell^{c}_{i})\): class \(c = c_m - 1\) joins};
  \node[right, font=\footnotesize, align=left] at (9.0, 0) {leaves of the largest class \(c_m\)};
  \node[right, font=\footnotesize, align=left] at (9.0, 4.2) {upper levels; the cap is sent};
\end{tikzpicture}
\caption{A mixed-height tree for a set with two classes.  The query position \(q\) opens leaf \(q\) of class \(c_m\) and leaf \(q \gg 1\) of class \(c_m - 1\) along a single path.}
\label{fig:mixed}
\end{figure}
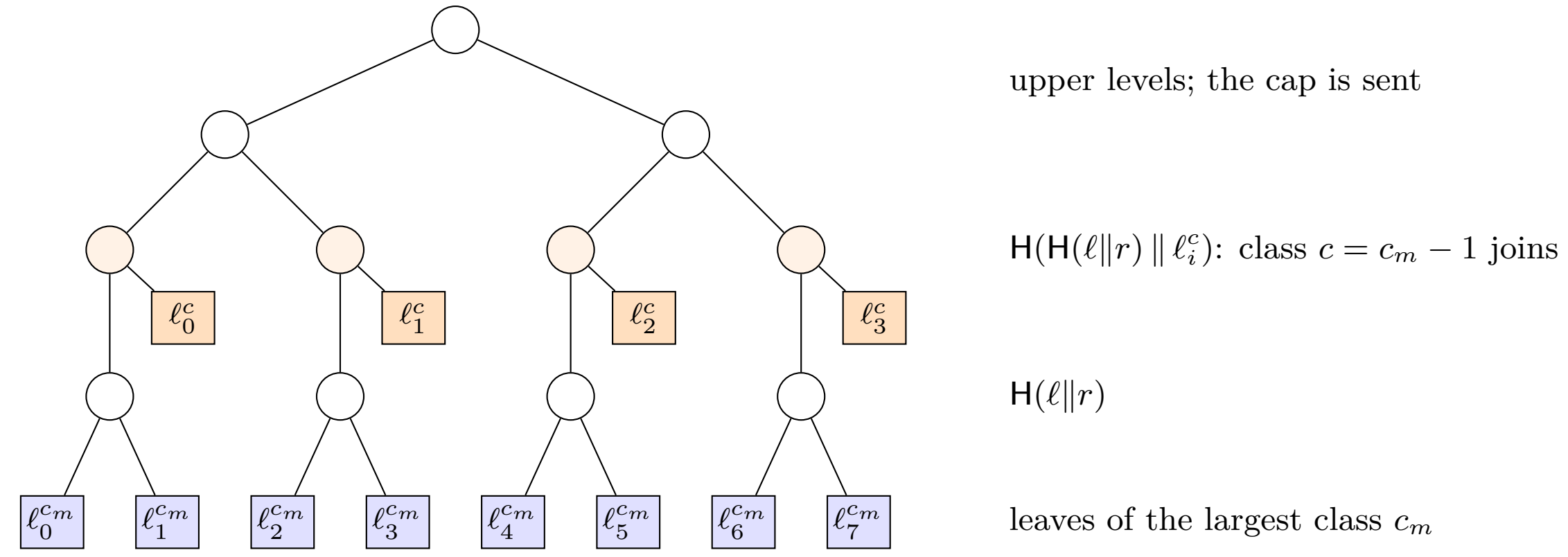

\paragraph{Mixed-height Merkle trees.}
The units of one layer form a \emph{set}, committed one after another with leaf digests chained across units.
A set's classes share one tree, built over the leaves of its largest class \(c_m\) (\cref{fig:mixed}).
The leaf digest of a smaller class \(c\) enters at the level that has \(2^{c - 1}\) nodes: a node there is \(\Hash(\Hash(\ell \| r) \,\|\, \mathrm{leaf}_c)\).
A query position \(q\) at the largest class corresponds to position \(q \gg (c_m - c)\) at class \(c\), so one path serves every class of the set.
Classes with fewer than \(2^{7}\) leaves lie entirely inside the cap and keep separate trees.
Each tree sends its cap of \(2^7\) nodes.

\paragraph{Opening a unit.}
\label{sec:pcs-open}
When a unit's claims are complete, the prover reduces them, per class, to a single point: one sumcheck in \(\Fq\) proves
\(\sum_x \sum_i \eq(z_i, x)\, g_i(x)\) equal to the combined claim, where claim \(i\) has point \(z_i\) and \(g_i\) is a random combination of the columns it touches.
At the final point \(\rho\), the prover sends \(\mle f_k(\rho)\) for every column of the class, the verifier checks the reduction, draws \(\alpha'_k \in \Fo\), and records the claim \(y = \mle g(\rho)\) for \(g = \sum_k \alpha'_k f_k\).

\paragraph{Batching claims at different points.}
Let \(N_V\) be the largest class plus one, and \(x_{[c]}\) the top \(c\) coordinates of \(x \in \bits^{N_V}\).
Given claims \((\rho_t, y_t)\) on combined columns \(g_t\) of classes \(c_t\), the verifier draws \(\lambda_t \in \Fo\), and a sumcheck over \(N_V\) variables proves
\begin{equation}
\label{eq:batchpts}
\sum_{x \in \bits^{N_V}} \sum_t \lambda_t\, \eq(\rho_t, x_{[c_t]})\, \mle g_t(x_{[c_t]}) = \sum_t 2^{N_V - c_t} \lambda_t\, y_t.
\end{equation}
It ends at a point \(r\) with the claim \(\sum_c H_c(r_{[c]}) = V\), where \(H_c = \sum_{t \in \text{class } c} \lambda_t \eq(\rho_t, r_{[c]})\, g_t\) is a fixed linear combination of the committed columns of class \(c\).
The prover runs its first 12 rounds on the vectors \(g_t(\cdot, \rho_{t,\mathrm{hi}})\) of at most \(2^{12}\) entries, kept from the unit openings; the remaining rounds need \(g_t(r_{\mathrm{lo}}, \cdot)\), and forming \(H_c\) needs every \(g_t\) in full.

\paragraph{Per-layer reduction.}
\label{sec:segments}
Applied once to all units of the proof, \eqref{eq:batchpts} would need every \(g_t\) at the end of the proof, when the witness is no longer resident, and the prover would have to rebuild every layer's witness twice more.
Instead, the units of one layer (its groups and its weights) form a \emph{segment}.
When the segment's last unit has been opened, at the end of the layer's prove pass, the prover still holds every \(g_t\) of the layer, and runs \eqref{eq:batchpts} over the segment's units alone.
This is a sumcheck over \(\Fo\), like every instance of \eqref{eq:batchpts}; it is a different step from the point reduction of a unit opening above, which runs over \(\Fq\) and precedes the draw of the \(\alpha'_k \in \Fo\).
It sends \(y_{s,c} = H^{(s)}_c(r^{(s)}_{[c]})\) for each class, and the verifier checks \(\sum_c y_{s,c} = V_s\).
The prover keeps \(H^{(s)}_c\), one vector per class instead of one per unit, in pinned host memory.

\paragraph{The global evaluation proof.}
After the prove pass, \eqref{eq:batchpts} runs once more over the units outside segments (the output gadgets, the embedding and output head, the fixed units and the multiplicity columns) and one pseudo-unit per (segment, class), with claim \((r^{(s)}_{[c]}, y_{s,c})\) on \(H^{(s)}_c\).
The prover reads the pseudo-units' vectors back from host memory, so no layer's witness is rebuilt for this step.
The result is again a claim \(\sum_c H_c(r_{[c]}) = V\) with \(H_c\) a fixed linear combination of committed columns: the coefficient of a column of a segment unit \(t\) is the product of the two levels' factors \(\lambda' \eq(r^{(s)}_{[c]}, r_{[c]}) \cdot \lambda_t \eq(\rho_t, r^{(s)}_{[c]}) \cdot \alpha'_{t,k}\).
BaseFold proves this claim by folding a single codeword, to which class \(c\)'s combined codeword is added at level \(N_V - c\), as in batched FRI for mixed degrees.
The level at which each class enters, like the class sizes themselves, is computed by the verifier from the shape in the statement (and so bound by \(h_{\mathrm{circ}}\)); the prover cannot choose it.
A fold oracle is committed at every second level, and the last message of \(2^5\) symbols is sent in the clear.
After a 30-bit proof of work, \(153\) query positions are drawn; class \(c\) opens the leaf at \(q \gg (N_V - c)\), and the oracle at level \(L\) opens the leaf at \(q \gg L\).
\cref{lem:pcs} states the soundness of this construction.

\paragraph{Weights and fixed units.}
\(C_W\) is a SHA-256 hash of the caps of the weight units.
Weights are committed once per model and shape; each proof reopens them as part of the global evaluation proof, so a proof pays for reducing and opening the weights while their encoding is reused.
The fixed units are checked the same way against digests computed once per model and shape.
The verifier therefore never evaluates the \(2^{26}\)-row table or the RoPE tables itself.
It receives their evaluations as claims that the commitment opens.

\subsection{Transcript and statement}
\label{sec:transcript}

The transcript starts by absorbing the statement: a protocol tag, the policy, \(C_W\), a digest \(h_{\mathrm{circ}}\) of the circuit description (model configuration, shapes, and every constant of \(\FO\) outside \(C_W\): the exponents, shifts and biases of every layer and the contents of \(\tau_{\exp}\), \(\tau_{\mathrm{SiLU}}\), the RoPE tables and \(\tau_{\exp,\theta}\)), \(\Sha(x)\), \(\Sha(y)\), \(n_c\), the length \(S\), and a 256-bit salt chosen by the prover and given to the verifier.
The salt has no role in soundness, which holds for every salt under the bound on random-oracle queries; it is reserved for a hiding mode, in which it makes the transcript of a repeated statement unpredictable, and our proofs set it to zero.
Challenges are drawn in this order, each after everything it tests has been absorbed:
\begin{enumerate}
\item caps of every unit, the chunk assignment and the multiplicity caps; an 11-bit proof-of-work nonce;
\item \(Z\) and the tuple coefficients \(\beta = (\beta_1, \ldots, \beta_8)\);
\item per domain and sumcheck: GKR and sumcheck round challenges after each round message, and the batching coefficients of \cref{sec:domains};
\item per unit opening: point-reduction coefficients in \(\Fq\) and column coefficients \(\alpha'\) in \(\Fo\);
\item \(\lambda_t\), the global sumcheck and fold challenges in \(\Fo\), a 30-bit proof-of-work nonce, and the query positions.
\end{enumerate}
The final query openings (leaf values and Merkle paths) are appended to the proof without being absorbed, since no challenge is drawn after them.
A challenge coordinate is sampled by rejection: 31-bit words of the hash output are drawn until one is below \(p\), so the number of compressions per challenge depends on the data.
The host verifier, the host prover and the warp-level device transcript run the same rejection loop on the same bytes, so they draw identical challenges; the tests check that the host and device transcripts produce byte-identical proofs.

\subsection{The verifier}
\label{sec:verifier}

The verifier recomputes the seed and the sampling values \(u_t\), checks that every token lies in the vocabulary, checks the policy, the stopping rule on \(y\) and that the generated tokens are the fed-back ones, and replays the transcript.
The cheap checks come first: the statement and policy checks before the transcript starts, and the caps of the weight and fixed units against \(C_W\) and the fixed digests before any sumcheck, so a proof for other weights or tables fails before any expensive work.
It checks every sumcheck and GKR round, the per-chunk and per-set fraction sums, the claims on public columns (index columns, causal masks, one-hot and prefix indicators of the targets, all in closed form with \(O(\log)\) work), the caps of the weight and fixed units against \(C_W\) and the fixed digests, the unit openings, the global sumcheck, the folds, both proof-of-work nonces, and every query.
Some model constants (biases, per-layer shift constants, the bounds \(K_{qs}\) and \(K_{gm}\)) enter claims that the verifier evaluates directly, since opening them would add columns to every query; \(h_{\mathrm{circ}}\) binds them to the statement.
The proof is a sequence of byte strings and arrays of field elements in a plain length-prefixed format; the verifier checks every length, shape and Merkle path depth against the values the protocol determines, and every element against \(p\), before using it.

\section{Security}
\label{sec:security}

We analyze the interactive oracle proof underlying Sable and obtain the non-interactive argument by the Fiat--Shamir transform.
The transcript hash (BLAKE3's compression function in sequential mode, \cref{sec:prelim-fs}) is modeled as a random oracle, and BLAKE3 is assumed collision resistant for the Merkle trees, as in deployed hash-based arguments.

\subsection{Tuple compression}

\begin{lemma}[Tuple compression]
\label{lem:tuple}
Let \(\mathcal{A}\) be a multiset of tuples in \(\Fp^{k+1}\) with \(|\mathcal{A}| < p\), let \(\mathcal{T}\) be a set of tuples, and let \(m : \mathcal{T} \to \Fq\) be arbitrary.
Let \(D\) be the number of distinct tuples in \(\mathcal{A} \cup \mathcal{T}\), and for a tuple \(u\) write \(\phi_u(Z, \beta) = Z - (u_0 + \beta u_1 + \cdots + \beta^k u_k)\).
If some tuple of \(\mathcal{A}\) is not in \(\mathcal{T}\), then for uniform \((Z, \beta) \in \Fq^2\),
\[
\Pr\Big[\sum_{u \in \mathcal{A}} \frac{1}{\phi_u(Z, \beta)} = \sum_{v \in \mathcal{T}} \frac{m(v)}{\phi_v(Z, \beta)} \text{ and all denominators are nonzero}\Big] \le \frac{k(D-1)}{|\Fq|},
\]
and some denominator vanishes with probability at most \(kD/|\Fq|\).
With independent coefficients \(\beta_1, \ldots, \beta_k\) in place of the powers of \(\beta\), the bounds become \((D-1)/|\Fq|\) and \(D/|\Fq|\).
\end{lemma}

The proof (\cref{app:tuple}) treats the identity in the bivariate rational function field \(\Fq(Z, \beta)\), where the \(\phi_u\) for distinct tuples are distinct irreducible polynomials, and applies the Schwartz--Zippel lemma to the cleared identity of total degree at most \(kD\).
A chunk has \(D \le 2^{30} + 2^{26}\) distinct tuples, and our tuples have arity up to \(7\) (\(k = 6\)).
With powers of one challenge, the lemma would give \(\epsilon_{\mathrm{LU}} \le 2 \cdot 6 \cdot (2^{30} + 2^{26}) / |\Fq| \approx 2^{-90.3}\).
Sable draws one challenge per coordinate (\(\beta_1, \ldots, \beta_8\)), so \(\epsilon_{\mathrm{LU}} \le 2(2^{30} + 2^{26})/|\Fq| \approx 2^{-92.9}\), at no measurable cost: the coefficients are a handful of extension elements per proof.
The 11-bit proof of work before \((Z, \beta)\) multiplies the cost of each attempt by \(2^{11}\), which gives \(2^{-103.9}\) in the convention of \cref{tab:budget}.

\subsection{The commitment}

\begin{lemma}[Mixed-height trees]
\label{lem:mixed}
For a mixed-height tree, an algorithm that outputs two valid openings of the same query position of the same class to different leaf contents, under the same cap, yields a BLAKE3 collision.
\end{lemma}

\begin{proof}
Both openings recompute the path from the leaf digest to the cap, injecting the digests of the other classes at their fixed levels.
The cap node is the same, so walking down from it, the first level at which the two computations differ yields two different inputs with the same BLAKE3 output.
The level at which each class enters is a function of the class sizes alone, so both computations inject at the same levels.
If the leaf contents differ, then the chained leaf digests differ or collide, so a difference exists at some level.
\end{proof}

\begin{lemma}[Evaluation proof]
\label{lem:pcs}
Assume the trees are binding.
If some recorded claim \((\rho_t, y_t)\) is false for every codeword close to the committed oracles, then the verifier of the global evaluation proof accepts with probability at most
\(\epsilon_{\mathrm{prox}} + (\sqrt{\rho} + \eta)^{q} \cdot 2^{-30} + \epsilon_{\mathrm{sc},8}\),
with \(\rho = 1/2\), \(q = 153\) queries and a 30-bit proof of work, where \(\epsilon_{\mathrm{prox}}\) is the correlated-agreement error for affine spaces in the Johnson regime~\cite[Thm.~1.6]{bciks} over \(\Fo\) for codewords of at most \(2^{24}\) symbols and \(\epsilon_{\mathrm{sc},8}\) the error of the sumchecks over \(\Fo\).
\end{lemma}

\paragraph{Which theorem this uses.}
The original analysis of BaseFold~\cite{basefold} is in the unique-decoding regime.
The list-decoding regime, which our 153 queries need, is analyzed for BaseFold over Reed--Solomon codes by Hab\"ock~\cite{habock24basefold}, through a correlated-agreement theorem for linear subcodes of Reed--Solomon codes that generalizes the one of BCIKS.
Our evaluation proof differs from the protocol analyzed there in three ways, and the argument below rests on the following conditions.
\begin{enumerate}
\item \emph{Mixed lengths.} The combined codeword of class \(c\) enters the folding at level \(N_V - c\) with fresh coefficients \(\alpha'\) (and \(\lambda\)) in \(\Fo\), drawn after the class's columns are committed, as in batched FRI for mixed degrees. At such a level the word that is folded next is an affine combination, with independent uniform coefficients, of the folded word and the new class's columns, all of the same length, so the affine-space correlated-agreement bound applies to that level's words. At most \(19\) classes enter (\(2^5\) to \(2^{23}\)), so a union bound over the entry levels costs at most \(\log_2 19 < 4.3\) bits of the proximity term.
\item \emph{Two folds per oracle.} A fold oracle is committed at every second level, so the queries check a fold of four symbols by two challenges. We treat it as two consecutive binary folds, each by a uniform challenge drawn after the word it folds is fixed; the intermediate word is not committed but is a fixed function of committed oracles, and the query check verifies the composition. Each binary fold is a line through two words, to which the line version of correlated agreement applies.
\item \emph{Batching with the sumcheck.} The folding challenges are the challenges of the evaluation sumcheck over \(\Fo\), as in BaseFold, so the sumcheck terms (\(\epsilon_{\mathrm{sc},8}\)) and the proximity terms are separate rows of \cref{tab:budget}.
\end{enumerate}
We are not aware of a published theorem that covers exactly this combination of the list-decoding regime, mixed-length batching at later levels and two folds per oracle.
The argument below composes the BCIKS correlated-agreement theorems round by round, as analyses of batched FRI do, and it is the one step of our soundness analysis that is not a direct application of a stated theorem.
The proximity term has a wide margin (\(163.6\) bits against the \(103.4\) bits of the query term), so the conclusion is not sensitive to the union bounds above.
The per-layer reductions of \cref{sec:segments} insert one more sumcheck of degree 2 over \(\Fo\) between a unit's claims and the global evaluation proof: a false claim \((\rho_t, y_t)\) of a segment unit makes the segment's batched claim false except with probability \(1/|\Fo|\) over the \(\lambda_t\), and the sumcheck then ends in a false value \(y_{s,c}\) for some class except with probability \(2 N_V / |\Fo|\); that pseudo-unit claim is a claim on a fixed linear combination of committed columns, which the global proof tests like any other.
The independent coefficients \(\alpha'_k\) and \(\lambda_t\) in \(\Fo\) make the batched codeword far from the code whenever one of its constituents is, except with the correlated-agreement error, which is included in \(\epsilon_{\mathrm{prox}}\).
For \(\eta = 0.01\), \(m = \lceil \sqrt{\rho}/(2\eta) \rceil = 36\), and codewords of \(2^{24}\) symbols, correlated agreement over lines holds with error \(\epsilon_{\mathrm{line}} = (m + \tfrac12)^7 n^2 / (3\rho^{3/2}|\Fo|) \approx 2^{-163.6}\) for \(\delta \le 1 - \sqrt\rho - \sqrt\rho/(2m)\)~\cite[Thm.~5.1]{bciks}.
The batch draws an independent coefficient per column, so the combined words form an affine space.
The proof of correlated agreement over affine spaces~\cite[Thm.~1.6, Sec.~6.3]{bciks} applies the line theorem as a black box and needs, besides, fewer than \(|\Fo|\) codewords within distance \(\delta\) of a word, which holds in the Johnson regime; so it holds with the same error, and \(\epsilon_{\mathrm{prox}} = \epsilon_{\mathrm{line}} \approx 2^{-163.6}\) per entry level (above \(2^{-159}\) after the union bound of condition 1), and the queries give \(153 \cdot 0.4797 + 30 = 103.4\) bits.

\subsection{Soundness}

\begin{table}[t]
\centering\small
\begin{tabular}{@{}lll@{}}
\toprule
Term & Bound per occurrence & Bits \\
\midrule
Sumcheck or GKR round, degree \(\le 19\), \(\Fq\) & \(19/|\Fq|\) & \(119.7\) \\
Zerocheck point of a domain, \(\le 48\) variables & \(49/|\Fq|\) & \(118.3\) \\
Batching coefficients (domain, matmul, opening) & \(2/|\Fq|\) & \(123.0\) \\
LogUp, one chunk, independent tuple coefficients, 11-bit proof of work & \(2 D / |\Fq| \cdot 2^{-11}\) & \(103.9\) \\
Gathers, \(D \le 2^{30}\) (\(S \le 2^{20}\), group \(\le 8\)), same proof of work & \(2 D / |\Fq| \cdot 2^{-11}\) & \(104.0\) \\
Chunk combination \(\gamma_i\) & \(1/|\Fq|\) & \(124.0\) \\
Proximity gaps, affine spaces, Johnson regime, \(\Fo\), \(n \le 2^{24}\) & \(\approx 2^{-163.6}\) & \(163.6\) \\
Queries: \(153\) at rate \(1/2\), \(\eta = 0.01\), 30-bit proof of work & \((\sqrt{1/2} + 0.01)^{153} \cdot 2^{-30}\) & \(103.4\) \\
Sumchecks and folds over \(\Fo\) (global and per-layer) & \(\le 3/|\Fo|\) & \(> 240\) \\
\midrule
Round-by-round soundness error & & \(\mathbf{103.4}\) \\
\bottomrule
\end{tabular}
\caption{Soundness budget of the main proof.  Gathers: \(D\) counts the gathered key rows of an attention group of at most 8 query heads per key-value head (5 in Qwen2.5-32B, 8 in the Qwen3 models); the router's slot multiset has fewer.  Bits are \(\log_2(1/\epsilon)\) per occurrence; a prover making \(Q\) random-oracle queries succeeds with probability at most about \(Q \cdot 2^{-103.4}\).}
\label{tab:budget}
\end{table}

\begin{theorem}[Soundness]
\label{thm:soundness}
Model BLAKE3 as a random oracle.
Let \(\mathsf{st}\) be a statement whose \(C_W\) commits to a valid model (weights in \([-127, 127]\) and embedding rows in \(A\), as produced by the program of \cref{sec:quantize}), and such that no \(W\) committed by \(C_W\) satisfies \((\mathsf{st}, W) \in \mathcal{R}\).
Then any prover that makes at most \(Q\) random-oracle queries makes the Sable verifier accept \(\mathsf{st}\) with probability at most \(Q \cdot \epsilon_{\mathrm{rbr}} + O(Q^2 / 2^{256})\), where \(\epsilon_{\mathrm{rbr}} \le 2^{-103.4}\) is the maximum of the terms in \cref{tab:budget}.
Here a query counts once per evaluation of the transcript hash, and the proofs of work make each attempt at the two challenges they precede cost \(2^{30}\) and \(2^{11}\) evaluations: without them the two tightest terms are \(\epsilon_{\mathrm{query}} = (\sqrt{1/2} + 0.01)^{153} = 2^{-73.4}\) and \(\epsilon_{\mathrm{LU}} = 2^{-92.9}\) per attempt.
\end{theorem}

\begin{proof}[Proof sketch]
We show that the underlying IOP has round-by-round soundness error \(\epsilon_{\mathrm{rbr}}\); the theorem then follows from the analysis of the Fiat--Shamir transform for such IOPs~\cite{rbr,bcs16}, with the \(O(Q^2/2^{256})\) term accounting for hash collisions in the Merkle trees (\cref{lem:mixed}).

The state function of the round-by-round analysis is the following predicate.
Call a partial transcript \emph{doomed} if some claim it contains is false with respect to the committed oracles: a claim on a column's multilinear extension, a sumcheck or GKR claim, a fraction sum, or a recorded opening claim.
Every term of \cref{tab:budget} is the probability that one challenge turns a doomed transcript into one that is not, so it is a per-round escape probability; none of them is a union bound over a whole proof.
A \(1 \times 4\)K proof draws about \(5.1 \times 10^5\) challenges (\cref{tab:vcount}), and \(\epsilon_{\mathrm{rbr}}\) bounds the escape probability at each of them.
In particular, the LogUp test is a single round for all chunks: \(Z\) and the coefficients \(\beta\) are drawn once, after all multiplicities are committed, and a doomed transcript escapes only if the identity of a chunk that contains a bad lookup holds at the drawn point, which \cref{lem:tuple} bounds for that chunk.
By \cref{lem:nowrap,lem:det}, if every constraint, static lookup, gather and output constraint holds for the committed assignment, then the assignment is the execution of \(\FO\) on the committed weights, and \((\mathsf{st}, W) \in \mathcal{R}\).
Hence, for a false statement, the committed assignment violates at least one of them, and the statement checks of \cref{sec:verifier} reject any assignment that violates none.
The first challenge that tests the violated item then makes the transcript doomed except with probability bounded by one term of the budget: \cref{lem:domain} for constraints and leaf fractions, \cref{lem:chunks,lem:tuple} for lookups and gathers, \cref{lem:sumcheck} for each product and reduction sumcheck, and the zerocheck bound for the output constraints.
Each later round keeps a doomed transcript doomed except with probability at most the corresponding per-round term, since a false claim entering a sumcheck round yields a false claim leaving it unless the verifier's challenge hits a root of a nonzero polynomial of the stated degree.
At the end, a doomed transcript contains a false claim on a committed column, which the global evaluation proof rejects except with the probability of \cref{lem:pcs}.
\end{proof}

\paragraph{Weight binding.}
\(C_W\) is the digest of the weight units' caps, so by \cref{lem:mixed} a proof that verifies against \(C_W\) is bound to one committed weight assignment.
The proof does not check the ranges of the committed weights; they are a property of \(C_W\), established by whoever computes it.
For open weights, anyone recomputes \(C_W\) from the published checkpoint and quantization program, which also establishes the ranges.
For closed weights the party that computes \(C_W\) is the one the verifier distrusts, and with out-of-range weights \cref{lem:nowrap,lem:det} no longer hold; a one-time range proof of the committed weights per model release (one lookup per weight, about \(3.3 \times 10^{10}\) lookups, a small fraction of one 16K proof's \(7.2 \times 10^{11}\)) would remove the assumption, and we have not built it.

\paragraph{What is not claimed.}
The main proof is not zero-knowledge: the opened columns and claimed evaluations reveal information about the trace and about the weights.
The weight codewords are fixed and every proof opens them at fresh positions, so about \(1.9 \times 10^4\) proofs reveal enough codeword symbols to recover every weight column by linear algebra (\cref{sec:threat}); the system as built is for open-weight models.
The statement concerns \(\FO\): a sequence produced by other kernels will in general not satisfy \(\mathcal{R}\), because one changed rounding decision changes every later token.
Tokens injected by the server after the prompt (tool outputs, retrieved documents) are part of \(x\) only if the client can check their provenance.

\section{A single-GPU prover}
\label{sec:gpu}

The prover is written in Python on top of PyTorch, with CUDA kernels compiled at run time by NVRTC.
Python orchestrates the protocol, and all arithmetic on witness-sized data runs on the GPU.
This section describes the techniques that matter for its speed; \cref{sec:eval-ablation} measures them.

\subsection{Memory: streaming and recomputation}

The witness of a 16{,}384-token proof is several terabytes, so no pass keeps it.
The commit pass stores only the residual stream at layer boundaries (\(S \cdot d\) values of 32 and 8 bits per layer, in pinned host memory).
Each later phase that needs a layer's witness (the prove pass and the query phase) recomputes it from the stored boundary with the same integer kernels as the forward pass, about \(42\)\,ms per layer at 4{,}096 tokens.
The per-layer reduction of \cref{sec:segments} exists to keep this count at three builds per layer (commit, prove, query): without it, the global evaluation proof needed two more full passes over every layer, and removing them saved \(8.2\%\), \(11.5\%\) and \(12.5\%\) of the proving time at the three shapes (\cref{tab:ablation}).
Codewords are never stored either: the commit pass hashes them as it encodes, and the global evaluation proof re-encodes the combined columns.
Peak GPU memory is \(37\)\,GiB for one 4{,}096-token sequence and \(72\)\,GiB for one 16{,}384-token sequence.
Peak host memory is \(100\) and \(174\)\,GiB; the model's 8-bit weights, kept in host memory and streamed to the GPU layer by layer, take about \(33\)\,GiB of it, and the boundaries and the pinned pool of the per-layer reduction most of the rest.

\subsection{Generated sumcheck kernels}
\label{sec:gpu-kernels}

Every domain defines a polynomial program, the bracket of \eqref{eq:domain}, over its half-columns.
The prover turns each program into a CUDA kernel that computes one round of the sumcheck.
Each thread handles pairs of rows, evaluates the program at the round's interpolation points \(\{0, 2, \ldots, d\}\), and reduces the results across the grid.
Coefficients that depend on challenges are kernel arguments read from shared memory, and format constants that differ per layer come from a device buffer, so one compiled kernel serves every layer of a model.
Large programs (the MLP domain has dozens of columns and degree 10) use 224 to 255 registers per thread, and their rounds are limited by instruction latency while memory bandwidth stays far from saturated.
Moving the extension-field coefficients from registers to shared memory reduced spilling and cut the GPU time of a layer by \(3.6\%\).

\paragraph{Base-field first rounds.}
\label{sec:gpu-small}
All committed values are base-field integers.
The first round of every sumcheck reads base-field tables and evaluates the program with base-field arithmetic where its coefficients allow; the second round reads the base table again and folds it twice, with the first two challenges, before writing the quarter-size extension table that the third round reads.
The more aggressive small-value algorithms of~\cite{bdt24,ddbdt26} expand rounds 2 and 3 over Lagrange nodes and need the program's coefficients in the base field.
Ours are extension elements (challenge-weighted combinations of constraints and lookups), so those algorithms would roughly double the evaluations per pair; our estimate of their gain on top of the current scheme is about \(1\%\) of the proof.

\paragraph{Lookup leaves.}
A lookup with numerator one contributes \(a(q_0 + q_1) + a \lambda q_0 q_1\) to \eqref{eq:domain}.
We evaluate it as \(a\lambda (q_0 + \mu)(q_1 + \mu) - a\mu\) with \(\mu = 1/\lambda\), folding \(a\lambda\) into the coefficients of \(q_0\).
This halves the extension multiplications per lookup in the compute-bound first round.

\paragraph{GKR rounds.}
Large GKR rounds run at 83 to 88\% of the GPU's memory bandwidth and are close to the roofline for reading, folding and writing the layer.
Small rounds (at most \(2^{15}\) pairs) are latency-bound; we spread their (pair, lookup) work items over all blocks of the grid instead of looping over lookups in one thread, which cut the GPU time of a layer from \(1.01\) to \(0.90\)\,s at 4{,}096 tokens.

\subsection{Field arithmetic}
\label{sec:gpu-field}

Base-field products use Barrett reduction of 62-bit products.
For \(\Fq = \Fp[Y]/(Y^4 - 3)\) we use the schoolbook product with the factor \(3\) folded into one operand: each output coordinate is a sum of four products below \(p^2\), so the sum is below \(4p^2 < 2^{64}\) and needs a single reduction.
This raised the measured throughput of \(\Fq\) multiplications from \(255\) to \(371\) billion per second on our GPU (one reduction per coordinate against Karatsuba over \(\F_{p^2}\) with six reductions), and cut the MLP domain's sumcheck from \(40.8\) to \(36.6\)\,ms.
An instruction count had predicted no gain, because the 64-bit Barrett reduction costs more than the multiplications it saves.
\(\Fo\) uses Karatsuba over the faster \(\Fq\); a delayed reduction does not fit in 64 bits there, since eight products would be summed per coordinate.
The same delayed reduction in the linear-combination kernel of the commitment brings it to the memory bandwidth limit (\(4.79 \to 3.83\)\,ms for 128 columns of \(2^{23}\)).

\paragraph{Opening kernels.}
A unit opening needs, for every column, its value at the reduced point \(\rho\), and the combination \(\sum_k \alpha'_k f_k\) folded at the high coordinates of \(\rho\).
Computing the values as dot products with a full \(\eq(\rho, \cdot)\) table reads 16 bytes of table per 4-byte column element.
One kernel instead folds every column at the high coordinates with a table of at most \(2^{12}\) entries, reading each column once; the values and the folded combination follow from the folded vectors of at most \(2^{11}\) entries.
This cut the proving time of 8 layers by \(1.6\%\) at 4{,}096 tokens and \(2.4\%\) at 16{,}384 tokens.

\subsection{A device-resident transcript}
\label{sec:gpu-transcript}

A sumcheck round needs the transcript between two kernels: absorb the round message, squeeze the challenge, fold.
With the transcript on the host, each round costs a device-to-host copy and a synchronization, and the prove pass of one layer at 4{,}096 tokens has several thousand rounds.
Sable keeps the transcript state (\cref{sec:prelim-fs}) on the GPU.
The last block of each round kernel absorbs the round message and derives the challenge.
One warp performs each compression, with lane \(j < 16\) holding word \(j\) of the block, so that a block is filled in one step and the compression reads its message through warp shuffles.
This brought a round's transcript step from \(25.9\) to \(8.8\,\mu\)s.
The device transcript absorbs exactly the bytes the host transcript would and samples challenges by the same rejection loop, so proofs are byte-identical either way, which the tests check; the verifier always uses the host implementation.

The prover never needs the values of the claims it produces, since they matter only to the verifier's checks.
In the prove pass, claimed values, points and coefficients therefore stay on the device as lazy values, and the commitment's openings run on the same device transcript.
A whole layer's prove pass then runs without a host synchronization.

\subsection{Graph replay}
\label{sec:gpu-graph}

The prove passes of the middle layers launch the same kernels with the same shapes; only the data differs.
Sable captures the prove pass of layer \(L - 2\) once as a CUDA graph and replays it for layers \(L - 3, \ldots, 1\).
Before each replay, the inputs are written into fixed buffers: the layer's weights, its stored boundary, its format constants, and the incoming claims.
After each replay, the outputs are relabeled: claims on layer \(l - 1\) with shifted challenge indices, the opening states, and the transcript messages.
Three rules keep the replay identical to a plain run.
Every constant that depends on the layer index is read from a device buffer.
The transcript absorbs the layer index once at the start of the layer, and every later label omits it.
Claims that a layer makes on the fixed units are replayed per layer from a template.
The proof is byte-identical to that of the plain prover.
At 4{,}096 tokens, replay brings a 64-layer proof from \(133\) to \(106\)\,s, \(0.43\)\,s per replayed layer; at 16{,}384 tokens the prover is GPU-bound and the gain is \(1\%\).
The gain is larger than the \(0.27\)\,s of host-side Python time per layer that we measured on 8-layer proofs without replay; we did not isolate the remainder, which may come from the order of kernels in the captured graph.

\subsection{Other kernels}

Reed--Solomon encoding uses a radix-\(2\) NTT in two passes for codewords up to \(2^{24}\) symbols: the last pass covers up to 13 contiguous bits, the others up to 10, in tiles of 8{,}192 symbols.
Encoding 26 codewords of \(2^{24}\) symbols takes \(6.2\)\,ms.
Softmax multiplicities come from a histogram that the attention witness kernel accumulates in shared memory, which avoids rebuilding the tile domain's columns and saved \(3\%\) at 16{,}384 tokens.
Multiplicity columns stay on the host as bytes with a sparse list of larger entries, and are compressed on a side stream.

\section{Toward recursive compression}
\label{sec:recursion}

A proof of 134.8\,MiB is acceptable for an auditor who verifies it once, but too large to return with every response or to publish.
The intended form for every client is a recursive proof: a proof, in the same proof system, that the Sable verifier accepts the main proof.
Its outer layers would be zero-knowledge, and the final layer would use a low-rate code for a short proof.
Its cost is set by what the main verifier computes, so this section first reduces the verifier's work and then gives the first circuit component, BLAKE3.

\subsection{What the verifier computes}

We count the verifier's operations by instrumenting the reference verifier.
For the full 64-layer model on one 4{,}096-token sequence, the verifier originally performed \(17.8\) million BLAKE3 compressions: \(10.6\) million in the transcript and \(7.2\) million for Merkle leaves and paths.
It also performed about \(11\) million \(\Fq\) multiplications and, through the public columns it evaluated by folding, about \(4.7 \times 10^9\) folded elements per proof.
A recursive circuit that pays about \(4096\) trace cells per compression and \(16\) per extension operation would cost more than a third of the main proof.
We removed this work in the following steps, each a protocol change with its own tests.

\begin{enumerate}
\item \emph{Closed-form public columns.}
Every structured public column has an \(O(\log n)\) closed-form multilinear extension: index columns \(i\) and \((i \gg s) \ll s\) are \(\sum_j 2^j x_j\) over the kept bits, the causal tile mask \([c \le r]\) is a comparator evaluated from the top bit, and the vocabulary indicators \([v < y_t]\), \([v = y_t]\) and one-hots are comparators against the constant \(y_t\), weighted by \(\eq\) over the rows.
\item \emph{Committed fixed tables.}
The \(2^{26}\)-row table \(\Tab\) and the RoPE tables are committed once per model and shape as fixed units.
The prover sends the table's evaluations, which become claims on these units, and the verifier compares their caps with precomputed digests.
On a small test model this cut the verifier's folded elements from \(4.7 \times 10^9\) to \(3 \times 10^4\).
\item \emph{Recorded final openings.}
After the query positions are drawn, nothing else is drawn, so absorbing the leaf values and paths into the transcript binds nothing; they are now recorded in the proof without being hashed.
For one layer this reduced the absorbed bytes from \(11.0\) to \(1.5\)\,MiB.
\item \emph{Fewer, wider columns.}
Every query opens every column, so the number of columns, more than the number of committed elements, sets the leaf hashing.
Tensors with few small blocks share columns, with padding allowed up to \(2^{20}\) elements; the attention probabilities are packed into wide columns after the prover sends each block's value at the shared point; the weights are packed flat.
Columns fell from \(84.8\)K to \(32.6\)K at a cost of \(2.5\%\) more committed elements.
\item \emph{Shared Merkle paths.}
A layer's units share trees, first one per class and then one mixed-height tree per layer (\cref{sec:pcs}).
\end{enumerate}

\begin{table}[t]
\centering\small
\begin{tabular}{@{}p{0.46\linewidth}rrrr@{}}
\toprule
& Proof & \multicolumn{3}{c}{BLAKE3 compressions} \\
\cmidrule(l){3-5}
Configuration & (MiB) & Transcript & Merkle & Total \\
\midrule
Before the reductions & \(360.0\) & \(10.56\)M & \(7.20\)M & \(17.77\)M \\
\(+\) fixed units, recorded openings, trees per class, small-tensor packing & \(224\) & \(\approx 2.1\)M & \(4.1\)M & \(\approx 6.2\)M \\
\(+\) packing of \(P\) and the weights & \(154.9\) & \(2.15\)M & \(3.09\)M & \(5.24\)M \\
\(+\) mixed-height trees & \(132.5\) & \(2.12\)M & \(2.48\)M & \(4.60\)M \\
\(+\) per-layer reduction, sequential-mode transcript & \(134.1\) & \(1.72\)M & \(2.50\)M & \(4.22\)M \\
\bottomrule
\end{tabular}
\caption{Proof size and verifier BLAKE3 compressions for Qwen2.5-32B, 64 layers, one 4{,}096-token sequence.
In the first four rows the transcript used BLAKE3's tree mode, and its compressions are counted as absorbed bytes over 64 plus two per challenge, a lower bound, since every challenge then needs one compression per level of the chunk tree; the second row's count is estimated from its absorbed bytes and its number of challenges.
Two range lookups per layer that close no-wrap gaps, and the row clamps of the gate shift (\cref{sec:g-gate}), bring the final proof to \(134.8\)\,MiB.
The last row is an exact count of the sequential mode (\cref{sec:prelim-fs}).}
\label{tab:verifier}
\end{table}

\cref{tab:verifier} gives the totals.
The remaining transcript work is driven by challenges: \(506\)K challenges, mostly GKR rounds, of which a tree of depth \(n\) has \(n(n+1)/2\), each cost one output compression, and their round messages and re-absorbed outputs make up most of the \(1.21\)M absorbing compressions.
The remaining Merkle work is mostly leaf hashing.
The final verifier also performs \(11.0\) million \(\Fq\) multiplications, \(4.6\) million \(\Fq\) additions, \(0.26\) million \(\Fq\) inversions and \(0.22\) million \(\Fo\) multiplications.
Under the cell counts above, layer 1 of the recursion would have about \(1.8 \times 10^{10}\) cells, which the cost model of \cref{sec:eval-model} prices at about \(10\%\) of the main proof.
The Python reference verifier takes \(94\)\,s for this proof (\cref{sec:eval}).

\subsection{BLAKE3 in a domain}

The recursive verifier is a computation over \(\Fp\) and its extensions plus BLAKE3, and we prove it with the same domains, lookups and commitment as the main proof.
BLAKE3's compression function applies the function \(G\) 56 times (7 rounds of 8 calls) to a state of sixteen 32-bit words, with two message words per call:
\begin{align*}
s_1 &= a + b + m_x, & x_1 &= d \oplus s_1, & d_1 &= x_1 \ggg 16,\\
s_2 &= c + d_1, & x_2 &= b \oplus s_2, & b_1 &= x_2 \ggg 12,\\
s_3 &= s_1 + b_1 + m_y, & x_3 &= d_1 \oplus s_3, & d_2 &= x_3 \ggg 8,\\
s_4 &= s_2 + d_2, & x_4 &= b_1 \oplus s_4, & b_2 &= x_4 \ggg 7,
\end{align*}
with outputs \((s_3, b_2, s_4, d_2)\) and additions modulo \(2^{32}\).

\paragraph{The G row.}
A 32-bit word does not fit in \(\Fp\), so every word is held as four bytes.
One domain row proves one call of \(G\) with 80 columns: the bytes of the inputs \(a, b, c, d, m_x, m_y\); the bytes of \(s_1, \ldots, s_4\); the carries \(k_{i,1..4}\) of the four additions; the bytes of \(x_1\) and \(x_3\); the nibbles \(h_j, l_j\) of the bytes of \(x_2\); and the top bit \(t_j\) and low seven bits \(r_j\) of the bytes of \(x_4\).
\begin{itemize}
\item \emph{Additions}: per byte \(j\), a linear constraint such as \(a_j + b_j + m_{x,j} + k_{1,j} - s_{1,j} - 256\, k_{1,j+1} = 0\) (with \(k_{i,0} = 0\)), and the carry range \(k(k-1)(k-2) = 0\) for the three-operand additions and \(k(k-1) = 0\) for the others.
\item \emph{XOR}: lookups into three tables of \(2^{16}\) rows: \((u, v, u \oplus v)\) for \(x_1\) and \(x_3\); \((u, v, (u \oplus v) \gg 4, (u \oplus v) \mathbin{\&} 15)\) for \(x_2\); and \((u, v, (u \oplus v) \gg 7, (u \oplus v) \mathbin{\&} 127)\) for \(x_4\).
The tables contain only bytes, so these lookups also bound every byte they read.
\item \emph{Rotations}: by 16 and 8 bits they permute bytes: \(d_{1,j} = x_{1,(j+2) \bmod 4}\), \(d_{2,j} = x_{3,(j+1) \bmod 4}\).
By 12 and 7 bits they recombine the split parts: \(b_{1,j} = h_{(j+1) \bmod 4} + 16\, l_{(j+2) \bmod 4}\) and \(b_{2,j} = 2\, r_{(j+1) \bmod 4} + t_j\).
\end{itemize}
A row has 32 constraints of degree at most 3 and 16 lookups.
All sums are below \(2^{11}\), so the constraints are wrap-free.

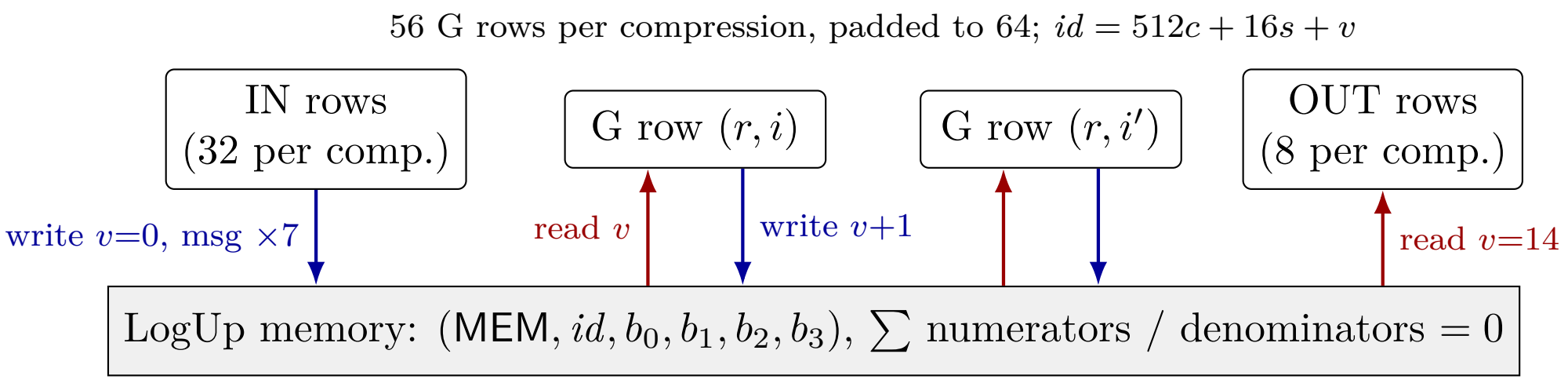
\begin{figure}[t]
\centering
\begin{tikzpicture}[
  row/.style={draw, rounded corners=2pt, minimum width=2.2cm, minimum height=0.65cm, font=\small, align=center},
  mem/.style={draw, fill=black!6, minimum width=9.8cm, minimum height=0.75cm, font=\small},
  wr/.style={-{Latex[length=2mm]}, thick, blue!60!black},
  rd/.style={-{Latex[length=2mm]}, thick, red!60!black}]
  \node[mem] (M) at (0,0) {LogUp memory: \((\mathsf{MEM}, \mathit{id}, b_0, b_1, b_2, b_3)\), \(\sum\) numerators \(/\) denominators \(= 0\)};
  \node[row] (IN) at (-4.2, 1.7) {IN rows\\(32 per comp.)};
  \node[row] (G1) at (-1.0, 1.7) {G row \((r, i)\)};
  \node[row] (G2) at (2.0, 1.7) {G row \((r, i')\)};
  \node[row] (OUT) at (4.8, 1.7) {OUT rows\\(8 per comp.)};
  \draw[wr] (IN.south) -- node[left, font=\scriptsize] {write \(v{=}0\), msg \(\times 7\)} ($(M.north)+(-4.2,0)$);
  \draw[rd] ($(M.north)+(-1.4,0)$) -- node[left, font=\scriptsize] {read \(v\)} ($(G1.south)+(-0.4,0)$);
  \draw[wr] ($(G1.south)+(0.4,0)$) -- node[right, font=\scriptsize] {write \(v{+}1\)} ($(M.north)+(-0.6,0)$);
  \draw[rd] ($(M.north)+(1.6,0)$) -- ($(G2.south)+(-0.4,0)$);
  \draw[wr] ($(G2.south)+(0.4,0)$) -- ($(M.north)+(2.4,0)$);
  \draw[rd] ($(M.north)+(4.8,0)$) -- node[right, font=\scriptsize] {read \(v{=}14\)} (OUT.south);
  \node[font=\scriptsize, align=center] at (0.5, 2.55) {56 G rows per compression, padded to 64; \(\mathit{id} = 512c + 16 s + v\)};
\end{tikzpicture}
\caption{Wiring of BLAKE3 compressions.  Blue arrows are writes (numerator \(-\mu\)), red arrows are reads (numerator \(+1\)).}
\label{fig:b3wire}
\end{figure}

\paragraph{Wiring.}
The rows are connected by a LogUp \emph{memory} (\cref{fig:b3wire}).
Each word is a tuple \((\mathsf{MEM}, \mathit{id}, b_0, b_1, b_2, b_3)\) in a dynamic set.
The row that produces a word writes it with numerator \(-\mu\), where \(\mu\) is the number of times the word will be read, and each consumer reads it with numerator \(+1\).
For compression \(c\), state slot \(s\) at version \(v\) has \(\mathit{id} = 512c + 16s + v\); each half-round writes every slot once, so versions run from \(0\) to \(14\).
Message word \(w\) has \(\mathit{id} = 512c + 256 + w\) and is read seven times.
G rows read their six inputs and write their four outputs.
\emph{Input rows} write the initial state (chaining value, IV, counter, block length, flags) at version \(0\) and the message words with multiplicity \(7\).
\emph{Output rows} read slots \(i\) and \(i + 8\) at version \(14\), XOR them, and write output word \(i\) under a separate tag.
The ids come from public columns, a per-compression base \(512c\) and a per-position offset shared by all compressions, both with closed forms, and the multiplicities are constants of the layout, so no multiplicity column is committed.
Bytes written to the memory are bounded by the XOR lookups (G rows) or by range lookups (input rows), so reads need no range checks of their own.

\begin{lemma}[Memory]
\label{lem:memory}
Suppose the ids written by the layout are distinct, every id is written once with multiplicity equal to the number of reads of that id in the layout, and all multiplicities are below \(p\).
If the memory set's fractions sum to zero as a rational function, then every read returns the bytes written under its id.
\end{lemma}

\begin{proof}
Group the tuples by id.
By \cref{lem:logup} applied to the multiset of reads and the written tuples with their multiplicities, the multiset of read tuples equals the multiset of written tuples, each counted with its multiplicity.
Since each id is written exactly once, every read tuple with id \(i\) must equal the unique written tuple with id \(i\), and the counts match by assumption on the layout.
\end{proof}

Together with the row constraints, \cref{lem:memory} implies that the output rows hold the BLAKE3 compression of the input rows.
We tested the gadget against a reference implementation and two forgeries: a forged XOR byte with every arithmetic constraint repaired, and a message word in the input rows that differs from the one the G rows read.
Both are rejected.

\paragraph{Cost.}
A compression has \(64 \times 80 + 32 \times 4 + 8 \times 12 = 5{,}344\) committed cells, \(30\%\) above the \(4{,}096\) of the design estimate.
We measured the full cost of a block of compressions on our GPU: commitment of its trace, the domain proofs of its G, input, output and table rows, and the openings of its units at the domains' points (the global evaluation proof, shared by all blocks of a recursion layer, is not included).
A block of \(2^{16}\) compressions (\(2^{22}\) G rows) takes \(0.16\)\,s, \(2.4\,\mu\)s per compression: \(13\)\,ms of commitment, \(0.13\)\,s of domain proofs and \(15\)\,ms of openings.
A block of \(2^{18}\) compressions (\(2^{24}\) rows) takes \(0.33\)\,s, \(1.3\,\mu\)s per compression, because the upper GKR layers are latency-bound and barely grow with the block.
At that rate the \(4.2\) million compressions of a 64-layer \(1 \times 4\)K verifier cost about \(5.5\)\,s, \(6\%\) of the main proof, and the \(9.2\) million of a \(16\)K verifier about \(12\)\,s, \(3\%\).
The extension-field arithmetic of the verifier (\(11\) to \(17\) million \(\Fq\) multiplications) is not yet a circuit and is not included.

\subsection{What remains}

The recursive layer also needs gadgets for \(\Fq\) and \(\Fo\) arithmetic, a recorder that turns a run of the verifier into the layer's witness (with chaining between the transcript, Merkle and leaf compressions), segmentation into fixed-size circuits, outer layers, and a final layer at a low code rate.
The outer layers must be zero-knowledge, which needs hiding commitments and masked sumcheck and GKR messages for multilinear arguments~\cite{cfs,libra}; this mode is not yet built.

\section{Evaluation}
\label{sec:eval}

\subsection{Setup}
\label{sec:eval-setup}

\paragraph{Hardware and software.}
All measurements use one NVIDIA RTX PRO 6000 Blackwell Server Edition GPU (96\,GB GDDR7, compute capability 12.0) in a host with an Intel Xeon 6776P (24 cores available) and 214\,GB of memory.
The code is that of the public release (\url{https://github.com/moonmath-ai/Sable}), whose README lists the command behind each table.  Software: Linux 6.11, NVIDIA driver 580.159 with CUDA 13.0, PyTorch 2.14 and CuPy 14.2 (NVRTC).
The GPU was shared with another user's jobs; every reported run was taken while the GPU was otherwise idle, as checked before and after the run.

\paragraph{Model and workloads.}
Qwen2.5-32B-Instruct quantized to \(\FO\) (\cref{sec:quantize}), all 64 layers; \cref{sec:eval-qwen3} adds Qwen3-32B (64 layers) and Qwen3-30B-A3B (48 layers).
Three shapes: one sequence of 4{,}096 tokens (\(1 \times 4\)K), four sequences of 4{,}096 tokens in one proof (\(4 \times 4\)K), and one sequence of 16{,}384 tokens (\(16\)K).
Each sequence consists of a prompt of uniformly random token ids followed by 16 greedily generated tokens, which the proof covers together with the stopping rule; the cost depends on the total length and hardly on the split between prompt and generation.
The prover's work does not depend on the token values, since every gadget commits and checks every row whatever its value.

\paragraph{What is timed.}
Wall-clock time of the prover from the statement to the last proof message, with the weight commitment computed beforehand (it is a property of the model release, computed once) and all kernels compiled.
The kernels are compiled at run time and cached on disk; with an empty cache, the first \(1 \times 4\)K proof took \(1{,}530\)\,s against \(92\)\,s for a second proof in the same process, so compilation is a one-time cost of about 24 minutes per machine and GPU type, which we exclude.
The prover keeps a pool of pinned host memory for the per-layer reduction (\cref{sec:segments}); it is allocated by the first proof of a process, which also captures the CUDA graphs and is therefore slower (\(103\), \(328\) and \(427\)\,s at the three shapes).
The end-to-end results (\cref{tab:e2e,tab:runs}) are the mean of three warm runs that follow such a warm-up run in the same process.
Other numbers, measured during development to compare variants of the prover, are single warm runs; repeated warm runs of the same configuration differed by up to about \(2\%\).
Runs use the protocol of \cref{sec:protocol,sec:security}.

\subsection{End-to-end proving}
\label{sec:eval-e2e}

\begin{table}[t]
\centering\small
\setlength{\tabcolsep}{4.5pt}
\begin{tabular}{@{}lrrrrrrr@{}}
\toprule
Shape & Tokens & Prove (s) & Overhead & GPU (GiB) & Host (GiB) & Proof (MiB) & Verify (s) \\
\midrule
\(1 \times 4\)K & 4{,}096 & \(92.5 \pm 0.0\) & \(50\) & \(36.8\) & \(101\) & \(134.8\) & \(18.8\) \\
\(4 \times 4\)K & 16{,}384 & \(253.1 \pm 0.0\) & \(45\) & \(63.5\) & \(155\) & \(239.0\) & \(26.8\) \\
\(16\)K & 16{,}384 & \(397.5 \pm 0.3\) & \(41\) & \(71.8\) & \(175\) & \(345.7\) & \(40.2\) \\
\bottomrule
\end{tabular}
\caption{Proving Qwen2.5-32B (64 layers) on one RTX PRO 6000: mean and standard deviation of three warm runs.  Overhead: proving time over the time our \(\FO\) serving engine takes for the same work (prefill and 16 generated tokens).  GPU, Host: peak memory, including the warm-up run.  Verify: the verifier with a compiled core on one CPU core, without a GPU (\cref{sec:eval-verify}).  Columns per proof: \(32.6\)K, \(59.1\)K and \(90.9\)K; committed elements \(1.17 \times 10^{11}\), \(3.55 \times 10^{11}\) and \(6.38 \times 10^{11}\), including \(3.4 \times 10^{10}\) weight elements and \(4.7 \times 10^{8}\) in the fixed units.}
\label{tab:e2e}
\end{table}

\begin{table}[t]
\centering\small
\begin{tabular}{@{}llrr@{}}
\toprule
Shape & Prompt and sampling & Warm-up (s) & Timed runs (s) \\
\midrule
\(1 \times 4\)K & random ids, greedy & \(103.2\) & \(92.47\), \(92.44\), \(92.47\) \\
\(1 \times 4\)K & WikiText-2, greedy & \(96.2\) & \(92.31\), \(92.26\) \\
\(1 \times 4\)K & random ids, temperature \(1.0\) & \(95.0\) & \(92.20\), \(92.38\), \(92.22\) \\
\(4 \times 4\)K & random ids, greedy & \(327.7\) & \(253.13\), \(253.08\), \(253.15\) \\
\(16\)K & random ids, greedy & \(426.6\) & \(397.92\), \(397.27\), \(397.38\) \\
\bottomrule
\end{tabular}
\caption{Individual proving times behind \cref{tab:e2e}, 64 layers, each row in a fresh process whose first proof is the warm-up.}
\label{tab:runs}
\end{table}

\cref{tab:e2e} gives the end-to-end results and \cref{tab:runs} every timed run.
The \(4 \times 4\)K proof covers four times the tokens of the \(1 \times 4\)K proof in \(2.74\) times the time: weights, the table side and the evaluation proof are shared, and the larger domains use the GPU better.
The \(16\)K proof takes \(1.57\) times as long as \(4 \times 4\)K for the same number of tokens, because attention grows quadratically with the sequence length: a \(16\)K sequence has \(4\) times as many causal score entries as four sequences of \(4\)K.
Proving time per token is nearly the same at \(1 \times 4\)K and \(16\)K (\(22.6\) and \(24.3\)\,ms): at 4K the prover is close to its fixed per-token cost, and attention catches up by \(16\)K.
The prover's work does not depend on the prompt: a \(1 \times 4\)K proof of a WikiText-2 prompt takes the same time as one of random ids (\(92.3\) against \(92.5\)\,s), with the same \(1.0765 \times 10^{11}\) lookups.
At temperature \(1.0\), the sampling gadget (\cref{sec:g-sampling}) adds \(304\) lookups and \(0.08\)\,MiB, its proving time is within the variation between processes (\(92.3\) against \(92.5\)\,s), and the proof verifies.
The proofs make \(1.08 \times 10^{11}\), \(4.31 \times 10^{11}\) and \(7.20 \times 10^{11}\) static lookups, in \(103\), \(416\) and \(726\) chunks of at most \(2^{30}\).

Before the proof-size reductions of \cref{sec:recursion}, the same proofs took \(101.3\), \(298.1\) and \(474.3\)\,s and were \(360.0\), \(468.0\) and \(549.6\)\,MiB long.
With the final layout and soundness bounds, and a global evaluation proof that rebuilt every layer's witness in three passes, they took \(108.2\), \(304.7\) and \(479.1\)\,s; the optimizations of \cref{tab:ablation} then brought them to the times of \cref{tab:e2e}, mostly through the per-layer reduction, which removed two of the three rebuild passes.

\paragraph{Overhead and cost.}
Our serving engine for \(\FO\) (\cref{sec:discussion}) computes the same work as the \(1 \times 4\)K proof, prefill and 16 generated tokens, in \(1.86\)\,s (time to first token \(1.30\)\,s, then \(37.4\)\,ms per token), the \(4 \times 4\)K work in \(5.60\)\,s and the \(16\)K work in \(9.60\)\,s.
A proof therefore takes 41 to 50 times the inference it attests.
Against FP8 in SGLang, which prefills 4{,}096 tokens in \(0.46\)\,s and decodes 15 more in about \(0.6\)\,s, the factor is about 90.
At a public on-demand price of 2.50 dollars per GPU-hour for this GPU~\cite{coreweave}, a \(1 \times 4\)K proof costs \(6.4\) cents, a sequence in a \(4 \times 4\)K proof \(4.4\) cents, and a \(16\)K proof \(28\) cents.
Several sequences per proof amortize the shared work, but the GPU's 96\,GB bound the shapes.
We probed larger ones with 4-layer proofs, whose peak GPU memory is set by one layer as in a 64-layer proof: a proof of eight 4K sequences runs out of GPU memory in the prove pass, and one of a 32K sequence in its first witness build.

\paragraph{Where the time goes.}
\label{sec:eval-breakdown}
We profiled 8-layer proofs with Nsight Systems, tracing the kernels inside replayed graphs, first in the configuration before the per-layer reduction.
Per replayed layer, the prove pass keeps the GPU busy \(94\%\), \(97\%\) and \(97\%\) of the time at the three shapes (\(0.90\), \(2.63\) and \(4.22\)\,s of span), so the prover is GPU-bound there.
Over the whole \(16\)K proof, the prove pass took about \(55\%\) of the time, the three rebuild passes at the end \(26\%\) and the commit pass \(16\%\); the per-layer reduction removed most of the second item, and in the final prover the remaining query pass takes \(11\), \(32\) and \(53\)\,s at the three shapes.
Within the prove pass of one layer at \(1 \times 4\)K, sumcheck round kernels take \(512\)\,ms: \(4{,}600\) rounds below \(30\,\mu\)s cost \(80\)\,ms in total, and \(630\) rounds above \(100\,\mu\)s cost \(400\)\,ms.
By protocol step, the MLP domains take \(133\)\,ms, the softmax tile domains \(119\)\,ms, the unit openings' point reductions \(100\)\,ms, and data preparation outside sumchecks (witness, column packing, the reductions' tables) \(268\)\,ms.
At \(16\)K the tile domains take \(1.25\)\,s of the prove pass per layer, close to the cost model's charge (\cref{sec:eval-model}) for their \(5.4 \times 10^9\) lookup leaves per layer.
The large GKR rounds already run at 83 to 88\% of memory bandwidth; the domain rounds are compute-bound (\cref{sec:gpu-kernels}).

\begin{table}[t]
\centering\small
\begin{tabular}{@{}lrrrrrr@{}}
\toprule
& \multicolumn{2}{c}{Commit pass} & \multicolumn{2}{c}{Prove pass (replayed)} & \multicolumn{2}{c}{Final passes} \\
\cmidrule(lr){2-3}\cmidrule(lr){4-5}\cmidrule(lr){6-7}
Shape & span & busy & span & busy & span & busy \\
\midrule
\(1 \times 4\)K & \(0.26\) & \(0.18\) & \(1.01\) & \(0.89\) & \(0.30\) & \(0.17\) \\
\(4 \times 4\)K & \(0.79\) & \(0.68\) & \(2.65\) & \(2.51\) & \(0.71\) & \(0.50\) \\
\(16\)K & \(1.22\) & \(1.05\) & \(4.18\) & \(4.02\) & \(1.10\) & \(0.83\) \\
\bottomrule
\end{tabular}
\caption{Seconds per layer of the final prover by phase (Nsight Systems, 8-layer proofs, kernels inside replayed graphs traced).  Span: wall time of the phase on the GPU; busy: GPU busy time, during which some kernel was running.  Final passes: the global evaluation proof and the query pass.}
\label{tab:breakdown}
\end{table}

\cref{tab:breakdown} gives the same breakdown for the final prover.
The prove pass is \(88\%\) to \(96\%\) busy; the commit pass and the final passes, which are not replayed as graphs, \(56\%\) to \(87\%\).
In the final passes at \(16\)K, Reed--Solomon encoding takes \(298\)\,ms per layer and leaf hashing \(71\)\,ms, the rest being witness reconstruction and column packing; keeping the Merkle levels in host memory from the commit pass and encoding only the queried tiles saves about \(1\%\) of the \(1 \times 4\)K proof, but the stored levels exceed the host memory of our machine at the larger shapes, so the final prover recomputes them.
The proofs make \(1.08 \times 10^{11}\), \(4.31 \times 10^{11}\) and \(7.20 \times 10^{11}\) static lookups, in \(103\), \(416\) and \(726\) chunks of at most \(2^{30}\).

\subsection{Proving Qwen3}
\label{sec:eval-qwen3}

\begin{table}[t]
\centering\small
\setlength{\tabcolsep}{4pt}
\begin{tabular}{@{}llrrrrrr@{}}
\toprule
Model & Shape & Prove (s) & Overhead & GPU (GiB) & Host (GiB) & Proof (MiB) & Verify (s) \\
\midrule
Qwen3-32B & \(1 \times 4\)K & \(121.8 \pm 0.0\) & \(53\) & \(36.5\) & \(83\) & \(165.1\) & \(26.5\) \\
 & \(4 \times 4\)K\(^{*\dagger}\) & \(485.7 \pm 25.3\) & \(68\) & \(68.6\) & \(140\) & \(298.0\) & \(36.2\) \\
 & \(16\)K\(^{*}\) & \(668.7 \pm 2.2\) & \(50\) & \(82.0\) & \(173\) & \(476.7\) & \(55.8\) \\
\midrule
Qwen3-30B-A3B & \(1 \times 4\)K & \(84.7 \pm 0.9\) & \(17\) & \(18.1\) & \(72\) & \(106.2\) & \(20.2\) \\
 & \(4 \times 4\)K & \(181.9 \pm 5.7\) & \(17\) & \(49.4\) & \(101\) & \(152.6\) & \(23.0\) \\
 & \(16\)K & \(265.7 \pm 6.3\) & \(21\) & \(67.7\) & \(110\) & \(219.5\) & \(29.5\) \\
\bottomrule
\end{tabular}
\caption{Proving the Qwen3 models on one RTX PRO 6000, with the columns of \cref{tab:e2e}: mean and standard deviation of three warm runs after a warm-up run in the same process.  \(^{*}\)Without graph replay, whose captured graphs do not fit in GPU memory at these shapes.  \(^{\dagger}\)The three runs took \(462\), \(482\) and \(512\)\,s with the same time in every timed phase of the prover, so the spread lies outside the timed phases, and we did not locate it; the same prover on this model calibrated without the Q/K split took \(464.7 \pm 5.3\)\,s.}
\label{tab:e2e-qwen3}
\end{table}

\Cref{tab:e2e-qwen3} gives the end-to-end results for the two Qwen3 models, with the shapes, workloads and protocol of \cref{tab:e2e}.
Qwen3-32B has the depth and width of Qwen2.5-32B, but 64 query heads instead of 40, QK-norm, and a slightly narrower MLP.
Its proofs make \(1.39 \times 10^{11}\), \(5.56 \times 10^{11}\) and \(1.02 \times 10^{12}\) static lookups, \(1.29\), \(1.29\) and \(1.42\) times those of Qwen2.5-32B, and take \(1.32\), \(1.92\) and \(1.68\) times as long.
The gap is largest at the two larger shapes, where attention takes a larger share of the proof and where the prover runs without graph replay, because the captured graphs no longer fit in 96\,GB beside the larger attention witness.
Graph replay saved \(4\%\) and \(1\%\) of the Qwen2.5 proofs at these shapes (\cref{tab:ablation}), so its absence explains little of the difference; the larger attention explains part of it, and at \(4 \times 4\)K we did not locate the rest.

Qwen3-30B-A3B has 48 layers of width 2{,}048, 32 query heads, and 128 experts of width 768 per layer, of which each token uses 8.
The proof binds all \(30.5 \times 10^{9}\) weights through the weight commitment but proves only the experts each token selects, so it makes \(4.65 \times 10^{10}\), \(1.86 \times 10^{11}\) and \(3.60 \times 10^{11}\) lookups, about a third of those of Qwen3-32B, and takes \(84.7\), \(181.9\) and \(265.7\)\,s.
The \(4 \times 4\)K proof covers four times the tokens of the \(1 \times 4\)K proof in \(2.15\) times the time (\(2.74\) for Qwen2.5-32B): the work on the weights is shared between the sequences, and it is a larger part of the proof when each token uses a tenth of them.
The overhead factor of this model is lower because our serving engine decodes it in \(201\)\,ms per token, five times the \(38\)\,ms of Qwen3-32B: it runs the experts through the exact integer grouped product of the prover's witness, which a decoding step of one token uses poorly.
Against prefill alone, the \(1 \times 4\)K proof takes \(41\) times the \(2.06\)\,s our engine needs to prefill 4{,}096 tokens.

On one CPU core, the verifier checks the six proofs in \(20.2\) to \(55.8\)\,s; on the GPU, run in a fresh process, it takes \(21.4\) to \(53.1\)\,s, no faster, since the GPU path runs the protocol logic in Python without the compiled core.
For serving, our engine generates 512 tokens after a 4{,}096-token prompt at \(26.6\) tokens per second for Qwen3-32B and \(5.0\) for Qwen3-30B-A3B in a batch of one, and at \(412\) and \(131\) tokens per second in a batch of 32.

\subsection{Comparison with prior systems}
\label{sec:eval-prior}

\begin{table}[t]
\centering\small
\begin{tabular}{@{}lrrlrr@{}}
\toprule
System & Parameters & Tokens & GPU & Prove (s) & ps/pt \\
\midrule
zkLLM~\cite{zkllm} & 13.0B & 2{,}048 & A100 40GB & \(803\) & \(30.2\) \\
FlashProve~\cite{flashprove} & 16.2B\(^{\dagger}\) & \(24 \times 1{,}024\) & H100 80GB & \(223\) & \(0.56\) \\
Sable, Qwen2.5-32B, \(1 \times 4\)K & 32.5B & 4{,}096 & RTX PRO 6000 & \(92.5\) & \(0.69\) \\
Sable, Qwen2.5-32B, \(4 \times 4\)K & 32.5B & 16{,}384 & RTX PRO 6000 & \(253.1\) & \(0.48\) \\
Sable, Qwen2.5-32B, \(16\)K & 32.5B & 16{,}384 & RTX PRO 6000 & \(397.5\) & \(0.75\) \\
Sable, Qwen3-30B-A3B, \(1 \times 4\)K & 3.3B\(^{\dagger}\) & 4{,}096 & RTX PRO 6000 & \(84.7\) & \(6.27\) \\
Sable, Qwen3-30B-A3B, \(4 \times 4\)K & 3.3B\(^{\dagger}\) & 16{,}384 & RTX PRO 6000 & \(181.9\) & \(3.36\) \\
Sable, Qwen3-30B-A3B, \(16\)K & 3.3B\(^{\dagger}\) & 16{,}384 & RTX PRO 6000 & \(265.7\) & \(4.91\) \\
\bottomrule
\end{tabular}
\caption{Proving time against zkLLM's largest reported configuration (LLaMa-2-13B, its Table~1) and FlashProve's (Llama~4 Maverick, batch 24), normalized per parameter and token (ps/pt: picoseconds per parameter and token; per active parameter for mixtures of experts).  zkLLM is zero-knowledge, works over BLS12-381 and approximates non-linear functions; it also commits the model once, in \(986\)\,s.  FlashProve: a batch of 24 prompts of 1{,}024 tokens, each proof checking one token; \(^{\dagger}\)parameters used per token of a mixture of experts (30.5B in total for Qwen3-30B-A3B, 400.7B for FlashProve's model); it is zero-knowledge and commits the model once, in \(2{,}324\)\,s.}
\label{tab:prior}
\end{table}

\cref{tab:prior} normalizes proving time by model size and number of tokens.
zkLLM~\cite{zkllm} takes \(30.2\)\,ps per parameter and token for LLaMa-2-13B on an A100; Sable takes \(0.48\) to \(0.75\)\,ps on Qwen2.5-32B, 40 to 64 times less.
Part of this factor is hardware: our GPU has \(3.5\) times as many CUDA cores as the A100, at a higher clock.
The systems also prove different statements: zkLLM hides the weights and works over a pairing-friendly field with approximate non-linear functions, while Sable is not zero-knowledge and fixes every rounding of \(\FO\).
FlashProve~\cite{flashprove} proves a much larger mixture of experts on an H100 with 24 CPU cores in \(0.56\)\,ps per active parameter and token, within the range of Sable's dense proofs and 6 to 11 times less than Sable's proofs of Qwen3-30B-A3B, whose cost per active parameter is high partly because attention and the commitment of all 30.5B weights do not shrink with the experts a token uses. FlashProve works at a context of 1{,}024 tokens, where attention is a small part of the work, and includes zero knowledge, which for Sable is future work.
zkGPT~\cite{zkgpt} proves GPT-2 (124M parameters) in \(21.8\)\,s on a 16-core CPU and reports zkLLM at \(15.8\)\,s on the same model; it does not state the number of tokens, so we do not normalize it.

\subsection{Comparison with the cost model}
\label{sec:eval-model}

Before implementing the prover, we built a cost model that counts, for every gadget, the committed elements, lookup leaves and constraint terms per element of the trace, and prices them from hardware rates: \(3.9 \times 10^{-10}\) H100-seconds per committed element (\(0.6\times\) that at rate \(1/2\)), \(1.5 \times 10^{-10}\) per lookup leaf, \(5.5 \times 10^{-11}\) per constraint term, with factors of \(0.7\) and \(0.5\) for base-field first rounds, plus the forward passes and the weight opening.
To price the same work on our GPU, a second model splits each unit cost into the resources it uses (32-bit multiplies, other integer operations, DRAM traffic, INT8 tensor operations and host transfers), from operation and byte counts of the algorithm behind it, and scales each share by the ratio of the two GPUs' throughputs for that resource, from published architecture figures.
It uses no measurement of the prover.
Its central scenario takes the RTX PRO 6000 at its boost clock with 96 of its 128 integer lanes per multiprocessor effective for the operation mix and its full memory bandwidth; a low and a high scenario bracket the clock, the integer issue rate and the effective bandwidth.
Per unit, commitment then costs \(0.87\) times its H100 time, a lookup leaf \(1.03\) times and a constraint term \(1.10\) times, so the model prices our GPU at about one H100.
\cref{tab:model} compares its predictions with the measurements.

\begin{table}[t]
\centering\small
\begin{tabular}{@{}lrrrrrr@{}}
\toprule
& \multicolumn{2}{c}{Cost model (RTX-s)} & Measured & & \multicolumn{2}{c}{Committed elements} \\
\cmidrule(lr){2-3}\cmidrule(lr){6-7}
Shape & Central & Band & s & Ratio & Model & Sable (trace) \\
\midrule
\(1 \times 4\)K & \(37.7\) & \(29.1\)--\(44.2\) & \(92.5\) & \(2.45\) & \(7.87 \times 10^{10}\) & \(8.29 \times 10^{10}\) \\
\(4 \times 4\)K & \(146.4\) & \(112.6\)--\(171.5\) & \(253.1\) & \(1.73\) & \(3.15 \times 10^{11}\) & \(3.21 \times 10^{11}\) \\
\(16\)K & \(254.2\) & \(194.2\)--\(298.5\) & \(397.5\) & \(1.56\) & \(5.84 \times 10^{11}\) & \(6.03 \times 10^{11}\) \\
\bottomrule
\end{tabular}
\caption{Cost model for the RTX PRO 6000 against measurement (64 layers).  Band: the model's high and low hardware scenarios.  Ratio: measured over central.  The model's stated accuracy is \([0.5\times, 2\times]\).  Committed elements of Sable are the trace units of the final layout, without the weights and fixed units, which the model charges separately.}
\label{tab:model}
\end{table}

The model's count of committed trace elements is within \(5\%\) of the implementation's, padding included, so the gadget accounting is accurate.
The prover takes \(2.45\), \(1.73\) and \(1.56\) times the central prediction, falling as the shapes grow; at \(1 \times 4\)K this is outside the model's stated factor of two, and against the low scenario the ratios are \(2.10\), \(1.48\) and \(1.34\).
At \(1 \times 4\)K the prover is partly limited by latency: it runs about 3{,}800 single-block sumcheck rounds per layer, each taking \(12\) to \(18\,\mu\)s of kernel time regardless of its size, most of it the round's transcript step, and the GPU idles between domains while the host computes small claims.
At \(16\)K the remaining gap is concentrated in attention: the final sumcheck of a softmax tile domain runs compute-bound at one thread block per multiprocessor (its kernels need over 200 registers per thread), and field additions with their conditional subtractions are about half of its instructions, which the model's per-term multiplication counts do not capture.
The model charges fixed costs per element and has no term for latency, which is where the small shape loses most.
Its remaining uncertainty is the integer issue rate of the RTX PRO 6000, which the scenarios bracket; the gap at \(16\)K is larger than that range.

\subsection{Engineering ablations}
\label{sec:eval-ablation}

\begin{table}[t]
\centering\footnotesize
\begin{tabular}{@{}p{0.33\linewidth}p{0.29\linewidth}rr@{}}
\toprule
Change & Metric & Before & After \\
\midrule
Merge same-shape domains (about \(2\%\) faster) & 8 layers, \(1 \times 4\)K / \(4 \times 4\)K / \(16\)K (s) & n/a & \(22.0 / 53.8 / 85.6\) \\
Shared-memory softmax histogram & 8 layers, \(16\)K (s) & \(85.6\) & \(82.0\) \\
Remove implicit host synchronizations & 8 layers, \(1 \times 4\)K / \(4 \times 4\)K / \(16\)K (s) & \(22.0 / 52.9 / 82.0\) & \(20.8 / 48.6 / 78.0\) \\
Multiplicity compression on a side stream & 8 layers, \(4 \times 4\)K / \(16\)K (s) & \(48.4 / 77.6\) & \(46.4 / 73.5\) \\
Factored lookup leaves (\cref{sec:gpu-kernels}) & 8 layers, three shapes (s) & \(19.7 / 46.4 / 71.8\) & \(19.6 / 45.8 / 70.8\) \\
Two-pass NTT up to \(2^{24}\) & 26 codewords of \(2^{24}\) (ms) & \(8.5\) & \(6.2\) \\
Warp-cooperative transcript & transcript step of a round (\(\mu\)s) & \(25.9\) & \(8.8\) \\
Openings on the device transcript & 8 layers, \(1 \times 4\)K / \(16\)K (s) & \(19.0 / 69.6\) & \(18.5 / 67.3\) \\
Device transcript for the whole prove pass & 8 layers, three shapes (s) & \(18.1 / 44.3 / 67.0\) & \(16.7 / 41.4 / 64.1\) \\
Graph replay of middle layers & 64 layers, three shapes (s) & \(133 / 322 / 503\) & \(106 / 310 / 497\) \\
Flat small GKR rounds & GPU time per layer, \(1 \times 4\)K (s) & \(1.013\) & \(0.902\) \\
Second round folds twice from base table & GPU time per layer, \(1 \times 4\)K (s) & \(0.901\) & \(0.890\) \\
Coefficients in shared memory & GPU time per layer, \(1 \times 4\)K (s) & \(0.889\) & \(0.857\) \\
Delayed-reduction \(\Fq\) multiply & GPU time per layer, \(1 \times 4\)K (s) & \(0.857\) & \(0.830\) \\
Delayed reduction in column combination & 128 columns of \(2^{23}\) (ms) & \(4.79\) & \(3.83\) \\
Fused unit opening (\cref{sec:gpu-field}) & 8 layers, \(1 \times 4\)K / \(16\)K (s) & \(16.42 / 62.39\) & \(16.16 / 60.90\) \\
Per-layer reduction (\cref{sec:segments}) & 64 layers, three shapes (s) & \(108.2 / 304.7 / 479.1\) & \(99.3 / 269.6 / 419.3\) \\
Claim preparation and point reductions on the device & 64 layers, three shapes (s) & \(99.3 / 269.6 / 419.3\) & \(97.0 / 264.3 / 411.3\) \\
Evaluation points by additions & 64 layers, three shapes (s) & \(97.0 / 264.3 / 411.3\) & \(96.6 / 265.2 / 410.9\) \\
Sequential-mode transcript (\cref{sec:prelim-fs}) & 64 layers, three shapes (s) & \(96.6 / 265.2 / 410.9\) & \(93.1 / 259.3 / 404.9\) \\
Leaf kernel writes the next GKR layer & 64 layers, three shapes (s) & \(93.1 / 259.3 / 404.9\) & \(92.4 / 254.0 / 397.1\) \\
\bottomrule
\end{tabular}
\caption{Effect of individual optimizations, in the order they were applied, each measured against the state just before it.  Baselines of consecutive rows differ slightly because of run-to-run variation (about \(2\%\)) and small unrelated changes in between.}
\label{tab:ablation}
\end{table}

\cref{tab:ablation} lists the measured effect of the main optimizations.
Removing host round trips mattered most: implicit synchronizations (pageable copies, data-dependent output sizes, scalar reads), the host transcript, and the Python overhead of launching each layer's work.
They took an 8-layer \(1 \times 4\)K proof from about \(23\)\,s to \(16\)\,s, and with graph replay the 64-layer \(1 \times 4\)K proof went from \(134\)\,s to \(101\)\,s.
Kernel-level changes then brought the GPU time of a replayed layer from \(1.01\) to \(0.83\)\,s.

Several attempts gave no gain and were reverted: a register cap on the round kernels (spills doubled), splitting the off-diagonal attention tiles into a separate lower-degree domain (the extra GKR and reductions cost more than the lower degree saved), a fused single-kernel sumcheck tail, graph replay of the commitment passes (already GPU-bound), splitting each (pair, point) of the round polynomial into its own thread (\(40.8 \to 52.6\)\,ms for the MLP domain), compiling the witness with \texttt{torch.compile} (faster witness alone, \(42.6 \to 34.2\)\,ms per layer, but slower proofs and \(9.5\)\,GiB more memory), and keeping the commitment's combined column to skip one witness rebuild (1.3 to 4.4\,GiB per layer).

\paragraph{Comparison with an external library.}
We compared three kernels with the SP1-GPU library (version 6.9.0, which builds on sppark) on the same GPU and shapes.
A coset low-degree extension of 26 columns from \(2^{23}\) to \(2^{24}\) took \(10.2\)\,ms there and \(6.1\)\,ms in Sable.
A product sumcheck over \(2^{24}\) took \(1.17\)\,ms there (kernels only, no transcript) and \(1.50\)\,ms in Sable (with the device transcript).
A LogUp-GKR layer took \(3.54\)\,ms there against \(3.28\)\,ms for \(2^{24}\) leaves, and \(0.27\) against \(0.61\)\,ms for \(2^{20}\), where our per-round transcript dominates.
No library we tested would make the prover as a whole faster.

\subsection{Proof size and verification}
\label{sec:eval-verify}

For the final prover the proofs are \(134.8\), \(239.0\) and \(345.7\)\,MiB at the three shapes (\(1.37\), \(1.78\) and \(1.86\) million messages).
At \(1 \times 4\)K, the leaf data of the query openings (\(32.6\)K columns at 4 positions for 153 queries) is \(76\)\,MiB; Merkle paths, unit openings and the round messages of sumchecks and GKR make up the rest.
The Python reference verifier, single-threaded with NumPy, takes \(94\), \(129\) and \(145\)\,s; its time is dominated by interpreter overhead.
\cref{tab:vcount} counts its operations.
Verification grows slowly with the shape: from \(1 \times 4\)K to \(16\)K the trace grows \(5.4\times\), the BLAKE3 work \(2.2\times\) and the \(\Fq\) multiplications \(1.5\times\).

\begin{table}[t]
\centering\small
\begin{tabular}{@{}lrrrrrr@{}}
\toprule
& \multicolumn{3}{c}{BLAKE3 compressions} & & & \\
\cmidrule(lr){2-4}
Shape & transcript & Merkle & total & Challenges & \(\Fq\) mult. & \(\Fo\) mult. \\
\midrule
\(1 \times 4\)K & \(1.72\)M & \(2.50\)M & \(4.22\)M & \(506\)K & \(11.0\)M & \(0.22\)M \\
\(4 \times 4\)K & \(2.46\)M & \(4.35\)M & \(6.81\)M & \(761\)K & \(15.3\)M & \(0.31\)M \\
\(16\)K & \(2.70\)M & \(6.49\)M & \(9.19\)M & \(825\)K & \(16.5\)M & \(0.40\)M \\
\bottomrule
\end{tabular}
\caption{Operations of the verifier (64 layers).  Transcript compressions are exact counts in the sequential mode; Merkle compressions include leaf hashing.}
\label{tab:vcount}
\end{table}

The reference verifier also runs on the host alone with a compiled core: the sumcheck and LogUp-GKR verifiers, equality tables, multilinear extensions and the product checks of the domain programs run in C++ on the transcript state and the proof's messages, while the protocol logic (claim routing, opening bookkeeping) stays in Python.
On one core of a Xeon 6776P, with no GPU, it verifies the three proofs in \(18.8\), \(26.8\) and \(40.2\)\,s, of which \(1.3\), \(3.5\) and \(10.2\)\,s are spent in the compiled core.
The remainder is interpreter work on \(1.4\) to \(1.9\) million messages and the leaf hashing of the query openings through a Python binding, which a verifier compiled as a whole would remove.
Its operation counts (\cref{tab:vcount}) at single-core rates of a few hundred nanoseconds per BLAKE3 compression and tens of nanoseconds per \(\Fq\) multiplication put such a verifier at a few seconds per proof.

\subsection{Accuracy of \texorpdfstring{\(\FO\)}{F\_O}}
\label{sec:eval-acc}

\(\FO\) is a different function from the BF16 model, and its accuracy determines whether a deployment can serve it.
We compare it with BF16 Qwen2.5-32B-Instruct and with two deployed 8-bit quantizations of the same checkpoint, simulated by quantizing and dequantizing in BF16: FP8 (E4M3, per-tensor scales, activations quantized dynamically) and SmoothQuant W8A8 INT8~\cite{smoothquant} (smoothing strength \(0.5\) calibrated on 32{,}768 tokens of WikiText-2 training text, per-channel weights, per-token activations).
\(\FO\) is evaluated with an engine that computes the same integer function as the prover's witness generator, vectorized over heads and sequences, and that we checked to produce identical logits and greedy tokens.
WikiText-2 (raw) is tokenized as one string and cut into 145 non-overlapping windows of 2{,}048 tokens covering the whole test split, each evaluated from its first token (296{,}815 predictions); MMLU uses the 5-shot prompt of the original benchmark on all 14{,}042 test questions, scoring the answer letter with the largest logit; GSM8K uses the first 8 training problems with their worked answers as 8-shot prompt, greedy decoding of at most 256 tokens, and exact match of the final number on all 1{,}319 test problems.
Intervals are 95\% percentile intervals from 2{,}000 bootstrap resamples of the WikiText-2 windows; since all variants see the same windows, differences between two variants are resampled jointly (paired), which gives much narrower intervals than those of single perplexities.

\begin{table}[t]
\centering\small
\begin{tabular}{@{}lrrrrrr@{}}
\toprule
& \multicolumn{4}{c}{WikiText-2} & & \\
\cmidrule(lr){2-5}
Variant & Perplexity & over BF16 & Top-1 & Agreement & MMLU & GSM8K \\
\midrule
BF16 & \(5.151\) & & \(62.0\%\) & n/a & \(83.2\%\) & \(88.6\%\) \\
FP8 (W8A8, per tensor) & \(5.186\) & \(+0.7\%\) & \(61.9\%\) & \(95.5\%\) & \(83.0\%\) & \(88.4\%\) \\
SmoothQuant INT8 (W8A8) & \(5.231\) & \(+1.6\%\) & \(61.7\%\) & \(93.9\%\) & \(83.2\%\) & \(88.2\%\) \\
\(\FO\) & \(5.580\) & \(+8.3\%\) & \(60.4\%\) & \(86.3\%\) & \(81.8\%\) & \(88.0\%\) \\
\bottomrule
\end{tabular}
\caption{Accuracy of \(\FO\) and of two 8-bit quantizations against BF16 Qwen2.5-32B-Instruct.  Over BF16: excess perplexity; paired 95\% intervals are \(0.6\)--\(0.8\%\), \(1.4\)--\(1.8\%\) and \(7.3\)--\(9.5\%\).  Single perplexities have intervals of about \(\pm 0.33\), mostly from the varying difficulty of the windows.  Agreement: the fraction of WikiText-2 positions at which the greedy next token equals BF16's.}
\label{tab:acc}
\end{table}

\cref{tab:acc} gives the results.
\(\FO\) loses more next-token accuracy than the deployed 8-bit schemes: its perplexity is \(8.3\%\) above BF16's, against \(0.7\%\) and \(1.6\%\) for FP8 and INT8, and its greedy token differs from BF16's at \(13.7\%\) of the positions.
On MMLU, \(\FO\) is \(1.4\) points below BF16 (\(81.8\%\) against \(83.2\%\); \(82.7\%\) against \(84.2\%\) averaged over subjects), where FP8 and INT8 lose at most \(0.2\); on GSM8K, where answers depend on a chain of greedy decisions, it is \(0.6\) points below BF16, against \(0.2\) and \(0.4\) for FP8 and INT8.
The left-shift regime of \cref{rem:leftshift} is common on real inputs: over the first 16 WikiText-2 windows, \(4.1\%\) of the rows of the attention output and down projections have a shift \(k \le 0\), and \(1.7\%\) a true left shift (down to \(k = -8\)); the logits never do.
Honest proofs therefore exercise the input bound of \cref{sec:g-shiftround}.

\paragraph{Where the accuracy goes.}
\label{sec:eval-attrib}
To find which of \(\FO\)'s integer steps cost accuracy, we evaluated variants of \(\FO\) that each make one step more precise and keep the weights, the calibration and every other step (\cref{tab:attrib}):
a \emph{float softmax} computes the attention probabilities and their normalization in double precision from the same integer scores, instead of as 10-bit values with 16-bit denominator mantissas;
\emph{12-bit queries and keys} shift by 4 bits less in the \(\gad{ShiftQ8S}\) that produces them and saturate at \(16 \cdot 127\), with RoPE and the scores kept at this precision and the softmax scale divided by \(2^8\);
a \emph{SiLU with 4 more bits} evaluates SiLU in double precision on the same integer input and rounds it with 4 more fractional bits;
a \emph{16-bit up projection} rounds the up projection to 16 bits instead of 8.
In the last two, the shift after the gate product grows by the added bits.
We also rebuilt \(\FO\) with the same program calibrated on 8 windows of 1{,}024 tokens from the WikiText-2 training split, instead of our technical text.

\begin{table}[t]
\centering\small
\begin{tabular}{@{}lrrrrr@{}}
\toprule
Variant of \(\FO\) & Perplexity & over BF16 & Top-1 & Agreement & Gap removed \\
\midrule
\(\FO\) & \(5.580\) & \(8.3\%\) & \(60.39\%\) & \(86.3\%\) & \(0\%\) \\
float softmax & \(5.588\) & \(8.5\%\) & \(60.38\%\) & \(86.3\%\) & \(-2\%\) (\(-4\) to \(0\)) \\
12-bit queries and keys & \(5.415\) & \(5.1\%\) & \(61.07\%\) & \(88.6\%\) & \(37\%\) (\(35\) to \(40\)) \\
SiLU with 4 more bits & \(5.585\) & \(8.4\%\) & \(60.48\%\) & \(86.4\%\) & \(-1\%\) (\(-4\) to \(1\)) \\
16-bit up projection & \(5.568\) & \(8.1\%\) & \(60.49\%\) & \(86.6\%\) & \(3\%\) (\(0\) to \(5\)) \\
all four & \(5.422\) & \(5.3\%\) & \(61.08\%\) & \(88.8\%\) & \(36\%\) (\(33\) to \(39\)) \\
calibration on WikiText-2 train & \(5.576\) & \(8.3\%\) & \(60.45\%\) & \(86.5\%\) & \(1\%\) (\(-2\) to \(3\)) \\
\bottomrule
\end{tabular}
\caption{Accuracy attribution on the WikiText-2 test split.  Gap removed: the share of \(\FO\)'s excess log perplexity over BF16 that the variant removes, with its paired 95\% interval.  None of these variants is provable as it stands.}
\label{tab:attrib}
\end{table}

Only the queries and keys matter: with 12 bits they remove \(37\%\) of the gap to BF16 and raise agreement to \(88.6\%\).
The other three changes each move perplexity by less than \(3\%\) of the gap, and all four together do no better than the queries and keys alone.
Recalibrating on WikiText-2 text changes nothing, so the choice of calibration text does not explain the gap.
The remaining \(63\%\) lies in steps we did not vary; the 8-bit activations entering every projection and the 8-bit values \(V\) are the obvious candidates.

The variants are not provable as they stand, because the extra bits break bounds of the constraint system.
12-bit queries and keys make scores reach \(2^{30}\), far outside the \(2^{23}\)-entry exponential table, so a provable version would round each score by a shift before the table lookup, adding committed values and lookups to every causal score entry; 16-bit probabilities would make a \(PV\) block sum reach \(2^{35}\), above \(p\), and would need each probability committed as two 8-bit limbs.
The cost model of \cref{sec:eval-model} puts the first change at \(1.30\) times the proving time at 4K and \(1.68\) times at \(16\)K, the second at \(1.15\) and \(1.35\) times, and both at \(1.45\) and \(2.02\) times: precision is bought where the proof is already most expensive.
Since the float softmax does not improve accuracy, only the first is worth building.

\paragraph{Qwen3.}
\label{sec:eval-acc-qwen3}
\cref{tab:acc-qwen3} repeats the comparison for the two Qwen3 models of \cref{sec:qwen3}, with the same WikiText-2 windows and MMLU prompts.
For GSM8K we raise the generation limit to 1{,}024 tokens for every variant of these models.
The Qwen3 checkpoints are hybrid reasoning releases, and under a few-shot prompt without a chat template they sometimes open an answer in a step-by-step style (``Okay, let's see\ldots'') that runs past 256 tokens before its final number; at that limit the score would mostly count how often a variant switches style.
Generation stops after the first complete final answer (\texttt{\#\#\#\# N} followed by a non-digit), which cannot change the extracted answer, since scoring reads the first such marker.
\(\FO\) is served by the engine of \cref{sec:discussion} with continuous batching, which computes each sequence exactly as it would alone.

\begin{table}[t]
\centering\small
\setlength{\tabcolsep}{4.5pt}
\begin{tabular}{@{}llrrrrrr@{}}
\toprule
& & \multicolumn{4}{c}{WikiText-2} & & \\
\cmidrule(lr){3-6}
Model & Variant & Perplexity & over BF16 & Top-1 & Agreement & MMLU & GSM8K \\
\midrule
Qwen3-32B & BF16 & \(7.444\) & & \(56.3\%\) & n/a & \(81.8\%\) & \(93.9\%\) \\
& FP8 & \(7.480\) & \(+0.5\%\) & \(56.2\%\) & \(94.8\%\) & \(81.9\%\) & \(94.3\%\) \\
& SmoothQuant INT8 & \(7.572\) & \(+1.7\%\) & \(56.0\%\) & \(92.4\%\) & \(81.8\%\) & \(94.2\%\) \\
& \(\FO\) & \(7.933\) & \(+6.6\%\) & \(54.9\%\) & \(82.4\%\) & \(80.4\%\) & \(92.2\%\) \\
\midrule
Qwen3-30B-A3B & BF16 & \(8.455\) & & \(56.1\%\) & n/a & \(79.6\%\) & \(92.7\%\) \\
& FP8 & \(8.484\) & \(+0.3\%\) & \(56.1\%\) & \(95.9\%\) & \(79.3\%\) & \(92.2\%\) \\
& SmoothQuant INT8 & \(8.425\) & \(-0.4\%\) & \(56.1\%\) & \(95.5\%\) & \(79.5\%\) & \(92.1\%\) \\
& \(\FO\) & \(8.618\) & \(+1.9\%\) & \(55.4\%\) & \(88.7\%\) & \(78.0\%\) & \(91.4\%\) \\
\bottomrule
\end{tabular}
\caption{Accuracy of \(\FO\) and of two 8-bit quantizations of the Qwen3 models, as in \cref{tab:acc}, with GSM8K generations of up to 1{,}024 tokens.  Paired 95\% intervals of the excess perplexity: \(0.4\)--\(0.6\%\), \(1.6\)--\(1.9\%\) and \(6.3\)--\(6.9\%\) for Qwen3-32B, \(0.2\)--\(0.4\%\), \(-0.5\)--\(-0.2\%\) and \(1.6\)--\(2.3\%\) for Qwen3-30B-A3B; single perplexities have intervals of about \(\pm 0.35\) and \(\pm 0.40\).  The checkpoints are prompted as base models, without a chat template.}
\label{tab:acc-qwen3}
\end{table}

On WikiText-2, \(\FO\) is \(6.6\%\) above BF16 in perplexity on Qwen3-32B and \(1.9\%\) on Qwen3-30B-A3B, and its greedy token agrees with BF16's at \(82.4\%\) and \(88.7\%\) of the positions.
The split of the score exponent between queries and keys (\cref{sec:quantize}) accounts for much of this: the same models calibrated with all of the slack given to the keys are \(8.6\%\) and \(4.1\%\) above BF16, with agreement \(79.3\%\) and \(85.3\%\), so the split removes \(23\%\) (\(19\) to \(26\)) and \(53\%\) (\(46\) to \(60\)) of the excess log perplexity, at no cost in the proof.
On MMLU, \(\FO\) is \(1.4\) points below BF16 on Qwen3-32B and \(1.6\) on Qwen3-30B-A3B; the split gained \(2.6\) and \(3.0\) points (McNemar \(z = 8.9\) and \(10.3\)).
On GSM8K it is \(1.7\) and \(1.4\) points below BF16, with \(54\) against \(31\) and \(50\) against \(32\) problems solved by only one of the two (McNemar \(z = -2.5\) and \(-2.0\)); FP8 and SmoothQuant INT8 differ from BF16 by at most \(0.5\) points (\(|z| \le 1.2\)).

The losses on the tasks are larger than on Qwen2.5 for two reasons that perplexity does not show.
First, Qwen3 is less certain of its MMLU answers: on a subsample of every seventh question, measured on models calibrated with all of the slack given to the keys, the median margin between BF16's best and second answer logit is \(5.6\) and \(5.8\) nats for Qwen3-32B and Qwen3-30B-A3B against \(8.3\) for Qwen2.5, while \(\FO\)'s error on that margin is \(1.8\) to \(2.3\) nats on all three models, so errors of the same size flip more answers.
Applying the Qwen3 flip rate per margin to Qwen2.5's margins accounts for about \(85\%\) of the excess flips of Qwen3-30B-A3B and \(60\%\) of those of Qwen3-32B.
Second, on GSM8K the loss of Qwen3-32B is mostly a change of style: \(527\) of \(\FO\)'s 1{,}319 answers open in the step-by-step style, against \(92\) for BF16, \(43\) for FP8 and \(324\) for SmoothQuant INT8.
In every variant, answers in that style are right less often (\(89.2\%\) of them for \(\FO\), \(84.8\%\) for BF16) than direct answers, of which \(\FO\) gets \(94.2\%\) right and BF16 \(94.6\%\).
Qwen3-30B-A3B never switches style under this prompt.

We repeated the attribution of \cref{sec:eval-attrib} on both Qwen3 models over the first 24 WikiText-2 windows, with paired intervals (\cref{tab:attrib-qwen3}).
On Qwen3-30B-A3B, 12-bit queries and keys remove all of the remaining gap and more (\(145\%\), \(119\) to \(186\)); every other step, including the router in floating point and an exact merge of the experts, is within noise of zero, except the float softmax, which makes the gap larger.
On Qwen3-32B the input of the down projection matters most: leaving it unrounded removes \(57\%\) (\(47\) to \(67\)), and rounding it to 9 bits instead of 8 removes \(46\%\) (\(33\) to \(57\)); 12-bit queries and keys remove \(21\%\), an unrounded MLP input \(12\%\), and a float QK-norm, a float softmax or a SiLU with 4 more bits nothing distinguishable from zero.
Smoothing the down input with power-of-two channel scales, which is provable, makes the gap several times larger.

\begin{table}[t]
\centering\small
\begin{tabular}{@{}lrr@{}}
\toprule
& \multicolumn{2}{c}{Gap removed (paired 95\% interval)} \\
\cmidrule(l){2-3}
Variant of \(\FO\) & Qwen3-32B & Qwen3-30B-A3B \\
\midrule
12-bit queries and keys & \(21\%\) (\(11\) to \(34\)) & \(145\%\) (\(119\) to \(186\)) \\
float softmax & \(-6\%\) (\(-17\) to \(6\)) & \(-35\%\) (\(-85\) to \(-6\)) \\
float QK-norm & \(3\%\) (\(-5\) to \(10\)) & \\
unrounded attention input & \(-3\%\) (\(-12\) to \(6\)) & \(8\%\) (\(-20\) to \(29\)) \\
unrounded MLP input & \(12\%\) (\(1\) to \(21\)) & \(10\%\) (\(-20\) to \(38\)) \\
unrounded attention output & \(-4\%\) (\(-14\) to \(4\)) & \(13\%\) (\(-16\) to \(39\)) \\
SiLU with 4 more bits & \(-5\%\) (\(-18\) to \(4\)) & \(9\%\) (\(-17\) to \(30\)) \\
16-bit up projection & \(8\%\) (\(-3\) to \(17\)) & \(-10\%\) (\(-55\) to \(24\)) \\
unrounded down input & \(57\%\) (\(47\) to \(67\)) & \(-16\%\) (\(-70\) to \(19\)) \\
9-bit down input & \(46\%\) (\(33\) to \(57\)) & \\
float router & & \(-3\%\) (\(-52\) to \(33\)) \\
exact expert merge & & \(-2\%\) (\(-50\) to \(32\)) \\
all of the above except 9 bits & \(87\%\) (\(82\) to \(93\)) & \(45\%\) (\(21\) to \(64\)) \\
\bottomrule
\end{tabular}
\caption{Accuracy attribution on Qwen3, on the first 24 WikiText-2 windows: the share of \(\FO\)'s excess log perplexity over BF16 (\(6.4\%\) and \(2.0\%\) on these windows) that each variant removes, as in \cref{tab:attrib}.  Unrounded: the row enters the next product with every entry exact, at the same direction and scale.  On Qwen3-30B-A3B the unrounded down input also makes the merge of the experts exact.  None of these variants is provable as it stands.}
\label{tab:attrib-qwen3}
\end{table}

The 9-bit down input is provable with the existing gadgets: \(\gad{Requant8}\) with bound 255 and the down projection's \(\gad{ShiftRound}\) stay wrap-free while \(d_{\mathrm{ff}} \cdot 127 \cdot 255 + 2^{30} + 2^{23} < p\), which holds for Qwen3-32B.
We did not adopt it: on the model calibrated without the split, and with GSM8K generations limited to 256 tokens, it improved perplexity and MMLU but lowered GSM8K, a drop we could not separate from the change of style above.
12-bit queries and keys need the score rounding of \cref{sec:eval-attrib}; by the cost model they multiply the proving time of Qwen3-32B by \(1.39\) at 4K and \(1.78\) at 16K.

\subsection{Where \texorpdfstring{\(\FO\)}{F\_O} is undefined}
\label{sec:eval-abort}

We ran \(\FO\) with the three conditions of \cref{sec:fo-undefined} recorded instead of enforced: a SiLU input \(|c| > 2^{14}\) or a gate shift outside \([1, 15]\); an attention-output or down-projection shift outside \([-16, 24]\); and a left-shifted output outside \(A\).
The scan runs the prefill of the evaluation engine, so it covers every prompt position and no generated one.
\cref{tab:abort} lists the inputs: every position of the WikiText-2 test split, all GSM8K prompts and 2{,}000 random MMLU prompts (3.45 million tokens of real text), and 160 adversarial sequences of 4{,}096 tokens.

\begin{table}[t]
\centering\small
\begin{tabular}{@{}lrrrrr@{}}
\toprule
& & & \multicolumn{3}{c}{Undefined} \\
\cmidrule(l){4-6}
Input & Sequences & Tokens & Qwen2.5 & Qwen3-32B & 30B-A3B \\
\midrule
WikiText-2 test, 2{,}048-token windows & 145 & 296{,}960 & 0 & 0 & 0 \\
GSM8K, 8-shot prompts & 1{,}319 & 1{,}754{,}311 & 0 & 0 & 0 \\
MMLU, 5-shot prompts (random 2{,}000) & 2{,}000 & 1{,}398{,}750 & 0 & 0 & 0 \\
One token repeated 4{,}096 times & 32 & 131{,}072 & 0 & 0 & 0 \\
Uniformly random token ids & 32 & 131{,}072 & 0 & 0 & 0 \\
Random digit strings & 32 & 131{,}072 & 0 & 0 & 0 \\
Random bytes (Latin-1 text) & 32 & 131{,}072 & 0 & 0 & 0 \\
Repeated injection-style phrases & 32 & 131{,}072 & 0 & 0 & 0 \\
Every vocabulary token at position 0 & & 151{,}936 & & 0 & 0 \\
\bottomrule
\end{tabular}
\caption{Sequences on which \(\FO\) is undefined.  The adversarial inputs are 4{,}096 tokens long.  The Qwen3 tokenizer gives the same token counts; the last row counts first tokens.}
\label{tab:abort}
\end{table}

For the Qwen3 models, the scan uses the serving engine (\cref{sec:discussion}), which evaluates every condition of \(\FO\), including the QK-norm and router shifts, and flags each sequence that reaches one; we also ran every vocabulary token as a one-token prompt.
No input reached any condition, so the honest prover would not have aborted on any of them.
Without the row clamps of \cref{sec:qwen3}, 7 first tokens of Qwen3-32B and 2 of Qwen3-30B-A3B left the domain, all of them rare tokens whose first-position residual has a very small norm; the clamps define these rows and change no other row: on both models calibrated without the split, WikiText-2 predictions and GSM8K answers were identical with and without them.
Calibration leaves margin: the SiLU range, for instance, is set \(25\%\) above the largest gate input seen.
This does not show that no input reaches a condition: an adversary who knows the weights could search for one, and a server could then refuse it.
For open weights, the client can confirm such a refusal by evaluating \(\FO\) itself (\cref{sec:threat}).

\subsection{Correctness and soundness testing}
\label{sec:eval-tests}

The test suite runs every protocol step on a small model with the shape of Qwen2 (two layers, width 256, four query and two key-value heads).
Besides honest proofs, it checks rejection in five ways.
Every message of a proof is corrupted in turn, and each corrupted proof must be rejected.
A proof made with different weights must not verify against \(C_W\).
A proof whose sampled tokens do not follow from the seed must be rejected at temperature \(\theta > 0\).
Targeted forgeries of the final openings (25 messages and 9 oracle elements) must be rejected.
Forgeries of the no-wrap bounds must be rejected: a \(\gad{ShiftQ8}\) row quantized at a too-large exponent with a wrapped minimality witness, rows whose \(\gad{ShiftQ8S}\) or gate shift exceeds its bound, and left-shift rows built from the residues of \cref{rem:leftshift}; so must a proof checked against model constants that differ from the prover's, and tokens outside the vocabulary. A \(\gad{NormScale}\) row normalized by a wrapped left shift is rejected, and the same proof verifies once the bound \((16, [E < 0]\,\Sigma)\) is removed from prover and verifier; for both signed normalizations, every wrapped left shift in a randomized search satisfies the field relation and is rejected by the input bound.
For Qwen3, small models with QK-norm and with a mixture of 8 experts (4 per token) prove and verify end to end, with the prover's witness checked against \(\FO\); a head row's requantization off by one entry, a router weight off by one, and a slot routed to an expert that was not selected are rejected.
Every row clamp is exercised by models re-quantized so that some rows clamp and some sit exactly at the edge, and a flag forged on an edge row, which gives the same shift, is rejected.
Two forgeries target wraps that the bounds of \cref{sec:qwen3-proof} exclude: QK-norm head rows claimed at exponent 22 through the unquartered minimality relation, which changes every generated token of the small model, and a router logit claimed saturated at \(k_r = 15\) (\cref{sec:qwen3-proof}).
Both are rejected, and both would be accepted with the unquartered relation or a router shift bound of 15.
The device transcript and the graph-replayed prover are checked to produce byte-identical proofs to the host transcript and the plain prover.
For the BLAKE3 circuit, a forged XOR byte with every arithmetic constraint repaired and a message word that differs between input rows and G rows are both rejected (\cref{sec:recursion}).

\section{Discussion}
\label{sec:discussion}

\paragraph{The price of exactness.}
Sable proves the function a server runs, so the server must run \(\FO\).
\(\FO\) uses 8-bit matrix products with exact 32-bit accumulation, which integer tensor cores support, but its fused rounding steps, the 9-bit outputs of \(\gad{RoPE8}\), and the blocked \(PV\) product need their own kernels.
We wrote a serving engine for it that computes \(\FO\) bit for bit (checked token by token against the reference implementation): attention is an exact integer kernel in Triton that never forms the score matrix (products of 9-bit integers in fp16 with fp32 accumulation are exact while every partial sum stays below \(2^{24}\), which the blocked \(PV\) product guarantees), the integer element-wise work of each layer is compiled into fused kernels, and each decoding step is replayed as one CUDA graph.
\cref{tab:serve} compares it with BF16 and FP8 (W8A8, per-tensor scales) versions of the same model in SGLang~\cite{sglang} on the same GPU.

\begin{table}[t]
\centering\small
\begin{tabular}{@{}lrrrrr@{}}
\toprule
& \multicolumn{2}{c}{Batch 1} & \multicolumn{3}{c}{Batch 32} \\
\cmidrule(lr){2-3}\cmidrule(l){4-6}
Engine & tok/s & TTFT (s) & tok/s & TTFT (s) & end to end \\
\midrule
\(\FO\), this engine & \(27.5\) & \(1.24\) & \(428.5\) & \(38.8\) & \(212.9\) \\
FP8, SGLang & \(25.3\) & \(0.46\) & \(542.0\) & \(14.6\) & \(365.7\) \\
BF16, SGLang & \(21.0\) & \(0.72\) & \(269.7^{*}\) & \(23.2\) & \(195.5\) \\
\bottomrule
\end{tabular}
\caption{Serving Qwen2.5-32B on one RTX PRO 6000: prompts of 4{,}096 tokens, 512 generated tokens per sequence, greedy decoding.  tok/s: decoding throughput after the first token; TTFT: time to the first token of the whole batch; end to end: generated tokens over total time.  \(^{*}\)BF16 weights leave room for 89{,}581 cached tokens, fewer than the \(147{,}456\) of 32 sequences, so SGLang ran about 19 at a time.}
\label{tab:serve}
\end{table}

At batch 1, decoding is bound by reading the weights, and \(\FO\), with 8-bit weights, matches FP8 and is \(30\%\) faster than BF16.
At batch 32, \(\FO\) decodes at \(79\%\) of the FP8 rate: its exact attention and integer element-wise work cost more than FP8's, and its 8-bit matrix products run on an older-generation cuBLASLt kernel (a tile configuration of the Ampere generation) than SGLang's FP8 path.
The largest gap is prefill, \(2.7\) times slower than FP8 at batch 32, because our engine prefills two sequences at a time with exact integer attention while SGLang batches them through tuned kernels.
These are costs of any exact scheme: a proof about a floating-point computation would first have to fix the order of every reduction, which current kernels do not do~\cite{he2025nondeterminism}.

\paragraph{Where the time goes, and what would reduce it.}
Commitment, lookups and constraint terms account for almost all of the prover's work.
At long contexts, attention dominates, and the \(16\)K proof already uses 72 of the GPU's 96\,GiB.
Every causal score entry costs one committed probability, one lookup and two constraint terms, and the probability must be committed because it is a lookup output.
Sliding-window attention in most layers, as some current models use, would remove most of these entries.
At short contexts, the prover is partly latency-bound; batching the many small sumchecks of a layer into fewer, larger ones, and moving the remaining host-side claim arithmetic to the GPU, are the next steps there.
Univariate skip for the first round~\cite{gruen} would save an estimated \(2\%\) to \(4\%\).

\paragraph{Soundness parameters.}
The tightest terms of \cref{tab:budget} are the queries (\(103.4\) bits, of which 30 are proof of work) and the lookups (\(103.9\) bits, of which 11 are proof of work).
That margin is what a composition with up to four recursive layers needs to stay above 100 bits overall.
The lookup term depends on drawing one coefficient per tuple coordinate: with the powers of a single challenge, which is a common choice, tuples of arity seven would lose \(\log_2 6 \approx 2.6\) bits (\cref{lem:tuple}) and the term would drop to \(101.3\) bits.
Likewise, an exact left shift in a 31-bit field needs a bound on its input as well as on its output (\cref{rem:leftshift}).

\paragraph{Proof size and recursion.}
A main proof of 106 to 477\,MiB suits an auditor; returning one with every response is impractical.
Reducing it has so far reduced the recursion's input as well: the verifier's hashing fell by \(4.2\times\).
The remaining verifier work is dominated by GKR challenges and leaf hashing; fewer GKR rounds (wider fraction trees, or fewer separate trees per domain) and larger work units would reduce both.
Our BLAKE3 circuit is within \(30\%\) of the cell count assumed by the design, but the commitment of its trace is not yet measured, and an algebraic hash would make the recursion several times cheaper at the cost of a further assumption.

\paragraph{Zero knowledge.}
The main proof reveals evaluations of the trace and is meant for the requesting client.
Multi-tenant proofs, closed weights that must stay secret, and publication of proofs all need zero knowledge; without it, about \(1.9 \times 10^4\) proofs reveal the weights (\cref{sec:threat}), and a batch reveals its prompts to every verifier of the batch.
For multilinear arguments this requires hiding commitments (random symbols in every column, salted leaves), masked sumcheck and GKR messages in the style of~\cite{cfs,libra}, and lookup roots sent only per chunk.
Each piece is standard; their combination with a streamed, chunked prover needs its own analysis.

\paragraph{Accuracy.}
\(\FO\)'s perplexity is \(8.3\%\) above BF16's, against \(0.7\%\) for FP8.
Wider queries and keys would recover about a third of the gap at an estimated 30 to 68\% more proving time (\cref{sec:eval-attrib}); we have not located the rest.
On Qwen3, a better split of the existing score precision between queries and keys removed about a quarter to a half of the gap at no proving cost, which suggests that calibration still has room before formats must widen; finding integer formats that close the gap at a small proving cost is the main open question for \(\FO\).
Task benchmarks deserve the same scrutiny as perplexity: on Qwen3 a few nats of logit error matter more on MMLU than on Qwen2.5 because the model's own margins are smaller, and on GSM8K most of the loss of Qwen3-32B came from a change of answer style.

\paragraph{Generality.}
Besides Qwen2.5-32B, we evaluated two Qwen3 models: a dense one with QK-norm and a mixture of experts with 128 experts per layer (\cref{sec:qwen3}).
Both reuse the protocol and add gadgets: QK-norm is the existing requantization and normalization applied per head row, and the experts are selected by a threshold argument and a dynamic multiset and multiplied by gathered sumchecks (\cref{sec:qwen3-proof}).
Other attention variants change the tile list.
Other architectures change the gadget composition and keep the protocol.

\section{Conclusion}
\label{sec:conclusion}

Sable shows that bit-exact proofs of production-size language model inference are feasible on one GPU.
The full 64-layer Qwen2.5-32B is proven in 92.5 seconds for 4{,}096 tokens and in under seven minutes for 16{,}384 tokens, 41 to 50 times the time of the inference it attests and 40 to 64 times less time per parameter and token than zkLLM.
The same prover handles QK-norm and a mixture of experts: it proves 4{,}096 tokens of Qwen3-32B in 121.8 seconds and of Qwen3-30B-A3B, whose tokens each use 8 of 128 experts per layer, in 84.7 seconds, and on both models \(\FO\) stays within 1.6 points of BF16 on MMLU and 1.7 points on GSM8K.
Three ingredients make this possible.
An integer function fixes every rounding decision of the model.
A constraint system over a 31-bit field never wraps, and its chunked lookups remain sound beyond \(p\) lookups.
A streamed prover recomputes the witness where it would otherwise store it, and replays identical layers as GPU graphs.
The same proof system proves BLAKE3 compressions at a few microseconds each, which is the basis for compressing the main proof recursively.
Served by an engine that computes it exactly, \(\FO\) decodes as fast as FP8 at batch 1 and at \(79\%\) of its rate at batch 32.
The remaining costs are a proof that takes about 50 times the inference it attests, an integer function that loses accuracy against BF16, and, until the recursive layers with zero knowledge exist, a proof suited to auditors of open-weight models.
The next steps are those layers, a verifier compiled as a whole, and integer formats that recover the lost accuracy.

\section*{Acknowledgements}
We thank Omer Shlomovits, Miki Asa, Karthik Inbasekar and Suyash Bagad for useful discussions and comments.

\bibliographystyle{alpha}
\bibliography{refs}

\appendix
\section{Proofs}
\label{app:proofs}

\subsection{LogUp over a prime field (\texorpdfstring{\cref{lem:logup}}{Lemma})}

For each value \(w\) that occurs among the \(a_i\) or in \(T\), let \(c(w) = |\{i : a_i = w\}|\), an integer in \([0, N] \subset [0, p)\), and let \(m(w) = 0\) for \(w \notin T\).
The hypothesis says \(\sum_w (c(w) - m(w))/(Z - w) = 0\) in \(\F(Z)\), where the sum runs over distinct \(w\).
Multiplying by \(\prod_{w'} (Z - w')\) and evaluating at \(Z = w_0\) leaves \((c(w_0) - m(w_0)) \prod_{w' \ne w_0} (w_0 - w') = 0\), and the product is nonzero, so \(m(w_0) = c(w_0)\) in \(\F\).
If \(w_0 \notin T\), then \(m(w_0) = 0\), so \(c(w_0) \equiv 0 \pmod p\), and \(0 \le c(w_0) < p\) gives \(c(w_0) = 0\).
Hence every \(a_i\) lies in \(T\). \qed

\subsection{Tuple compression (\texorpdfstring{\cref{lem:tuple}}{Lemma})}
\label{app:tuple}

Let \(U\) be the set of distinct tuples in \(\mathcal{A} \cup \mathcal{T}\), \(|U| = D\).
For \(u \in U\) let \(c_u = \mathrm{mult}_{\mathcal{A}}(u) - m(u) \in \Fq\), with \(m(u) = 0\) for \(u \notin \mathcal{T}\).
By hypothesis there is \(u_0 \in \mathcal{A} \setminus \mathcal{T}\), and \(c_{u_0} = \mathrm{mult}_{\mathcal{A}}(u_0) \in [1, p)\) is nonzero in \(\Fq\).

Write \(q_u(\beta) = \sum_i \beta^i u_i\), so \(\phi_u = Z - q_u(\beta)\).
Distinct tuples give distinct polynomials \(q_u\), hence distinct \(\phi_u\); each \(\phi_u\) is monic of degree 1 in \(Z\) and of total degree at most \(k\).
Let \(\Phi = \prod_{u \in U} \phi_u\) and \(R = \sum_{u \in U} c_u \prod_{w \ne u} \phi_w\), so that \(\sum_u c_u / \phi_u = R / \Phi\) in \(\Fq(Z, \beta)\).

\emph{\(R\) is nonzero.}
Consider the homomorphism \(\Fq[Z, \beta] \to \Fq[\beta]\) that substitutes \(Z = q_{u_0}(\beta)\).
It maps \(\phi_{u_0}\) to \(0\), so every term of \(R\) except the one for \(u_0\) maps to \(0\), and \(R\) maps to \(c_{u_0} \prod_{w \ne u_0} (q_{u_0}(\beta) - q_w(\beta))\).
Each factor is a nonzero polynomial because \(q_w \ne q_{u_0}\), and \(\Fq[\beta]\) is an integral domain, so the image is nonzero and \(R \ne 0\).

\emph{Bounds.}
\(R\) has total degree at most \(k(D - 1)\).
The event in the lemma, the identity holding with all denominators nonzero, implies \(R(Z, \beta) = 0\), which by the Schwartz--Zippel lemma has probability at most \(k(D-1)/|\Fq|\).
A denominator vanishes exactly when \(\Phi(Z, \beta) = 0\), and \(\Phi\) is a nonzero polynomial of total degree at most \(kD\).
With independent \(\beta_1, \ldots, \beta_k\), each \(\phi_u = Z - u_0 - \sum_i \beta_i u_i\) is linear, the same argument applies with the substitution \(Z = u_{0,0} + \sum_i \beta_i u_{0,i}\), and the degrees become \(D - 1\) and \(D\). \qed

\subsection{Domains (\texorpdfstring{\cref{lem:domain}}{Lemma})}
\label{app:domain}

Fix the columns \(f_j\).
Let \(p^\ast_\ell, q^\ast_\ell : \bits^{n-1} \to \Fq\) be the true layer above the leaves, computed from the columns, and \(\mle{p^\ast_\ell}, \mle{q^\ast_\ell}\) their multilinear extensions.
Since \(p^\ast_\ell(y)\) is a polynomial in the half-columns at \(y\), \(\sum_y \eq(r, y)\, p^\ast_\ell(y) = \mle{p^\ast_\ell}(r)\), and likewise for \(q^\ast_\ell\) and for the constraint halves \(C^h_k\).

\emph{Step 1.}
If some root fraction is false, then by the soundness of GKR for fractional sums~\cite{logupgkr}, except with probability \(\epsilon_{\mathrm{GKR}}\), the claims at \(r\) satisfy \((\pi_\ell, \kappa_\ell) \ne (\mle{p^\ast_\ell}(r), \mle{q^\ast_\ell}(r))\) for some \(\ell\).
If instead every root fraction is right but some constraint \(C_k\) is nonzero at a row \((y^\ast, h^\ast)\), then \(y \mapsto C^{h^\ast}_k(y)\) is a nonzero function on \(\bits^{n-1}\), its multilinear extension is a nonzero multilinear polynomial, and since \(r\) is uniform and independent of the columns, \(\mle{C^{h^\ast}_k}(r) \ne 0\) except with probability \((n-1)/|\Fq|\).

\emph{Step 2.}
Let \(\Delta(a, \lambda, \gamma, \eta) = \sum_\ell a_\ell \big((\pi_\ell - \mle{p^\ast_\ell}(r)) + \lambda (\kappa_\ell - \mle{q^\ast_\ell}(r))\big) - \sum_k \gamma_k \big(\mle{C^0_k}(r) + \eta\, \mle{C^1_k}(r)\big)\), the difference between the claimed right-hand side of \eqref{eq:domain} and its true value.
In either case of step 1, \(\Delta\) is a nonzero polynomial of total degree 2 in the fresh challenges, so \(\Delta \ne 0\) except with probability \(2/|\Fq|\), and then the claim of the sumcheck in step 2 is false.

\emph{Step 3.}
By \cref{lem:sumcheck}, a false sumcheck claim leads to a false final claim except with probability \(\epsilon_{\mathrm{sc}}\).
The final claim is the bracket evaluated at the half-column values the prover sends, so at least one sent value \(f^h_j(\rho)\) differs from the true \(\mle{f^h_j}(\rho)\).

\emph{Step 4.}
For that column, \(\mle{f_j}(\rho, t) = (1 - t)\mle{f^0_j}(\rho) + t\,\mle{f^1_j}(\rho)\) because \(\mle{f_j}\) is multilinear in its last variable.
The claimed value \((1 - t) f^0_j(\rho) + t f^1_j(\rho)\) is a different affine function of \(t\), so the two agree for at most one \(t\), and the claim at \((\rho, t)\) is false except with probability \(1/|\Fq|\).
Adding the four error terms proves the lemma.
With \(K > 1\) blocks, the block variables add rounds to the sumcheck and coordinates to \(r\), and the argument is unchanged. \qed

\subsection{Chunks (\texorpdfstring{\cref{lem:chunks}}{Lemma})}
\label{app:chunks}

Let \(i\) be a chunk that contains a lookup outside \(\Tab\).
Its lookups number at most \(2^{30} < p\), so by \cref{lem:tuple}, applied with \(\mathcal{A}\) the chunk's tuples, \(\mathcal{T} = \Tab\) and \(m = m_i\), the identity \(S_i = \sum_v m_i(v)/\phi_v(Z, \beta)\) holds at the drawn \((Z, \beta)\) with probability at most \(\epsilon_{\mathrm{LU}}\).
(Here \(S_i\) equals the true fraction sum of the chunk by the hypothesis on root fractions.)
If it fails, then \(\sum_i \gamma_i S_i \ne \sum_i \gamma_i \sum_v m_i(v)/\phi_v\) except with probability \(1/|\Fq|\) over the \(\gamma_i\), drawn after the \(S_i\) are fixed.
The table GKR then starts from a false claim, and by its soundness ends in a false claim on \(M = \sum_i \gamma_i m_i\) or on a table column, hence on some \(m_i\) or table column, except with probability \(\epsilon_{\mathrm{GKR}}\). \qed

\subsection{Row maximum}

\begin{lemma}[Row maximum]
\label{lem:rowmax}
Suppose the gather set of a key-value group sums to zero as a rational function, the range lookups \(m - j^\ast_m \in [0, 2^{24})\) and \(j^\ast_m - s_m \in [0, 2^{24})\) hold with \(s_m\) the first position of \(m\)'s sequence, \(M_m = \sum_c \hat Q_{mc} K^g_{mc}\), and \((\mathsf{EXP}, x_{mj}, P_{mj}) \in \Tab\) for every kept entry.
Then \(M_m = \max_{s_m \le j \le m} \mathrm{acc}_{mj}\) and \(P_{mj} = \tau_{\exp}(M_m - \mathrm{acc}_{mj})\) for every valid \(j\).
\end{lemma}

\begin{proof}
The gather set has fewer than \(p\) reads, so by \cref{lem:logup} every read tuple \((j^\ast_m, c, K^g_{mc})\) equals a table-side tuple \((j, c, \hat K_{jc})\).
The table side has exactly one tuple per pair \((j, c)\), so its tuples are distinct, as \cref{lem:logup} requires, and \(K^g_{mc} = \hat K_{j^\ast_m, c}\), and \(M_m = \mathrm{acc}_{m, j^\ast_m}\).
The range lookups, over integers below \(2^{24}\), give \(s_m \le j^\ast_m \le m\), so \(j^\ast_m\) is a valid key.
For every valid \(j\), the softmax lookup has input \(M_m - \mathrm{acc}_{mj}\) (with \(\nu_{mj} = 1\) on diagonal tiles); both scores are integers below \(2^{22}\) in absolute value, so the field value \(M_m - \mathrm{acc}_{mj}\) lies in the table's domain \([0, 2^{23})\) only if the integer difference does.
Hence \(M_m \ge \mathrm{acc}_{mj}\) for every valid \(j\), with equality at \(j^\ast_m\).
The table has one row per input, so \(P_{mj}\) is its output.
\end{proof}

\subsection{No wrap and determinism}
\label{app:nowrap}

\begin{table}[h]
\centering\small
\begin{tabular}{@{}lll@{}}
\toprule
Relation & Largest term & Bound \\
\midrule
Linear-layer accumulator & \(2^{15} \cdot 127^2\) & \(< 2^{29}\) \\
MoE down accumulator & \(k d_e \cdot 127 \cdot 1023\), \(k d_e = 6144\) & \(< 2^{29.6}\) \\
Attention score & \(128 \cdot 180^2\) & \(< 2^{22}\) \\
\(\gad{ShiftRound}\), \(k \ge 1\) & \(|\mathrm{acc}| + 2^{k-1} + |2^k c| + u\), \(|2^k c| \le 2^{30}\) & \(< 2^{30.9}\) \\
\(\gad{Requant8}\) & \(2|a| + s + 2 s \cdot 127\), \(s \le 2^{17}\) & \(< 2^{26}\) \\
\(\gad{ShiftQ8}\) & \(|\mathrm{acc}| + s/2 + 127 s\), \(s \le 2^{22}\) & \(< 2^{30.1}\) \\
\(\gad{Requant8}\) minimality & \(4|a| + 255 s + 1\); limbs \(w < 2^{30}\) & \(< 2^{30.1}\) \\
\(\gad{ShiftQ8}\) minimality & \(|\mathrm{acc}| + 255 s/4 + 1\); limbs \(w < 2^{28}\) & \(< 2^{30}\) \\
\(\gad{ShiftQ8S}\), \(o = 0\) & \(|\mathrm{acc}| + 2^{k-1} + 2^k(127 + |b|) + u\), \(u < 2^k\), \(k \le k_{\max}(b)\) & \(< p\) \\
\(\gad{ShiftQ8S}\), \(o = \pm 1\) & \(|\mathrm{acc}| + 2^k(|b| + 128)\); limbs \(w < 2^{30}\) & \(< p\) \\
QK-norm \(\gad{ShiftQ8S}\) & \(|q'' G| < 2^{14}\), \(b = 0\), \(k \le 22\): \(2^{14} + 2^{21} \cdot 255 + 2^{30}\) & \(< 2^{30.6}\) \\
Router \(\gad{SatShift}\) & \(|\mathrm{acc}| < 2^{25}\), \(k \le 14\): \(2^{25} + 2^{13}(2^{16} - 1) + 1\); limbs \(w < 2^{30}\) & \(< 2^{30.6}\) \\
Top-\(k\) selection & \(|z - \vartheta| + 1\), \(|j - j_t| + 1\) & \(< 2^{17}\) \\
Router weights & \(2^{15} p_s + \Sigma_p + 2\Sigma_p w_s\), \(\Sigma_p \le 2^{15}\), \(w_s \le 2^{14}\) & \(< 2^{30.2}\) \\
MoE requantization & \(2|a| + s + 2s \cdot 1023\); \(4|a| + 2047 s + 1\), \(|a| \le 2^{21}\), \(s \le 2^{17}\) & \(< 2^{28.1}\) \\
Row clamps & \(|K_{gl} - E_h|\), \(|v| + 25\) & \(< 2^{7}\) \\
\(\gad{NormScale}\) & \(\Sigma < 2^{29}\); \(2^{E^-}\Sigma\) with \(\Sigma < 2^{16}\) if \(E < 0\); \(2^{E^+}\delta + \rho\) & \(\le 2^{30}\) \\
\(\gad{NormScale}\) output & \(|q|\mu + 2^{14} + 2^{15}|q'|\) & \(< 2^{23}\) \\
\(\gad{RoPE8}\) & \(2 \cdot 127 \cdot 2^{15} + 2^{14} + 2^{15} \cdot 180\) & \(< 2^{24}\) \\
MLP gate & \(|\mathrm{acc}| + 2^{k-1} + 2^k \cdot 2^{14}\), \(k \le 15\) & \(< 2^{30.1}\) \\
MLP product & \(|a q_u| \le 2^{15} \cdot 127\) & \(< 2^{22}\) \\
Softmax input & \(|M - \mathrm{acc}|\) & \(< 2^{23}\) \\
Softmax denominator & \(2^{10} S\), \(S \le 2^{20}\) & \(\le 2^{30}\) \\
Its normalization & \(2^{E^-} D\) with \(D < 2^{22}\) if \(E < 7\); \(2^{E^+}\delta + \rho\) & \(\le 2^{30}\) \\
\(PV\) block & \(2^{12} \cdot 2^{10} \cdot 127\) & \(< 2^{29}\) \\
\(PV\) sum of rounded blocks & \(127 D_m / 2^{k_m} + 2^8\) & \(< 2^{23}\) \\
\(\gad{MulRound}\) & \(|x_l y| + 2^{12}|t| + v\); \(|x_h y| + |t|\) & \(< 2^{30}\); \(< 2^{29}\) \\
\(\gad{SatAdd}\) & \(|s| + 2^{23} + 1\) & \(< 2^{25}\) \\
Greedy output & \(|z_{t,y_t} - z_{tj}| + 1\) & \(< 2^{24}\) \\
Sampling sums & \(2^{12} |V|\), \(|V| = 152{,}064\) & \(< 2^{29.3}\) \\
Sampling limb product & \(3 \cdot 2^{20} + 2^{12} + 2^{10} + 2^{10} \cdot 2^{12}\) & \(< 2^{23}\) \\
Sampling inequalities & \(w_t + A_t + P_t + 1 + d_{t,i}\), \(d_{t,i} < 2^{30}\) & \(< 2^{30.6}\) \\
BLAKE3 G row & \(3 \cdot 255 + 2 + 256 \cdot 2\) & \(< 2^{11}\) \\
\bottomrule
\end{tabular}
\caption{Integer bounds behind \cref{lem:nowrap}.}
\label{tab:bounds}
\end{table}

\paragraph{Proof of \cref{lem:nowrap}.}
Each row of \cref{tab:bounds} bounds the sum of the absolute values of the terms of one constraint, or of the input of one lookup, given the ranges that the gadget's other lookups enforce and the ranges of its inputs established by earlier gadgets.
For instance, in \(\gad{ShiftRound}\) with \(k \ge 7\), the lookups give \(0 \le u < 2^k \le 2^{24}\) and \(|2^k c| \le 2^{30}\), and \(|\mathrm{acc}| < 2^{29.6}\) and \(2^{k-1} \le 2^{23}\), so \(|\mathrm{acc}| + 2^{k-1} + |2^k c| + u < 2^{30.9} < p\).
If an expression \(E\) over \(\Z\) with \(|E| < p\) satisfies \(E \equiv 0 \pmod p\), then \(E = 0\); and if a lookup input \(e\) with \(|e| < p\) is congruent to a table value \(v \in [0, 2^{24})\), then \(e = v\) over \(\Z\) as long as \(|e - v| < p\), which the bounds give. \qed

\paragraph{Proof of \cref{lem:det}.}
By \cref{lem:nowrap}, every relation can be read over \(\Z\).
We treat the gadgets in forward order; their inputs are determined by the earlier gadgets (induction), the public inputs and the committed weights.
\begin{itemize}
\item \(\gad{ShiftRound}\), \(k \ge 1\): \(0 \le \mathrm{acc} + 2^{k-1} - 2^k c < 2^k\) determines \(c = \lfloor (\mathrm{acc} + 2^{k-1})/2^k \rfloor\). The regime tuples are functions of \(k\) through the table.
For \(k \le 0\): the \(\mathsf{SRA}\) lookup gives \(|\mathrm{acc}| \le 2^{23 + k}\), the first \(\mathsf{SR}\) lookup gives \(\mathrm{acc} - 2^{k} c = 0\) in \(\Fp\), that is \(c - 2^{-k}\mathrm{acc} = 0\) in \(\Fp\), and \(|c - 2^{-k}\mathrm{acc}| \le 2^{23} + 2^{23} < p\), so \(c = 2^{-k}\mathrm{acc}\) over \(\Z\).
\item \(\gad{Requant8}\): \(0 \le 2a_i + s - 2sq_i < 2s\) gives \(q_i = \lfloor a_i/s + 1/2 \rfloor\); \(q_i \le 127\) gives \(2a_i < 255 s\) and \(q_i \ge -127\) gives \(2a_i \ge -255 s\), so the row fits at \(e\).
If \(e > e_0\), then \(f = 1\) and \(w \ge 0\) gives \(4 a_{j^\ast} \ge 255 s\) or \(4 a_{j^\ast} < -255 s\), that is \(2 a_{j^\ast} \notin [-255 \cdot 2^{e-1}, 255 \cdot 2^{e-1})\): the row does not fit at \(e - 1\).
If \(e = e_0\), it is the least exponent allowed.
So \(e\) is the least admissible exponent, and \(q\) is determined.
The selected value equals \(a_{j^\ast}\) because the bits are Boolean and \(j^\ast < n\).
\(\gad{ShiftQ8}\) is identical with the undoubled remainder, and its minimality witness \(w = \pm a_{j^\ast} - 255 s/4 - [\sigma = 0]\) is the one above divided by 4: \(w \ge 0\) gives \(a_{j^\ast} \ge 255 \cdot 2^{e-2}\) or \(a_{j^\ast} < -255 \cdot 2^{e-2}\), so again the row does not fit at \(e - 1\).
The relation is wrap-free because \(|a_{j^\ast}| < 2^{29}\), \(255 s/4 < 2^{28}\) and the 14-bit limbs give \(w < 2^{28}\) (\cref{tab:bounds}); the undivided form with \(|4a_{j^\ast}|\) up to \(2^{31}\) would not be.
\item \(\gad{NormScale}\): for an integer \(\Sigma \in [1, 2^{29})\), there is exactly one even \(E\) and one \(\delta \in [2^{14}, 2^{16})\), \(\rho \in [0, 2^{E^+})\) with \(2^{E^-}\Sigma = 2^{E^+}\delta + \rho\), since the mantissa interval spans two octaves.
The lookups make the relation wrap-free, including \((16, [E < 0]\,\Sigma)\) for a left shift (\cref{tab:bounds}), so it holds over \(\Z\) and the field admits no other witness.
\(\mu\) is the table's output, and \(q'_i\) is determined by its range lookup.
\item \(\gad{ShiftQ8S}\): let \(r = \lfloor \mathrm{acc}/2^k \rceil + b\).
The per-row lookup \((5, K_{qs} - E_h)\) gives \(k \le k_{\max}(b)\), so both cases are wrap-free (\cref{tab:bounds}).
If \(o = 0\), the lookups give \(q = r\) and \(|q| \le 127\).
If \(o = 1\), \(w \ge 0\) gives \(\mathrm{acc} + 2^k b \ge 255 \cdot 2^{k-1}\), hence \(r \ge 128\); if \(o = -1\), it gives \(\mathrm{acc} + 2^k b < -255 \cdot 2^{k-1}\), hence \(r \le -128\).
The three cases are exclusive and exhaustive, so \(o\) and \(q = \clip(r, -127, 127)\) are determined.
\item \(\gad{RoPE8}\) and the MLP gate: one range lookup determines each quotient, and the gate's table determines \(a\).
\item \(\gad{MulRound}\): the ranges of \(x_l\) and \(x_h\) determine the split of \(x\), and those of \(v\) and \(t\) determine \(t = \lfloor x_l y / 2^{12} \rfloor\).
Then \(x y = 2^{12} H + v\) with \(0 \le v < 2^{12}\), so for \(k' = k - 12 \ge 1\),
\(\lfloor (xy + 2^{k-1})/2^k \rfloor = \lfloor (H + 2^{k'-1} + v/2^{12})/2^{k'} \rfloor = \lfloor (H + 2^{k'-1})/2^{k'} \rfloor\),
because \(H + 2^{k'-1}\) is an integer and \(0 \le v / 2^{12} < 1\).
The \(\gad{ShiftRound}\) relation on \(H\) determines \(c\).
\item \(\gad{SatAdd}\): \(o = 0\) requires \(s \in A\), \(o = 1\) requires \(s \ge 2^{23}\), and \(o = -1\) requires \(s < -2^{23}\); the cases are exclusive and exhaustive, and in each \(h'\) is the clipped sum.
\item Attention: \cref{lem:rowmax} determines \(M\) and \(P\); \(D\) and \(N^{(b)}\) are exact sums (virtual); the signed normalization of \(D\) is unique as for \(\gad{NormScale}\), with \(\delta \in [2^{15}, 2^{16})\) spanning one octave and \((22, (1 - f_m) D_m)\) as the input bound of a left shift; \(k_m\), the reciprocal and \(o\) follow.
\item Row clamps: the flag's two lookups exclude each other's case (\cref{sec:g-gate}), so the flag and the clamped shift are determined by the unclamped value.
\item QK-norm: \(\gad{ShiftQ8}\), \(\gad{NormScale}\) and \(\gad{ShiftQ8S}\) per head row, as above, with the shift determined by the clamp.
\item Router and selection: \(\gad{SatShift}\) is \(\gad{ShiftQ8S}\) with bound \(2^{15} - 1\), so \(z\) is determined.
\(\vartheta_t\) is the logit at \(j_t\), and the selection lookup gives \(\sigma_{tj} = 1\) exactly for the logits above \(\vartheta_t\) and those equal to it at indices up to \(j_t\).
Since \(\sum_j \sigma_{tj} = k\), \((\vartheta_t, j_t)\) is the \(k\)-th entry in the order by logit, descending, then index, ascending, and the selected set is the first \(k\) entries: it is unique, and the multiset equality makes the slots a permutation of it carrying each expert's exponents.
The outputs of \(\FO\) do not depend on the order of the slots.
\item Router weights: \(z_{\max}\) is a slot's logit and at least every other, so it is the maximum; \(p_s\) is a table output; \(0 \le \rho < 2\Sigma_p\) gives \(w_s = \lfloor (2^{15} p_s + \Sigma_p)/(2\Sigma_p) \rfloor\); \(\epsilon\) is the maximum of the \(\epsilon_s\) likewise, and \(\omega_s\), the requantization and the expert gadgets follow as above.
\item Output: at \(\theta = 0\), the range lookups give \(z_{t, y_t} > z_{tj}\) for \(j < y_t\) and \(z_{t, y_t} \ge z_{tj}\) for \(j > y_t\), so the public \(y_t\) is the first maximum.
At \(\theta > 0\), the probabilities are table outputs, the limb relations hold over \(\Z\) and give \(w_t = \lfloor u_t Z_t/2^{30} \rfloor\) (\cref{sec:g-sampling}), and the two inequalities hold for exactly one index, because the prefix sums are nondecreasing and \(w_t < Z_t\).
\end{itemize}
Linear-layer accumulators, \(\Sigma\), \(D\), \(N^{(b)}\) and the MLP product are polynomials in determined values and equal their integer definitions by the bounds of \cref{tab:bounds}.
By induction over the layers, the residual stream, the logits and the sampling decisions equal those of \(\FO\). \qed

\subsection{Determinism of \texorpdfstring{\(\FO\)}{F\_O} (\texorpdfstring{\cref{lem:det-fo}}{Lemma})}

\(\FO\) is a fixed composition of the operations of \cref{sec:primitives}, each of which is a function of its inputs.
The ranges claimed for intermediate values follow from the same bounds as \cref{tab:bounds}: \(\gad{Requant8}\) and \(\gad{ShiftQ8}\) never clip, by the half-open fitting interval; \(\gad{ShiftQ8S}\) clips by definition; the softmax probabilities are at most \(2^{10}\), so \(D_m \le 2^{10} S\) and each \(PV\) block is below \(2^{29}\); the rounded blocks sum to less than \(2^{23}\) because \(k_m \ge E_m\) and \(D_m < 2^{E_m + 16}\); and \(\gad{SatAdd}\) returns a value in \(A\).
The quantities whose range \(\FO\) requires without implying it (the SiLU input \(|c| \le 2^{14}\), the shift ranges, and inputs of left shifts with \(|2^{-k}\mathrm{acc}| \le 2^{23}\)) are checked by the honest prover, which aborts if one fails; for such inputs \(\FO\) is undefined.
\cref{sec:eval-abort} measures how often this happens. \qed

\section{Parameters}
\label{app:params}

\begin{table}[h]
\centering\small
\begin{tabular}{@{}p{0.14\linewidth}p{0.32\linewidth}p{0.48\linewidth}@{}}
\toprule
Component & Parameter & Value \\
\midrule
Fields & base field & \(p = 2^{31} - 2^{24} + 1\) \\
& IOP challenges & \(\Fq = \Fp[Y]/(Y^4 - 3)\) \\
& commitment challenges & \(\Fo = \Fp[X]/(X^8 - 3)\) \\
Lookups & table \(\Tab\) & \(2^{26}\) rows, 7 columns, tags of \cref{tab:tags} \\
& chunk size & at most \(2^{30}\) static lookups per multiplicity column \\
& tuple compression & independent coefficients \(\beta_1, \ldots, \beta_8 \in \Fq\) \\
& proof of work before \((Z, \beta)\) & 11 bits \\
Commitment & code & Reed--Solomon, rate \(1/2\), subgroups of order \(\le 2^{24}\) \\
& column sizes & \(2^5\) to \(2^{23}\); padding \(\le \max(10\%, 2^{20})\), weights \(15\%\) \\
& leaf & 4 positions of every column of a class, BLAKE3, 64-column chunks \\
& Merkle cap & \(2^7\) nodes per tree \\
& fold oracles & every second level; final message \(2^5\) symbols \\
& queries & 153, after a 30-bit proof of work \\
Transcript & hash & BLAKE3 compression function, sequential mode; rejection sampling of 31-bit words \\
Model formats & residual & \(A = [-2^{23}, 2^{23})\), 10 fractional bits \\
& matrix-product inputs and weights & 8 bits, power-of-two exponents per row / column \\
& probabilities & \(\tau_{\exp} \in [0, 2^{10}]\), key blocks of \(2^{12}\) \\
& attention tiles & \(128 \times 128\), on or below the diagonal \\
& weight blocks & \(2^{10} \times 2^{10}\) \\
\bottomrule
\end{tabular}
\caption{Parameters of the evaluated prover.}
\label{tab:params}
\end{table}

\section{Notation}
\label{app:notation}

\begin{table}[h]
\centering\small
\begin{tabular}{@{}ll@{}}
\toprule
Symbol & Meaning \\
\midrule
\(\mle f\), \(\eq\) & multilinear extension; equality polynomial \\
\(\FO\) & the integer function proved (\cref{sec:function}) \\
\(C_W\) & commitment to the quantized weights \\
\(A\) & activation range \([-2^{23}, 2^{23})\) \\
\(\Tab\) & static lookup table \\
\(Z\), \(\beta_i\) & LogUp challenge and tuple-compression coefficients \\
\(S, B\) & sequence length, number of sequences \\
\(L, d, d_{\mathrm{ff}}, d_h\) & layers, width, MLP width, head dimension \\
\(\hat Q, \hat K, V\) & queries and keys after \(\gad{RoPE8}\), values \\
\(\mathrm{acc}\), \(M\), \(P\), \(D\), \(N\) & scores, row maxima, probabilities, denominators, numerators \\
\(k_m\), \(E_m\), \(\delta_m\) & per-row shift, exponent and mantissa of the softmax denominator \\
\(\epsilon_{\mathrm{GKR}}, \epsilon_{\mathrm{sc}}, \epsilon_{\mathrm{LU}}\) & error terms of GKR, sumcheck and lookups \\
\bottomrule
\end{tabular}
\end{table}

\end{document}